\documentclass[a4paper,11pt]{article} 
\pdfoutput=1 
\usepackage{jheppub} 
\usepackage{amsmath,graphicx,hyperref,bm}
\usepackage{xcolor}
\newbox\slashbox \setbox\slashbox=\hbox{$/$}
\newbox\Slashbox \setbox\Slashbox=\hbox{\large$/$}
\def\pFMslash#1{\setbox\@tempboxa=\hbox{$#1$}
  \@tempdima=0.5\wd\slashbox \advance\@tempdima 0.5\wd\@tempboxa
  \copy\slashbox \kern-\@tempdima \box\@tempboxa}
\def\pFMSlash#1{\setbox\@tempboxa=\hbox{$#1$}
  \@tempdima=0.5\wd\Slashbox \advance\@tempdima 0.5\wd\@tempboxa
  \copy\Slashbox \kern-\@tempdima \box\@tempboxa}

\newcommand{\euv}{\epsilon_{\mathrm{UV}}}
\newcommand{\eir}{\epsilon_{\mathrm{IR}}}

\mathchardef\mhyphen="2D
\newcommand{\gae}{\gamma_{\mathrm{E}}}
\newcommand{\mc}{\mathcal}
\newcommand{\mr}{\mathrm}

\newcommand{\be}{\begin{equation}} 
\newcommand{\ee}{\end{equation}} 
\newcommand{\bea}{\begin{eqnarray}} 
\newcommand{\eea}{\end{eqnarray}}

\newcommand{\dg}{\dagger}
\newcommand{\n}{\overline{n}}

\newcommand{\bl}[1]{{\bf{#1}}}

\newcommand{\blp}[1]{{\bf{#1}}_{\perp}}

\newcommand{\nnb}{\nonumber} 

\newcommand{\as}{\alpha_s} 

\newcommand{\eps}{\epsilon} 
\newcommand{\veps}{\varepsilon}

\newcommand{\UV}{\eps_{\mr{UV}}}
\newcommand{\IR}{\eps_{\mr{IR}}}
\newcommand{\mums}{\mu^{2\epsilon}_{\overline{\mathrm{MS}}}}

\newcommand{\cs}{\overline{cs}}

\begin{document}

\title{Consistent treatment of rapidity divergence \\ in soft-collinear effective theory}

\def\KU{Department of Physics, Korea University, Seoul 02841, Korea} 
 
\def\Seoultech{Institute of Convergence Fundamental Studies and School of Liberal Arts, 
Seoul National University of Science and 
Technology, Seoul 01811, Korea}

\author[a]{Junegone Chay}
\emailAdd{chay@korea.ac.kr}
\affiliation[a]{\KU}
\author[b]{Chul Kim}
\emailAdd{chul@seoultech.ac.kr}
\affiliation[b]{\Seoultech}
 
\abstract{ %\vspace{0.1cm}\baselineskip 3.0 ex 
In soft-collinear effective theory, we analyze the structure of rapidity divergence due to the collinear and soft modes 
residing in disparate phase spaces. 
%The approach of an effective theory is suggested to a system of collinear modes with large rapidity and soft modes 
%with small rapidity. It corresponds to integrating out the large-rapidity (collinear) modes to obtain the system with the small-rapidity (soft) modes. We can consider the matching at the rapidity boundary, 
%and the matching procedure is exactly the zero-bin subtraction.  
The idea of an effective theory is applied to a system of collinear modes with large rapidity and soft modes 
with small rapidity. The large-rapidity (collinear) modes are integrated out to obtain the effective theory for
the small-rapidity (soft) modes. The full SCET with the collinear and soft modes should be matched onto the soft theory at the rapidity boundary, 
and the matching procedure becomes exactly the zero-bin subtraction.  
The large-rapidity region is out of reach for the soft mode, which results in the rapidity divergence.  
The rapidity divergence in the collinear sector comes from the zero-bin subtraction, which ensures the cancellation of the rapidity divergences 
from the soft and collinear sectors. In order to treat the rapidity divergence, we construct the rapidity regulators consistently for all the modes.
They are generalized by assigning independent rapidity scales for different collinear directions.
The soft regulator  incorporates the correct directional dependence when the innate collinear directions are not back-to-back, 
which is discussed in the $N$-jet operator.  As an application, we consider  the Sudakov form factor for the back-to-back collinear current and 
the soft-collinear current, where the soft rapidity regulator for a soft quark is developed. We extend the analysis to the boosted heavy quark sector and exploit the delicacy with the presence of the heavy quark mass.
We present the resummed results of large logarithms in the 
form factors for various currents with the light and the heavy quarks, employing the renormalization group evolution on the renormalization and the rapidity scales. 
}

\maketitle

\section{Introduction}

Effective field theories enable us to understand important physics by extracting relevant ingredients and disregarding
the unnecessary remainder. Soft-collinear effective theory (SCET) \cite{Bauer:2000ew,Bauer:2000yr,Bauer:2001yt,Bauer:2002nz} is an effective theory for QCD, 
which is appropriate for high-energy processes with energetic particles immersed in the background of soft particles. We pick up the collinear and 
soft modes to describe high-energy processes, and all the other modes are either integrated out or neglected.  
The degrees of freedom in the effective theory are classified by the phase spaces they reside in. 

Since there are various modes in different phase spaces in SCET, it possesses an additional divergence known as 
the rapidity divergence~\cite{Chiu:2011qc,Chiu:2012ir} in addition to the traditional ultraviolet (UV) and infrared (IR) divergences. 
A boundary in the phase space to separate collinear and soft modes results in the rapidity divergence because these
modes are constrained in different phase spaces. In full QCD, there is no rapidity divergence because there is no kinematic constraint.  
Therefore in SCET there may be rapidity divergence in each sector, but when we sum over all the contributions 
from different sectors, there should be no rapidity divergence. 
This is a good consistency check for the  effective theory. In this respect, the rapidity divergence 
seems to be an artifact in SCET by dissecting the phase space, which disappears in the total contribution.
But it gives a richer structure of the effective theory and 
we can obtain deeper understanding of underlying physics. 

Here we consider a system of the collinear and soft modes with the same offshellness, in which they are distinguished by their rapidities. 
The soft modes have small rapidity,  while the collinear modes have large rapidity.
For the factorization into the soft and collinear parts in SCET, the central idea is to apply the conventional effective theory approach that is 
widely used for separating long- and short-distance physics. 

We first construct an effective theory for the soft mode out of the full SCET.  
The full SCET contains both collinear and soft modes, while the soft theory contains only soft modes with small rapidity.   
By requiring that physics be the same near the rapidity boundary,
the full SCET with both modes is matched onto the soft theory, and produces the Wilson coefficient.
The Wilson coefficient is obtained by subtracting the contribution of the soft theory
from that of the full SCET.  In the matching near the boundary, 
the contribution of the soft theory to be subtracted is exactly the zero-bin contribution in SCET.  
Therefore the collinear contribution with the zero-bin 
subtraction~\cite{Manohar:2006nz} can be interpreted as the Wilson coefficients for the soft theory.

Note that the soft mode cannot resolve the large rapidity region.  
And we expect that there appears the rapidity divergence as the soft momentum approaches 
the rapidity boundary.
Suppose that the collinear mode beyond the boundary is $n$-collinear. Then the rapidity divergence in the soft sector arises 
as the momentum component $\n\cdot k \to \infty$ (and $n\cdot k \to 0$).  On the other hand, the naive collinear contribution before the zero-bin 
subtraction has no corresponding rapidity divergence since $\n\cdot k$ is bounded from above. 
However the true collinear part after the matching, which includes the zero-bin subtraction, contains the rapidity divergence with the same origin as 
the soft part.  Therefore the rapidity divergences in the collinear and the soft sectors have the opposite sign, which ensures the cancellation of 
the rapidity divergence when both are combined. 

The naive collinear part might contain another type of the rapidity divergence as $\overline{n}\cdot k \rightarrow 0$.  
But that region is shared with the soft part and this type of rapidity divergence is cancelled in the matching. 
And the true rapidity divergence with $\n\cdot k \to \infty$ in the
collinear sector is recovered by the zero-bin subtraction, similar to the pullup mechanism~\cite{Manohar:2000kr,Hoang:2001rr}. 
As a result, the soft-collinear factorization with the zero-bin subtraction is identified as the matching of the full SCET onto the soft part, 
and the collinear part can be considered as the matching coefficient describing the large-rapidity region.

The main issue of this paper is to implement this idea for the consistent treatment of the rapidity divergence in SCET. The first step is to
establish a proper method for regulating the rapidity divergences in the collinear and the soft 
sectors. We prescribe the rapidity regulators in both sectors, originating from the same source.

In addition to the conventional regularization method~\cite{Chiu:2011qc,Chiu:2012ir}, 
there have been many suggestions to regulate the rapidity divergence, such as the use of the Wilson lines off the lightcone~\cite{Collins:2011zzd}, 
the $\delta$-regulator~\cite{Idilbi:2007ff,Idilbi:2007yi}, the analytic regulator~\cite{Becher:2011dz},  the exponential regulator~\cite{Li:2016axz}, 
and the pure rapidity regulator~\cite{Ebert:2018gsn}, to name a few. The gauge invariance and the consistency in power
counting to all orders have also been recent issues in selecting appropriate rapidity regulators~\cite{Chiu:2012ir,Chiu:2009yx,Ebert:2018gsn}. 
Here we present a new prescription for the rapidity regulator based on where the rapidity divergence arises, as discussed earlier. It may look like an addition to
a big list of regulators, but we hope to convince the reader that our prescription is on a sound physical basis.

The construction of the rapidity regulators is interwoven between the collinear and the soft sectors, and let us first look at the Wilson lines in SCET.
The emission of collinear or soft gluons from collinear, energetic particles is eikonalized and exponentiated to 
all orders to yield the collinear and soft Wilson lines by integrating out large offshell modes.  
The collinear and the soft Wilson lines in the lightlike $n$ direction are written as \cite{Bauer:2001yt}
\begin{equation} \label{origWil}
W_n = \sum_{\mathrm{perm.}} \exp\Bigl[-g \frac{\overline{n}\cdot A_n}{\overline{n}\cdot \mathcal{P}} \Bigr],  \, \, 
S_n =  \sum_{\mathrm{perm.}} \exp\Bigl[-g \frac{n\cdot A_s}{n\cdot \mathcal{P}} \Bigr],
\end{equation}
where $A_n$ ($A_s$) is the collinear (soft) gauge field.  The lightlike vectors $n$ and $\overline{n}$ satisfy $n^2 =\overline{n}^2 =0$, 
$n\cdot \overline{n}=2$, and $\overline{n}\cdot \mathcal{P}$ ($n\cdot \mathcal{P}$) is the operator extracting the incoming momentum 
component in the $n$ ($\overline{n}$) direction. 

Note that the nature of the eikonalization is different in both cases. For the collinear Wilson line $W_n$, we consider the emission
of the $n$-collinear gluons from all the remaining parts except the $n$-collinear particle in consideration, that is, 
from the $\overline{n}$-collinear field in a back-to-back current or from the heavy quark
in a heavy-to-light current. For the $N$-jet operator, we consider the $n$-collinear gluons emitted from all the collinear particles except the $n$-collinear particle.
Whatever the sources are, when the intermediate states are integrated out, and the leading terms are taken, 
we obtain the collinear Wilson line which depends on $\overline{n}\cdot \mathcal{P}$. On the other hand, the soft Wilson line $S_n$ is obtained
by the emission of soft gluons from an $n$-collinear field, and the intermediate states are integrated out. Note that the source of the gluon 
emission is different. Therefore to extract the rapidity divergences in the soft 
and the collinear sectors correctly, we should trace the $n$-collinear gluons  in the collinear sector,  
and also the soft gluons from the same source as in the collinear sector, and take the $n$-collinear limit. 
Simply put, the rapidity matching does not happen 
between $W_n$ and $S_n$, but happens between $W_n$ and $S_{\bar{n}}$ for the back-to-back current.  

Let us take an example of the back-to-back current $\overline{\xi}_{\bar{n}} W_{\bar{n}} S^{\dagger}_{\bar{n}} \gamma^{\mu} S_{n} 
W^{\dagger}_n \xi_n$, which will be generalized later, and consider a soft gluon emitted from the soft Wilson line, not from $S_n$, 
but from $S^{\dagger}_{\bar{n}}$. In order to consider the matching with the $n$-collinear sector, in which the $n$-collinear gluon is 
emitted from the $\overline{n}$-collinear sector, we take 
the limit in which the component $\overline{n}\cdot k$ of the soft momentum becomes large, compared to other components with $k_{\perp}^2$ fixed. 
And it is taken to infinity in the soft sector because the large scale is beyond reach of the soft particles.  Therefore the region 
$\overline{n}\cdot k\rightarrow \infty$ is where the possible rapidity divergence occurs and we apply the rapidity regulator to extract it.
We can choose the rapidity regulator in the form $(\nu/\overline{n}\cdot k)^{\eta}$, where $\nu$ is the rapidity scale introduced and the
rapidity divergence appears as poles in $\eta$. In order to be consistent, we also choose the same rapidity regulator in the $n$-collinear sector 
since the rapidity region modified by the regulator should be the same in the overlapping region. 

We can also include the collinear currents which are not back-to-back, or even the $N$-jet operator in which there are well-separated $N$ collinear 
directions. We emphasize that the same rapidity regulator should be employed both in the collinear and soft sectors. 
When the rapidity divergence in one collinear sector is to be matched, the soft gluons emitted from other collinear directions are selected and 
the collinear limit in the given collinear direction is taken. In this process, the directional dependence in the soft sector is correctly produced.
Furthermore, we can assign a different rapidity scale to each collinear direction if there is a hierarchy of scales in different directions. 
The total contribution is free of every rapidity divergence associated with each collinear direction.
The rapidity regulator for the soft-collinear current  can be also consistently constructed using this method. 

%When the collinear and the soft particles have different offshellness, the collinear and the soft particles are distinguished 
%by their offshellness, not by their rapidity. The effective theory in this situation is called $\mathrm{SCET}_{\mathrm{I}}$. There is no 
%rapidity divergence in each sector because collinear and (u)soft particles do not overlap. In practice, the rapidity divergences 
%in the virtual correction and in the real emission cancel. But in the Sudakov form factor or in the $N$-jet operator, in which there is
%no contribution from real emissions, the rapidity divergence can be present in each sector. When the offshellness of the collinear and 
%the soft particles  is of the same magnitude, they should be distinguished by their rapidities and it is described 
%by $\mathrm{SCET}_{\mathrm{II}}$.  The rapidity divergence appears in each sector, though their sum cancels. But it affects the evolution 
%of the collinear and the soft sectors, which is one of the main results of the paper.

Our analysis on the rapidity divergence can be extended to the boosted heavy quark sector. 
Here the energy of the heavy quark, $Q$, is still much larger than the quark mass $m$. Hence the decoupling of the soft interactions 
from the heavy quark can be implemented in the same manner as in the case of the light quark. Since the origin of the rapidity 
divergence is soft dynamics, the divergence in the heavy quark sector also comes from the zero-bin subtraction.
% and it is given by  the same one as the light quark case.      ---- I commented this out.
This is another new observation we make.  We  consider the resummation of large logarithms in the Sudakov form factor 
for the heavy-to-heavy current.  Like the light quark case, the rapidity scale evolution is necessary. 
And it enables us to fully resum the large logarithms of  $Q/m$.  

The structure of the paper is as follows: In Section~\ref{anrapdiv}, we discuss the idea of applying an effective theory to a system with
the collinear and soft modes, and explain that the rapidity divergence in the collinear part comes from 
the matching procedure that corresponds to the zero-bin subtraction.
And we show how to set up the rapidity regulator with the zero-bin subtraction in the collinear sector. 
The soft rapidity regulator is defined by employing the same principle for the collinear sector, but 
with the appropriate expression for the soft Wilson line. In Section~\ref{sudakovf}, the Sudakov form factor is analyzed 
for the back-to-back current. We suggest how to implement the rapidity regulator in  
the $N$-jet operator in Section~\ref{njet}. In Section~\ref{sccurrent}, we consider the Sudakov form factor for the soft-collinear current, 
which is compared to the result by boosting the back-to-back current.  In Section~\ref{sudheavy}, our analysis is extended to the heavy-to-heavy
and heavy-to-light currents, in which the heavy quark mass sets another hierarchy in factorizing the form factors.
Finally we conclude and describe the outlook in Section~\ref{con}.  
In Appendix~\ref{sudakov}, we resum the large logarithms in the Sudakov form factor for various currents 
using the renormalization group (RG) equation with respect to the renormalization and rapidity scales. 
The evolutions and the resummation are performed to next-to-leading logarithmic (NLL) accuracy. 
In Appendix~\ref{appB}, we show the details of the calculations for extracting the rapidity divergences in the boosted heavy quark sector. 
 
\section{Rapidity divergence and the zero-bin subtraction \label{anrapdiv}}

We start with the matrix element of the back-to-back collinear current in SCET 
\be
\label{EFTcur}
V_{\mr{SCET}}^{\mu} = \langle p |~\bar{\xi}_n W_n \tilde{S}_n^{\dagger}\gamma_{\perp}^{\mu} S_{\bar{n}} 
W_{\bar{n}}^{\dagger} \xi_{\bar{n}}~|p'\rangle,
\ee
where $p$ and $p'$ are the on-shell momenta of the collinear quarks in the $n$- and $\n$-directions respectively. 
The soft Wilson lines $\tilde{S}_n~(S_{\bar{n}})$ are present by redefining the collinear fields, which results in the decoupling of the soft 
interactions. The notations follow the convention used in Ref.~\cite{Chay:2004zn}.
%, depending on the originating collinear  particles.   --- I commented this out.
In higher-order corrections, the UV, IR, and rapidity divergences are produced, in which  
the UV and the IR divergences are regulated by dimensional regularization. If we introduce a nonzero gluon mass $M$, 
the IR divergence is regulated by the mass. However, a new regulator is needed to regularize the rapidity divergence.

The widely-used rapidity regulator has been suggested in Refs.~\cite{Chiu:2011qc,Chiu:2012ir}, by modifying 
the original collinear and soft Wilson lines as 
\begin{equation} \label{wno}
W_n = \sum_{\mathrm{perms}} \exp \Bigl[ -\frac{g}{\overline{n}\cdot \mathcal{P}} \frac{\nu^{\eta}}{|\overline{n}\cdot 
\mathcal{P}_g|^{\eta}} \overline{n} \cdot A_n\Bigr],  \
S_n = \sum_{\mathrm{perms}} \exp \Bigl[ -\frac{g}{n\cdot \mathcal{P}} \frac{\nu^{\eta/2}}{|2\mathcal{P}_{g3}|^{\eta/2}} n\cdot A_s 
\Bigr]. 
\end{equation} 
Here the rapidity scale $\nu$ is introduced and the rapidity divergence appears as poles in $\eta$. 
As described in Introduction, we use the rapidity regulator of the form $(\nu/\overline{n}\cdot k)^{\eta}$ for the $n$-collinear sector, which
is the same as the prescription in Eq.~\eqref{wno} for $W_n$.  
%However, here we construct the soft rapidity regulator which is the same as the collinear rapidity regulator, but applied to $S_{\bar{n}}$. 
Then we construct the soft rapidity regulator to be consistent with the collinear rapidity regulator. This is applied to $S_{\bar{n}}$. 
Therefore the soft rapidity regulator is different from that in Eq.~\eqref{wno}, and needs further
explanation. 

\subsection{Effective theory approach to treating the  rapidity divergence}

Let us first consider the general argument in treating the rapidity divergence. In $\mr{SCET_{II}}$ where the collinear and the soft modes have 
the same offshellness, we distinguish these modes by their rapidities. In radiative corrections, there appears the integral of the form
\be
\label{intrkp}
I=\int^{p_+}_{\mu_L} \frac{dk_+}{k_+}=\int^{\Lambda}_{\mu_L} \frac{dk_+}{k_+}+\int^{p_+}_{\Lambda} \frac{dk_+}{k_+},
\ee
where $p_+$ is a hard momentum, and $\mu_L$ is a soft scale. In the final expression, the integral
is divided into the soft and collinear integrals by an arbitray scale  $\Lambda$  which 
separates the soft and collinear regions. 

In the spirit of the zero-bin contribution, the integral $I$, by rearranging the second term, can be written as  
\be
\label{intsc1} 
I=\int^{\Lambda}_{\mu_L} \frac{dk_+}{k_+}+\left(\int^{p_+}_{0} \frac{dk_+}{k_+} - \int^{\Lambda}_{0} \frac{dk_+}{k_+}\right).
\ee
The first term in the parenthesis is the naive collinear contribution, and the second term is the zero-bin contribution. 
We clearly see that double counting is avoided in the collinear part.  And if there is any divergence in the naive collinear contribution as $k_+\to 0$, 
it is removed by the zero-bin subtraction. 
It guarantees the factorization to secure the independence of the collinear sector as stressed in Ref.~\cite{Chiu:2009yx}.
The cutoff $\Lambda$ appears in the soft integrals, and in the zero-bin contribution. 
And it can be taken to infinity as far as the soft modes are concerned.
Then with the rapidity regulator, Eq.~\eqref{intsc1} can be expressed as\footnote{
Some readers may wonder if the pole $1/\eta$ can (mathematically) regularize the pole at $k_+=0$  as well in  the zero-bin contribution 
in Eq.~\eqref{intsc2}. But the purpose of the $\eta$-regulator is to capture the divergence as $k_+\to \infty$, meaning that we take $\eta$ 
to be slightly positive, i.e., $\eta = +0$. It does not regularize  the divergence for $k_+=0$.  We may introduce another $\eta^{\prime}=-0$ 
to regulate the divergence as $k_+\to 0$ as in pure dimensional regularization. However,  the divergence for $k_+=0$ cancels and we 
employ the regulator only for $k_+ \to \infty$. } 
\be
\label{intsc2} 
I=\nu^{\eta}\int^{\infty}_{\mu_L} \frac{dk_+}{k_+^{1+\eta}}+\left(\nu^{\eta}\int^{p_+}_{0} \frac{dk_+}{k_+^{1+\eta}} -
 \nu^{\eta}\int^{\infty}_{0} \frac{dk_+}{k_+^{1+\eta}}\right).
\ee
The soft and collinear parts in Eq.~\eqref{intsc2} are given as 
\begin{align}
\label{Is}
I_S (\mu_L,\nu)  &=  \nu^{\eta}\int^{\infty}_{\mu_L} \frac{dk_+}{k_+^{1+\eta}} = \frac{1}{\eta} + \ln\frac{\nu}{\mu_L}, \\
\label{Ic} 
I_C (p_+,\nu)  &=  \nu^{\eta}\int^{p_+}_{0} \frac{dk_+}{k_+^{1+\eta}} - \nu^{\eta}\int^{\infty}_{0} \frac{dk_+}{k_+^{1+\eta}} 
= -\frac{1}{\eta} - \ln\frac{\nu}{p_+}.
\end{align}
The rapidity divergence and the $\nu$ dependence in the soft and collinear sectors cancel when they are combined. 
But  the evolutions with respect to the rapidity scale $\nu$ to $\mu_L$ for the soft sector and to $p_+$ 
for the collinear sector are necessary for the resummation of the large logarithms in $p_+/\mu_L$~\cite{Chiu:2011qc,Chiu:2012ir}.

Note that the separation into the soft and collinear parts in Eq.~\eqref{intsc2} is similar to separating the long- and short-distance physics 
in effective theories. The effective theory at lower energy is matched at the cutoff scale to the full theory, yielding the Wilson coefficients. 
The same mechanism applies to  Eq.~\eqref{intsc2}.  Consider all the modes with the same offshellness, and there is a cutoff rapidity which
distinguishes the soft modes with small rapidity and the collinear modes with large rapidity. We match the two regions near the cutoff, 
which is accomplished by the zero-bin subtraction. It yields the Wilson coefficients, which corresponds to the collinear contribution in this case,
when the contributions from 
the large rapidity (collinear) region and the small rapidity (soft) region are matched at the boundary. 
We emphasize that the idea of an effective theory is applied to the description of the rapidity divergence, though we do not explicitly construct an effective soft theory.

As a consequence, the rapidity divergence arises entirely due to the fact that the soft part cannot describe large-rapidity physics. And the collinear 
part contains the divergence through matching onto the soft theory with small rapidity, i.e., the zero-bin subtraction. 
There is always one-to-one correspondence for the rapidity divergences between the soft and the collinear sectors.

\subsection{Collinear contribution and the zero-bin subtraction} 

\begin{figure}[t]
\begin{center}
\includegraphics[height=4.5cm]{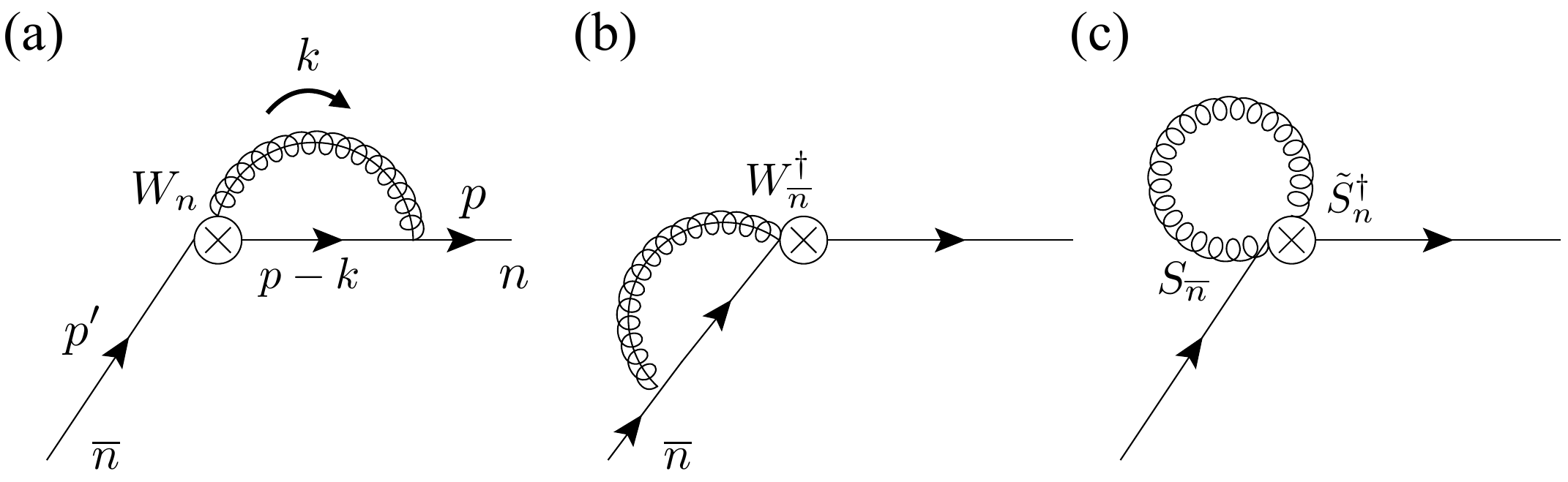}
\end{center}
\vspace{-0.5cm}
\caption{\label{fig1}
Feynman diagrams for the back-to-back collinear current  at one loop. 
}
\end{figure}

The matrix element in Eq.~\eqref{EFTcur} is factorized into the $n$-, $\n$-collinear and the soft parts. 
Let us first consider the $n$-collinear contribution at one loop. The corresponding Feynman diagram is shown in Fig.~\ref{fig1}-(a). 
The naive collinear contribution is given by
\begin{equation} \label{naivec}
\tilde{M}_n = 2ig^2 C_F \mums \nu^{\eta} \int \frac{d^D k}{(2\pi)^D} \frac{\overline{n}\cdot (p-k)}{(k^2 -2 p\cdot k+i\veps)  (k^2 -M^2+i\veps)
(\overline{n}\cdot k)^{1+\eta}},
\end{equation}
where $\mums = (\mu^2 e^{\gamma_{\mathrm{E}}}/4\pi)^{\eps}$ in the $\overline{\mathrm{MS}}$ scheme and we employ 
the rapidity regulator in Eq.~\eqref{wno}. The massless fermion is on shell ($p^2 =0$), and the nonzero gluon mass $M$ is inserted as 
an IR regulator, or the real gauge boson mass in electroweak processes. Performing the contour integral on $n\cdot k$, we obtain 
\begin{align} 
\tilde{M}_n &=   -\frac{\alpha_s C_F}{2\pi} \frac{(\mu^2e^{\gamma_{\mathrm{E}}})^{\eps}\nu^{\eta}}{\Gamma(1-\eps)}
\int^{p_+}_0 dk_+ k_+^{-1-\eta} \Bigl(1-\frac{k_+}{p_+}\Bigr) 
\int^{\infty}_0 \frac{d\blp{k}^2 (\blp{k}^2)^{-\eps}}{\blp{k}^2+\frac{p_+-k_+}{p_+}M^2} \nnb \\
&=  -\frac{\alpha_s C_F}{2\pi} e^{\gamma_{\mathrm{E}} \eps} \Gamma (\eps) \Bigl(\frac{\mu^2}{M^2}\Bigr)^{\eps} 
\Bigl(\frac{\nu}{p_+}
\Bigr)^{\eta} \int_0^1 dx (1-x)^{1-\eps} x^{-1-\eta},
\label{naivec2}
\end{align}
where $p_+\equiv \n\cdot p$ is the largest component of the external momentum $p$, and $x=k_+/p_+$.

In Eq.~\eqref{naivec2}, the pole in $\eta$ comes from the region $k_+ \to 0$. Since the momentum $k_+$ of the collinear gluon has an upper limit 
$p_+$, there is no rapidity divergence as $k_+ \to \infty$ in this naive collinear amplitude by itself. 
However, as we have explained in Introduction, it is cancelled when we subtract the zero-bin contribution because 
the soft sector shares the same phase space. Due to the zero-bin subtraction, the rapidity divergence in the collinear sector is pulled up to the
divergence for $k_+ \to \infty$.

The zero-bin contribution is obtained from Eq.~(\ref{naivec2}) by taking the limits $k_+\ll p_+$ or $x \rightarrow 0$ in the integrand, and 
the upper limit in the integral of $x$ to infinity. It is given by
\be \label{cozero}
M_n^{\varnothing} = -\frac{\alpha_s C_F}{2\pi} e^{\gamma_{\mathrm{E}} \eps} \Gamma (\eps) \Bigl(\frac{\mu^2}{M^2}\Bigr)^{\eps} 
\Bigl(\frac{\nu}{p_+} \Bigr)^{\eta}  \int_0^{\infty} dx  x^{-1-\eta}.  
\ee
Then the legitimate collinear contribution is obtained by subtracting the zero-bin contribution, Eq.~\eqref{cozero}, from the naive collinear 
contribution, Eq.~\eqref{naivec2}. It is given as 
\begin{align}
M_n &=  \tilde{M}_n - M_n^{\varnothing} \nnb\\
&=  -\frac{\alpha_s C_F}{2\pi} e^{\gae \eps} \Gamma (\eps) \Bigl(\frac{\mu^2}{M^2}\Bigr)^{\eps}
\Biggl[\int^1_0 \frac{dx}{x} \Bigl((1-x)^{1-\eps}-1\Bigr) - \Bigl(\frac{\nu}{p_+} \Bigr)^{\eta} \int^{\infty}_1 \frac{dx}{x^{1+\eta}}\Biggr] \nnb \\
\label{Man}
&= \frac{\alpha_s C_F}{2\pi}  \Bigl[\Bigl( \frac{1}{\eps} +\ln \frac{\mu^2}{M^2} \Bigr) \Bigl( \frac{1}{\eta} 
+\ln \frac{\nu}{\overline{n}\cdot p} +1\Bigr) + 1-\frac{\pi^2}{6}\Bigr].
\end{align}
In the second line, we divide the integration region of the zero-bin contribution into $x\in[0,1]$ and $x\in[1,\infty]$. Then the integral for 
 $x\in[0,1]$ is combined with the naive collinear contribution $\tilde{M}_{n}$. Note that there is no divergence in the integral 
with $x\in[0,1]$, and we put $\eta =0$. 
%The $\eta$-regulator is employed in the second integral, where $k_+$ (or $x$) goes to infinity.  
Finally the correct rapidity divergence in the collinear sector is captured through the zero-bin subtraction.

\subsection{Rapidity regulator in the soft sector}
\label{sftreg}
 
\begin{figure}[b]
\begin{center}
\includegraphics[height=3.8cm]{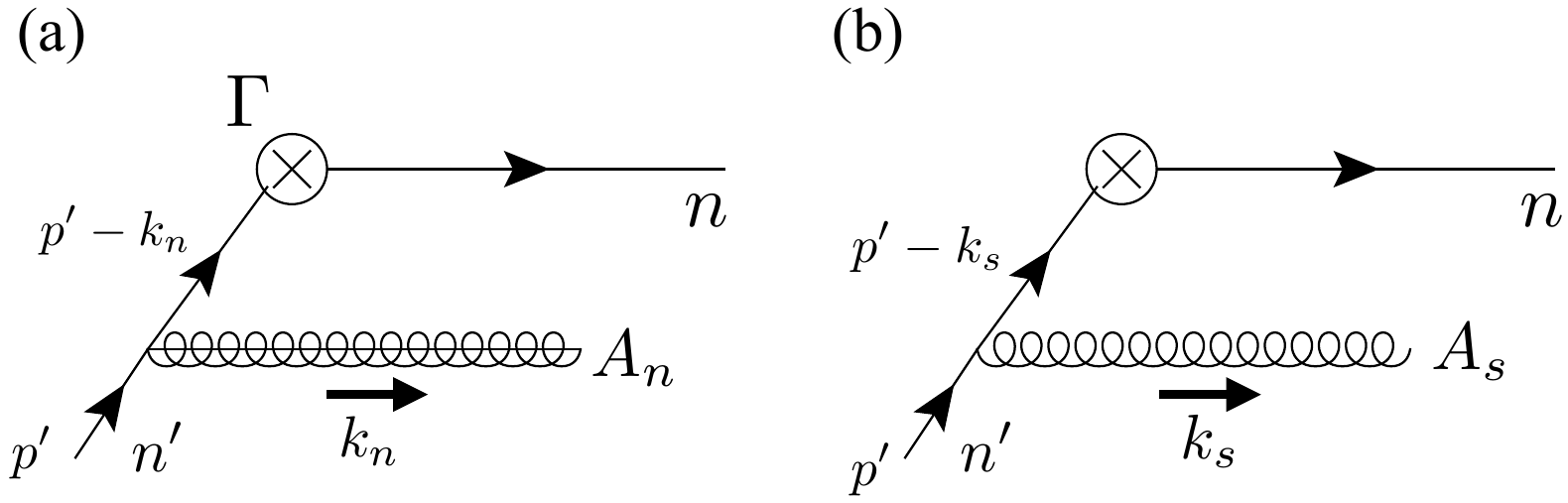}
\end{center}
\vspace{-0.5cm}
\caption{\label{fig2} \baselineskip 3.5 ex
(a) Collinear and (b) soft gluon emissions from the $n'$-collinear sector in the current $\bar{q}_{n} \Gamma q_{n'}$,
which yield the leading contributions to $W_n$ and $S_{n'}$ respectively.
If $\n\cdot k_s \to \infty$ and $n\cdot k_s \to 0$ in the soft phase space, $k_s$ becomes the soft version of the $n$-collinear momentum $k_n$, 
and the rapidity divergence arises when we separate the soft and collinear gluons in SCET. }
\end{figure}

We have to find a consistent rapidity regulator in the soft sector that conforms to the regulator in the collinear sector. Here we extend the
collinear current to $\overline{q}_n \Gamma q_{n'}$, where the lightcone directions $n$ and $n'$ are not necessarily back-to-back, 
but $n\cdot n' \sim\mc{O}(1)$.
Let us consider the configuration of Fig.~\ref{fig2}, in which a collinear or a soft gluon is emitted.
In Fig.~\ref{fig2}-(a), a collinear gluon from the $n'$-collinear quark $q_{n'}$ is emitted 
in the $n$ direction. It produces the collinear Wilson line $W_n$ 
at first order.
 
The same configuration is exhibited in Fig.~\ref{fig2}-(b) where a soft gluon is emitted,  producing  
the soft Wilson line $S_{n'}$ at first order. The momentum of the soft gluon scales as 
$(\n\cdot k, \blp{k},n \cdot k) = Q(\lambda,\lambda,\lambda)$ with a large scale $Q$. 
But the soft sector is an exact copy of QCD when $Q$ is taken to infinity. Therefore when the soft momentum is in the corner of phase 
space with $\n\cdot k \to \infty$, $n\cdot k \to 0$, it approaches the $n$-collinear momentum. It is
the region where the rapidity divergence occurs. 
It means that the soft rapidity regulator associated with the $n$-collinear sector should be implemented in $S_{n'}$, not $S_n$.
 
The collinear rapidity regulator for $W_n$ in Eq.~\eqref{wno} for Fig.~\ref{fig2}-(a) is given by
$(\nu/\n\cdot k)^{\eta}$. And the rapidity divergence shows up as poles of $1/\eta$ with $\n\cdot k \to \infty$. It will be consistent 
to use the same rapidity regulator for the soft part as $(\nu/\n\cdot k)^{\eta}$. However, the form of the Wilson lines $W_n$ and 
$S_{n'}$ is determined by the power counting of the collinear and soft momenta.  The collinear momentum and the collinear gauge field scale 
as $Q (1,\lambda, \lambda^2)$, while the soft momentum and the soft gauge field scale as $Q(\lambda, \lambda, \lambda)$. 
In order to take a consistent rapidity regulator in the collinear and the soft sectors, we choose the configuration of  Fig.~\ref{fig2}-(b) and 
take the $n$-collinear limit.

The soft Wilson line $S_{n'}$ is written as
\be
S_{n'} =  \sum_{\mathrm{perm.}} \exp\Bigl[-g \frac{1}{n'\cdot \mathcal{P}}n'\cdot A_s \Bigr],
\ee
where $n'\cdot \mc{P}$ returns the incoming momentum of the soft gluon with $p'\cdot k_s \sim \overline{n}'\cdot p' n'\cdot k_s$ and 
$p'\cdot A_s \sim \overline{n}'\cdot p' n'\cdot A_s$ at leading order, where $p'$ is the $n'$-collinear momentum. 
When the internal particle is integrated out in 
Fig.~\ref{fig2} (b), we obtain $gn'\cdot A_s/n'\cdot k$. On the other hand, if we take the $n$-collinear limit of 
the soft momentum $k$, it becomes $k^{\mu} \approx (\overline{n}\cdot k) n^{\mu}/2$. In this limit 
$n'\cdot k = (\overline{n}\cdot k)(n\cdot n')/2$.  Therefore the soft rapidity regulator to capture the divergence as 
$\overline{n}\cdot k \to \infty$ is given by
\be
\Bigl(\frac{\nu}{n'\cdot k} \cdot \frac{n\cdot n'}{2}\Bigr)^{\eta}
 \xrightarrow[\n\cdot k \to \infty]{} \Bigl(\frac{\nu}{\overline{n}\cdot k}\Bigr)^{\eta}. 
\ee 
The important point in taking this limit is to express the original regulator $(\nu/\n \cdot k)^{\eta}$ in terms of $n'\cdot k$, with which 
the soft Wilson line $S_{n'}$ is expressed.
As a consequence, we suggest that the soft rapidity regulator for Fig.~\ref{fig2}-(b) is given by
\be
\label{regol}
\Bigl(\frac{\nu}{n'\cdot k} \frac{n'\cdot n}{2}\Bigr)^{\eta},  
\ee
because it corresponds to the regulator $(\nu/\n\cdot k)^{\eta}$ in the limit $\n\cdot k \to \infty$. 
Accordingly, the soft Wilson line is modified as
\be
\label{mSnp}
S_{n'} =  \sum_{\mathrm{perm.}} \exp\Bigl[-g \frac{1}{n'\cdot \mathcal{P}} \Bigl(\frac{\nu}{|n'\cdot \mc{P}|} \frac{n'\cdot n}{2}\Bigr)^{\eta}
n'\cdot A_s \Bigr]. 
\ee
For $S_n$, we switch $n$ and $n'$. These soft Wilson lines appear in the collinear current  $\overline{q}_n \Gamma q_{n'}$.

Consider an $N$-jet operator with one $n$-collinear operator, and the remaining $(N-1)$ $n_i$-collinear operators ($i=1, \cdots, N-1$), 
in which we are interested in the rapidity divergence associated with the $n$ direction. For each $n_i$ direction, we can modify $S_{n_i}$ using
different rapidity regulators with $\eta_i$ and $\nu_i$ in the form
\be
\Bigl(\frac{\nu_i}{|n_i\cdot \mc{P}|} \frac{n\cdot n_i}{2}\Bigr)^{\eta_i}.
\ee
It properly captures the rapidity divergence in the $n$ direction when a soft gluon is radiated from 
the $n_i$-collinear sector to the $n$ direction in the limit $\overline{n}\cdot k \rightarrow \infty$. Note that each separate rapidity scale $\nu_i$
can be assigned to each $n_i$ direction, and the corresponding rapidity divergences are cancelled when the collinear and soft contributions are added.
We will discuss the $N$-jet operator in more detail in Section~\ref{njet}.

\section{Sudakov form factor for the back-to-back current\label{sudakovf}}

\subsection{Soft one-loop contribution to the back-to-back collinear current}

We now return to the back-to-back collinear current in Eq.~\eqref{EFTcur}, and consider its soft  contribution at one loop, shown 
in Fig.~\ref{fig1}-(c). Due to the presence of $\tilde{S}_{n}^{\dagger}$ and $S_{\bar{n}}$ in the current, the soft contribution
contains the factor 
\be
\label{fbtob}
\frac{1}{(n\cdot k) (\n\cdot k)}.
\ee 
It provides two types of rapidity divergences as $\n\cdot k \to \infty$, $n\cdot k \to 0$ in the $n$-collinear sector, and the rapidity divergence 
as $n\cdot k \to \infty$, $\n\cdot k \to 0$ in the $\n$-collinear sector, while  $k_L^2 \equiv \n\cdot k n\cdot k \sim \blp{k}^2$ remains fixed. 
The rapidity divergence is not regulated by the dimensional regularization because it appears irrespective of the UV  
$(\blp{k}^2 \to \infty)$, or the  IR $(\blp{k}^2 \to 0)$ limits. 
 
From Eq.~\eqref{mSnp}, the soft Wilson lines $\tilde{S}_{n}^{\dagger}$ and $S_{\bar{n}}$ with the rapidity regulator are written as
\bea
\label{mSndb}
\tilde{S}_{n}^{\dg} &=&  \sum_{\mathrm{perm.}} \exp\Bigl[-g n\cdot A_s \frac{1}{n\cdot \mathcal{P}^{\dg}+i\veps} 
\Bigl(\frac{\nu_-}{|n\cdot \mc{P}^{\dg}|} \Bigr)^{\eta_-}\Bigr],\\
\label{mSnbb}
S_{\bar{n}} &=&  \sum_{\mathrm{perm.}} \exp\Bigl[-g \frac{1}{\n\cdot \mathcal{P}+i\veps} \Bigl(\frac{\nu_+}{|\n\cdot \mc{P}|} 
\Bigr)^{\eta_+} \n\cdot A_s \Bigr],
\eea
with $n\cdot \n =2$. 
The $i\veps$-prescription follows from Ref.~\cite{Chay:2004zn}.  We introduce the independent rapidity regulator for 
each collinear direction. The rapidity regulator with $\eta_-$ in Eq.~\eqref{mSndb} regulates the divergence in the $\n$ direction, and 
the one with $\eta_+$ in Eq.~\eqref{mSnbb}  in the $n$ direction, with the corresponding two rapidity scales, $\nu_{\mp}$.  

The soft one-loop contribution, before the regulator is inserted, is given as 
\begin{align}
\tilde{M}_S &=  -2ig^2 C_F \mu_{\mr{\overline{MS}}}^{2\eps} \int \frac{d^Dk}{(2\pi)^D} 
\frac{1}{(k^2-M^2+i\veps)(k_+-i\veps)(k_- - i\veps)} \nnb \\
&=  -2 g^2 C_F \mu_{\mr{\overline{MS}}}^{2\eps} \int \frac{d^{D-1}k}{(2\pi)^{D-1}} 
\frac{1}{2k_0} \frac{1}{k_+ k_-} \Bigl|_{k_0 = \sqrt{{\bf k}^2 + M^2}}  \nnb \\
&=  -4\pi g^2 C_F \mu_{\mr{\overline{MS}}}^{2\eps} \int \frac{d^Dk}{(2\pi)^D} 
\frac{\delta(k^2-M^2)}{k_+ k_-} \Theta(k_0) \nnb \\
\label{MS0}
&=  -\frac{\as C_F}{2\pi} \frac{(\mu^2e^{\gamma_{\mathrm{E}}})^{\eps}}{\Gamma(1-\eps)} \int^{\infty}_0 dk_+ \int^{\infty}_{M^2/k_+}dk_-
\frac{(k_+ k_- - M^2)^{-\eps}}{k_+ k_-},
\end{align}
where we first perform the contour integral on $k_0$, with the relation
\be
\frac{1}{2\sqrt{{\bf k}^2 + M^2}} = \int dk_0 \delta(k^2-M^2) \Theta(k_0).
\ee
The rapidity regulators do not affect the pole structures of the contour integral, and are dropped for the moment.
\begin{figure}[t]
\begin{center}
\includegraphics[height=7cm]{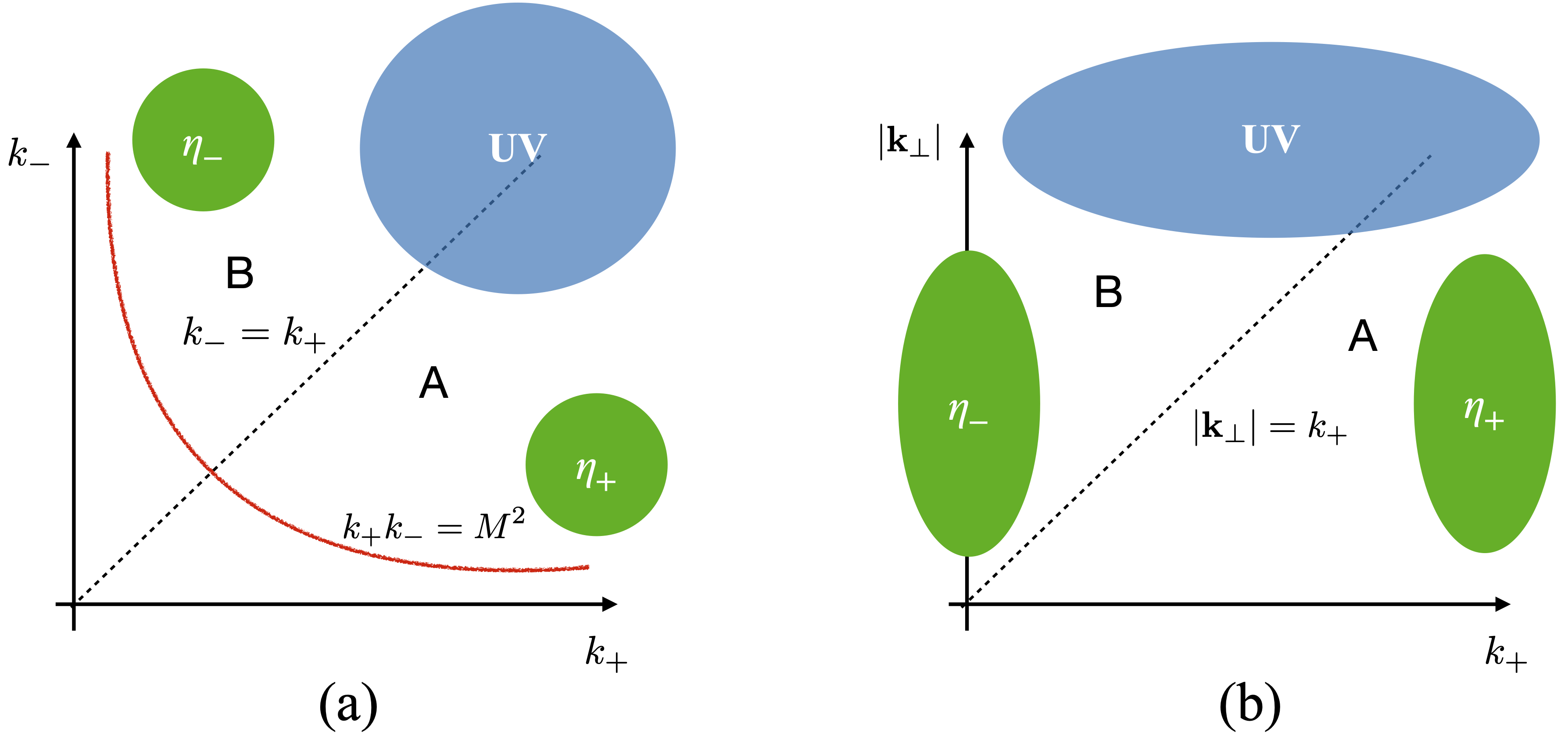}
\end{center}
\vspace{-0.5cm}
\caption{\label{fig3}\baselineskip 3.5ex
Structure of the phase space for the soft gluon in the back-to-back current. The rapidity divergence in the $n~(\n)$ direction arises
in the green region with $\eta_+~(\eta_-)$. (a) The phase space in the $k_+$-$k_-$ plane 
with nonzero gluon mass $M$. The (red) curve $k_+ k_- =M^2$ is the IR cutoff. (b)  The phase space 
in $k_+$-$|\blp{k}|$ plane with massless gluons. The IR divergence arises at $|\blp{k}|=0$. }
\end{figure}

Now we assign the rapidity regulators according to Eqs.~\eqref{mSndb} and \eqref{mSnbb}. 
As can be seen in Fig.~\ref{fig3} (a), the green regions in the phase space, where the rapidity divergences arise, are 
well separated. Therefore, for practical purposes, it is convenient to divide the phase space by the line $k_+=k_-$ in the $k_+$-$k_-$ plane. 
Then, we can employ the regulator from $S_{\bar{n}}$ only in the region for $k_+>k_-$ , while we employ the regulator from $S_{n}$ only 
in the region for $k_+<k_-$ because the omitted regulators produce no rapidity divergence. As a result, the
rapidity regulator at one loop can be written as\footnote{
The regulator in Eq.~\eqref{rsorig} is not new, and has been also proposed in Ref.~\cite{Jaiswal:2015nka}, 
where different jet vetoes along the beam directions have been considered.} 
\begin{equation} \label{rsorig}
R_S (\nu_+,\nu_- ) = \displaystyle \Bigl(\frac{\nu_+}{k_+}\Bigr)^{\eta_+} \Theta(k_+-k_-)
+\Bigl(\frac{\nu_-}{k_-}\Bigr)^{\eta_-} \Theta (k_- - k_+).
\end{equation}

It amounts to specifying independent rapidity scales for different collinear directions.
If we set $\eta=\eta_+=\eta_-$ and $\nu=\nu_+=\nu_-$, we obtain the same result 
using the regulator proposed in Refs.~\cite{Chiu:2011qc,Chiu:2012ir}, in which the soft regulator is written as 
\be
\label{rchiu}
R'_S (\nu) = \Bigl(\frac{\nu}{|k_+-k_-|}\Bigr)^{\eta}.
\ee
As $k_+\to \infty$, $k_-\to 0$, $k_+$ is dominant, while $k_-$ is dominant 
as $k_-\to \infty$, $k_+\to 0$. Dividing the full phase space by the line $k_+=k_-$, 
the soft rapidity regulator can be written as 
\be
\label{rchiu2}
R'_S (\nu) = \displaystyle \Bigl(\frac{\nu}{k_+}\Bigr)^{\eta} \Theta(k_+-k_-)
+\Bigl(\frac{\nu}{k_-}\Bigr)^{\eta} \Theta (k_- - k_+), 
\ee
which reduces to $R_S$ with $\nu_+=\nu_-=\nu$ at leading order, neglecting subleading corrections. 

Applying Eq.~\eqref{rsorig} to Eq.~\eqref{MS0}, the soft contribution can be written as 
\bea
M_S &=& -\frac{\as C_F}{2\pi} \frac{(\mu^2e^{\gamma_{\mathrm{E}}})^{\eps}}{\Gamma(1-\eps)} \left(\nu_+^{\eta_+}\int_A 
\frac{dk_+}{k_+^{1+\eta_+}} \frac{dk_-}{k_-} +\nu_-^{\eta_-}\int_B \frac{dk_+}{k_+} \frac{dk_-}{k_-^{1+\eta_-}} \right)(k_+k_--M^2)^{-\eps}, 
\eea  
where the phase space $A$ ($B$) is the region with $k_+>k_-$ ($k_+<k_-$),  both with $k_+k_->M^2$. [See Fig.~\ref{fig3}-(a).] 
The contribution from the region $A$ is given by
\bea
M_S^A &=& -\frac{\as C_F}{2\pi} \frac{(\mu^2e^{\gamma_{\mathrm{E}}})^{\eps}\nu_+^{\eta_+}}{\Gamma(1-\eps)}
\int^{\infty}_{M^2} \frac{dk_L^2}{k_L^2} (k_L^2 - M^2)^{-\eps}\int^{\infty}_{k_L} \frac{dk_+}{k_+^{1+\eta_+}} 
\nnb \\
&=& -\frac{\as C_F}{4\pi}  \Bigl(\frac{\mu^2e^{\gae}}{M^2}\Bigr)^{\eps} \Bigl(\frac{\nu_+}{M}\Bigr)^{\eta_+}
\frac{\Gamma(\eps+ \frac{\eta_+}{2})\Gamma(\frac{\eta_+}{2})\Gamma(1-\eta_+)}{\Gamma(1+\frac{\eta_+}{2})
\Gamma(1-\frac{\eta_+}{2})} \nnb \\
\label{MSA}
&=& \frac{\as C_F}{4\pi} \Biggl[\frac{1}{\eps^2} +\frac{1}{\eps}\ln\frac{\mu^2}{M^2} - 
\Bigl(\frac{1}{\eps}+\ln\frac{\mu^2}{M^2}\Bigr)\Bigl(\frac{2}{\eta_+}+\ln\frac{\nu_+^2}{M^2}\Bigr)
+\frac{1}{2} \ln^2\frac{\mu^2}{M^2}-\frac{\pi^2}{12}\Biggr],
\eea 
where we require that $\eta_+$ go to zero faster than $\eps^n$ with $n>0$. 
The contribution from the region $B$, $M_S^B$, is obtained from $M_S^A$ by switching $(\eta_+,\nu_+) \leftrightarrow (\eta_-,\nu_-)$. 
The complete soft contribution at one loop is given as 
\bea 
M_S = M_S^A+M_S^B &=&\frac{\as C_F}{4\pi} \Biggl[\frac{2}{\eps^2} +\frac{2}{\eps}\ln\frac{\mu^2}{M^2} 
+\ln^2\frac{\mu^2}{M^2}-\frac{\pi^2}{6}\nnb \\
\label{MSc} 
&&~~ - \Bigl(\frac{1}{\eps}+\ln\frac{\mu^2}{M^2}\Bigr)\Bigl(\frac{2}{\eta_+}+\ln\frac{\nu_+^2}{M^2}+\frac{2}{\eta_-}
+\ln\frac{\nu_-^2}{M^2}\Bigr)\Biggr]. 
\eea 
It is consistent with the result in Ref.~\cite{Chiu:2012ir} with a single $\eta$.   

\subsection{Factorization of the Sudakov form factor}

The $n$-collinear contribution at one loop in Fig.~\ref{fig1}-(a) is given by Eq.~\eqref{Man}. Combining it with 
the field strength renormalization and the residue for a light quark, 
\be
Z_{q}^{(1)} = -\frac{\as C_F}{4\pi} \frac{1}{\eps},~~~R_{q}^{(1)} = -\frac{\as C_F}{4\pi} \Bigl(\ln\frac{\mu^2}{M^2}-\frac{1}{2}\Bigr), 
\ee
we obtain the complete contribution to the $n$-collinear sector at one loop as 
\bea
C_n^{(1)} &=& M_n + \frac{1}{2}(Z_{q}^{(1)}+R_{q}^{(1)}) \nnb \\
\label{Mnol} 
&=& \frac{\as C_F}{4\pi} \Bigl[\Bigl(\frac{1}{\eps}+\ln\frac{\mu^2}{M^2}\Bigr) \Bigl(\frac{2}{\eta_+} + 2\ln\frac{\nu_+}{p_+}
+\frac{3}{2}\Bigr)+\frac{9}{4} - \frac{\pi^2}{3} \Bigr]. 
\eea
Replacing $(p_+,\eta_+,\nu_+)$ with $(p'_-,\eta_-,\nu_-)$, the bare one loop result for the $\n$-collinear sector is given by
\be
\label{Mnbol}
C_{\bar{n}}^{(1)} = \frac{\as C_F}{4\pi} \Bigl[\Bigl(\frac{1}{\eps}+\ln\frac{\mu^2}{M^2}\Bigr) \Bigl(\frac{2}{\eta_-} + 2\ln\frac{\nu_-}{p'_-}
+\frac{3}{2}\Bigr)+\frac{9}{4} - \frac{\pi^2}{3} \Bigr]. 
\ee

From the SCET current $V_{\mr{SCET}}^{\mu} = 
F_{\mr{SCET}}\cdot \bar{u}_n (p)\gamma_{\perp}^{\mu} u_{\n} (p')$ in Eq.~\eqref{EFTcur}, the Sudakov form factor is 
factorized as  
\be
\label{Sfscet} 
F_{\mr{SCET}} (Q^2,\mu;M^2) = C_{n} (p_+,\mu,\nu_+;M^2) C_{\bar{n}} (p'_-,\mu,\nu_-;M^2) S_{n\bar{n}}(\mu,\nu_+,\nu_-;M^2),
\ee
where $Q^2 =2p\cdot p'= p_+ p'_-$ is the momentum transfer squared to the current.
To next-to-leading order (NLO) in $\as$, the collinear and soft functions are given by 
$C_{n,\bar{n}}=1+C_{n,\bar{n}}^{(1)}$ and $S_{n\bar{n}}=1+M_S$ after the renormalization from Eqs.~\eqref{Mnol}, \eqref{Mnbol}, and \eqref{MSc}.
Each rapidity scale 
dependence in $C_{n}$ and $C_{\bar{n}}$ is cancelled by the soft function $S_{n\bar{n}}$. But the evolution of $\nu_+$ from $p_+$ to $M$ 
and that of $\nu_-$ from $p'_-$ to $M$ are needed to resum the large logarithms of $Q/M$. 
The full resummation of large logarithms with the evolution of $\nu_{\pm}$ as well as $\mu$ is thoroughly explained in Appendix~\ref{A1}. 

The advantage of introducing multiple rapidity scales $\nu_{\pm}$ in Eq.~\eqref{Sfscet} is that we can systematically deal with the cases, 
in which there is a hierarchy of scales between $p_+$ and $p'_-$. For example,  if $p_+ \gg p'_- \gg M$,  the range of the evolution in 
$\nu_-$ is smaller than the range of $\nu_+$. It is very interesting to consider the limit 
$p_+ \gg p'_- \sim M$, in which
we can directly describe the soft-collinear current\footnote{\
A similar situation has been discussed in Ref.~\cite{Chiu:2012ir} in the ``the lab frame''. But our approach 
here is different. We focus on the universality and the extension of the factorization introducing multiple rapidity scales. 
Starting from the factorized back-to-back current, we directly describe a new soft sector  through recombining one collinear 
sector and the soft sector, setting the relevant rapidity scales to be soft.} from Eq.~\eqref{Sfscet}. Identifying $\nu_- \sim M$  in  
$C_{\bar{n}}$ and $S_{n\bar{n}}$ in Eq.~\eqref{Sfscet}, we can combine the two functions into a new soft function to 
describe the soft sector. We refer to Section~\ref{njet} for more details.

\subsection{On-shell regularization with a massless gluon}
 
We can also employ pure dimensional regularization with a massless gluon,
in which the UV and IR divergences are expressed as poles in $\euv$ and $\eir$ respectively.  
The UV and IR divergences are  separated, and the problematic mixed divergence such as $(1/\euv)\cdot(1/\eir)$ does not appear. 
In this paper, we use two ways to treat the IR divergence. Firstly, a nonzero gluon mass $M$ is introduced to regulate the IR divergence. Secondly, the IR divergence is explicitly computed and treated as a pole in $\eps$ using the dimensional regularization. 
In this subsection, we introduce the latter, which is quite nontrivial when the rapidity divergence is involved. The technical detail is instructive, but it has not received adequate attention.

%{\color{red}  In the literature,  there have been three ways to treat the IR divergence.  Firstly, scaleless integrals are put to zero by setting $\euv =\eir$.  
%Secondly, a nonzero gluon mass $M$ is introduced to regulate the IR divergence. Finally, the IR divergence is explicitly computed. We follow the last method,
%which is quite nontrivial when the rapidity divergence is involved. The technical detail is instructive, but it has not received adequate attention. Here
%we describe the on-shell renormalization in detail. }

Compared to Eq.~\eqref{Man} with nonzero $M$, the $n$-collinear contribution with $M=0$ is given by
\bea 
M_n &=& \tilde{M}_n -M_n^{\varnothing}  \nnb \\
&=& - \frac{\alpha_s C_F}{2\pi} \frac{(\mu^2 e^{\gamma_{\mathrm{E}}})^{\eps}}{\Gamma (1-\eps)} \int^{\infty}_0
d\blp{k}^2 (\blp{k}^2)^{-1-\eps} \Biggl[\int^1_0 \frac{dx}{x} (1-x) - \Bigl(\frac{\nu_+}{p_+}\Bigr)^{\eta_+} \int^{\infty}_0 
\frac{dx}{x^{1+\eta_+}}\Biggr] \nnb \\
&=& - \frac{\alpha_s C_F}{2\pi}  
\left(\frac{1}{\euv}-\frac{1}{\eir}\right)
\Biggl[\int^1_0 \frac{dx}{x} \Bigl((1-x) -1\Bigr)- \Bigl(\frac{\nu_+}{p_+}\Bigr)^{\eta_+} \int^{\infty}_1 \frac{dx}{x^{1+\eta_+}}\Biggr] \nnb \\
\label{Mam0}
&=& \frac{\alpha_s C_F}{2\pi}\left(\frac{1}{\euv}-\frac{1}{\eir}\right)\left(\frac{1}{\eta_+}+\ln\frac{\nu_+}{p_+}+1\right).
\eea 
Here the integration over $\blp{k}^2$ in pure dimensional regularization is expressed as 
\be
\mu^{2\eps}\int^{\infty}_0 d\blp{k}^2 (\blp{k}^2)^{-1-\eps}= \frac{1}{\euv}-\frac{1}{\eir}.
\ee
The rapidity regulator in the integral over $x\in [0,1]$ is not necessary since there is no rapidity divergence. 
With the self-energy contribution where the residue is given by $R_{q}^{(1)} = \as C_F/(4\pi \eir)$, 
we obtain the $n$-collinear contribution at one loop as 
\be
\label{Mnol0}
C_n^{(1)} = M_n + \frac{1}{2}(Z_{q}^{(1)}+R_{q}^{(1)}) 
= \frac{\as C_F}{4\pi} \Bigl(\frac{1}{\euv}-\frac{1}{\eir}\Bigr) \Bigl(\frac{2}{\eta_+} + 2\ln\frac{\nu_+}{p_+}+\frac{3}{2}\Bigr). 
\ee
Similarly, the $\n$-collinear contribution is given by 
\be
\label{Mnbol0}
C_{\bar{n}}^{(1)}  
= \frac{\as C_F}{4\pi} \Bigl(\frac{1}{\euv}-\frac{1}{\eir}\Bigr) \Bigl(\frac{2}{\eta_-} + 2\ln\frac{\nu_-}{p'_-}+\frac{3}{2}\Bigr). 
\ee

The soft virtual contribution with a massless gluon, yet without the rapidity regulator, in Fig.~\ref{fig1}-(c) is given as
\bea 
\tilde{M}_S &=& -4\pi g^2 C_F \mu_{\mr{\overline{MS}}}^{2\eps} \int \frac{d^Dk}{(2\pi)^D} 
\frac{\delta(k^2)}{k_+ k_-} \Theta(k_0) \nnb \\
\label{MS00} 
&=&-\frac{\as C_F}{2\pi} \frac{(\mu^2 e^{\gamma_{\mathrm{E}}})^{\eps}}{\Gamma (1-\eps)}
\int \frac{dk_+}{k_+} \frac{dk_-}{k_-} d\blp{k}^2 (\blp{k}^2)^{-\eps} \delta(k^2) \Theta(k_0).
\eea  
Applying the rapidity regulator in Eq.~\eqref{rsorig}, we divide the soft phase space into the regions $A$ $(k_+>k_-)$ and $B$ 
$(k_+>k_-)$. To compute the contribution from the region $A$, it is useful to consider the phase space in $(k_+,|\blp{k}|)$ 
in Fig.~\ref{fig3}-(b).  

The contribution from the region $A$ in Fig.~\ref{fig3}-(b) can be written as 
\bea
\label{MSA0}
M_S^A &\equiv& M_S^{A1} + M_S^{A2}  \\
&=& - \frac{\as C_F}{2\pi} \frac{(\mu^2 e^{\gamma_{\mathrm{E}}})^{\eps}\nu_+^{\eta_+}}{\Gamma (1-\eps)} 
\Biggl[\int^{\infty}_{\Lambda^2} d\blp{k}^2 (\blp{k}^2)^{-1-\eps} + \int^{\Lambda^2}_0  d\blp{k}^2(\blp{k}^2)^{-1-\eps}\Biggr] 
\int^{\infty}_{|\blp{k}|} dk_+ k_+^{-1-\eta_+}, \nnb
\eea
where we divide the integration region for $\blp{k}^2$ into $[\Lambda^2, \infty]$ and $[0, \Lambda^2]$ in order to separate the UV 
and IR divergences. The dependence on the arbitrary scale $\Lambda^2$ cancels at the end of calculation.  
The two terms in Eq.~\eqref{MSA0} are labelled as $M_S^{A1}$ and $M_S^{A2}$, and are given by
\begin{align}
\label{MSA1} 
M_S^{A1} &= \frac{\as C_F}{4\pi} \Biggl[\frac{1}{\euv^2}+\frac{1}{\euv}\ln\frac{\mu^2}{\nu_+^2}-\frac{2}{\eta_+}
\Bigl(\frac{1}{\euv}+\ln\frac{\mu^2}{\Lambda^2}\Bigr)+\frac{1}{2}\ln^2\frac{\mu^2}{\nu_+^2}
-\frac{1}{2}\ln^2\frac{\nu_+^2}{\Lambda^2}-\frac{\pi^2}{12} \Biggr], \\
\label{MSA2} 
M_S^{A2} &= -\frac{\as C_F}{4\pi} \Biggl[\frac{1}{\eir^2}+\frac{1}{\eir}\ln\frac{\mu^2}{\nu_+^2}-\frac{2}{\eta_+}
\Bigl(\frac{1}{\eir}+\ln\frac{\mu^2}{\Lambda^2}\Bigr)+\frac{1}{2}\ln^2\frac{\mu^2}{\nu_+^2}
-\frac{1}{2}\ln^2\frac{\nu_+^2}{\Lambda^2}-\frac{\pi^2}{12} \Biggr].
\end{align}

Combining these two contributions, we have  
\be
\label{MSAf} 
M_S^A = \frac{\as C_F}{4\pi} \Biggl[\frac{1}{\euv^2}-\frac{1}{\eir^2}-\Bigl(\frac{1}{\euv}-\frac{1}{\eir}\Bigr)
\Bigl(\frac{2}{\eta_+}+\ln\frac{\nu_+^2}{\mu^2}\Bigr)\Biggl]. 
\ee
The contribution from the region $B$ can be obtained from 
Eq.~\eqref{MSAf} by switching $(\eta_+,\nu_+)\to (\eta_-,\nu_-)$. 
Finally the soft contribution at one loop using the pure on-shell dimensional regularization is given by 
\begin{equation} \label{msoft}
M_S = \frac{\alpha_s C_F}{2\pi} \Bigl[\frac{1}{\euv^2} -\frac{1}{\eir^2} -\Bigl(\frac{1}{\euv} -\frac{1}{\eir}\Bigr) 
\Bigl( \frac{1}{\eta_+} +\frac{1}{\eta_-}+\ln \frac{\nu_+ \nu_-}{\mu^2} \Bigr) \Bigr].
\end{equation}
The total contributions from Eqs.~\eqref{Mnol0}, \eqref{Mnbol0} and \eqref{msoft} are free of the rapidity scales  
and  the IR divergence of the full theory is reproduced. With $Q^2=p_+ p'_-$, the (bare) one loop correction to the form factor in SCET is given as  
\bea 
F_{\mr{SCET}}^{(1)} (Q^2,\mu) &=& C_{n}^{(1)}+C_{\n}^{(1)}+M_S \nnb \\
\label{sudbb}
&=& \frac{\alpha_s C_F}{2\pi}\Biggl[\frac{1}{\euv^2} -\frac{1}{\eir^2}+\Bigl(\frac{1}{\euv} 
-\frac{1}{\eir}\Bigr) \Bigl(\ln\frac{\mu^2}{Q^2}+\frac{3}{2}\Bigr)\Biggr].
\eea

\subsection{Soft contributions to timelike processes\label{timeS}}
 
So far we have considered the back-to-back collinear current with the spacelike momentum transfer.
For the current with the timelike momentum transfer as in Drell-Yan (DY) process, the current in SCET is given by 
\be
\label{DYcur}
V_{\mr{DY,SCET}}^{\mu} = \langle 0 |~\bar{\xi}_n W_n S_n^{\dagger}\gamma_{\perp}^{\mu} S_{\bar{n}} W_{\bar{n}}^{\dagger} 
\xi_{\bar{n}}~|pp'\rangle,
\ee 
where $p~(p')$ is the incoming $n$-$(\n$-)collinear momentum. Compared to Eq.~\eqref{EFTcur}, 
 $\tilde{S}_n^{\dagger}$ is replaced by $S_n^{\dagger}$~\cite{Chay:2004zn}. 
 The soft Wilson line $S_n^{\dagger}$ from the $n$-collinear antiquark is given by 
\be
S^{\dagger}_n = \exp\Bigl[ -gn\cdot A_s \displaystyle \frac{1}{n\cdot \mathcal{P}^{\dagger} -i\eps}\Bigr].
\ee

The matrix element of the full-theory current is schematically factorized as      
\be
\label{fcurt}
\langle 0 | \bar{q}_n \gamma^{\mu} q_{\bar{n}} | pp'\rangle = H_{\mr{DY}}(-Q^2,\mu) V_{\mr{DY,SCET}}^{\mu} (\mu) 
\sim H_{\mr{DY}} \cdot C_{n} \cdot C_{\bar{n}}\cdot S_{\mr{DY}} \cdot \bar{v}_{n}(p) \gamma_{\perp}^{\mu} u_{\bar{n}}(p'),  
\ee 
where $Q^2 = 2p\cdot p' = p_+ p'_-$, and $u_{\bar{n}}$ and $\overline{v}_{n}$ are the spinors for the $\n$-collinear quark and 
$n$-collinear antiquark respectively. The hard coefficient $H_{\mr{DY}}$ depends on $-Q^2$, in contrast to
$+Q^2$ for a spacelike process, and its anomalous dimension for $H_{\mr{DY}}$ at one loop is given by
\be
\gamma_{H,\mr{DY}} = -\frac{\as C_F}{4\pi}\Bigl(4\ln\frac{\mu^2}{-Q^2}+6\Bigr). 
\ee    
The minus sign in  the logarithm in $\gamma_{H,\mr{DY}}$ also shows up in $V_{\mr{DY,SCET}}^{\mu}$, 
and it appears specifically in the soft function $S_{\mr{DY}}$.
% since $C_{n}$ and $C_{\bar{n}}$ are the same.

\begin{figure}[h]
\begin{center}
\includegraphics[height=4.8cm]{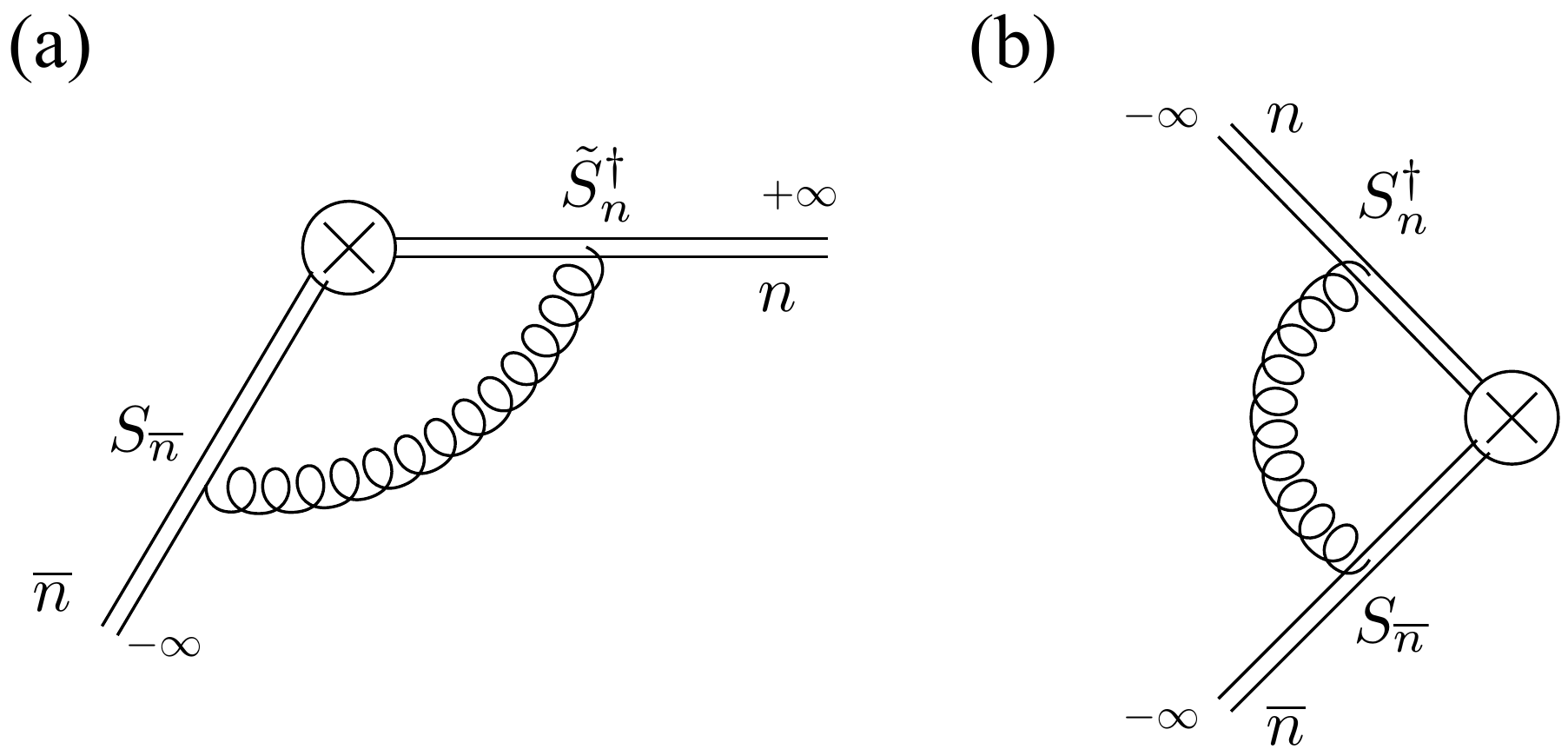}
\end{center}
\vspace{-0.6cm}
\caption{\label{fig4}\baselineskip 3.5ex
Soft Feynman diagrams for the virtual contributions to (a) the spacelike process and (b) the timelike (DY) process. 
Here the double line represents the path for a given soft Wilson line. }
\end{figure}

Fig.~\ref{fig4} shows the different paths of the soft Wilson lines for $S_{\mr{DY}}$ with respect to the spacelike process, which 
generates the relative minus sign in the logarithm. The amplitude for 
$S_{\mr{DY}} = \langle 0 | S_n^{\dg} S_{\bar{n}} | 0\rangle$ at one loop is written as 
\be
\label{MSDY0}
\tilde{M}_S^{\mathrm{DY}} =  -2ig^2 C_F \mums \int \frac{d^{D-1} k}{(2\pi)^D}~I_{\mr{DY}} (k), 
\ee
where $I_{\mr{DY}}$ is given by
\bea 
I_{\mr{DY}}(k) &=& \int^{\infty}_{-\infty} dk_0~ \frac{1}{k^2 +i\eps}\frac{1}{(k_+ -i\eps)(k_- +i\eps)} \nnb \\
\label{Ik}
&=& \int^{\infty}_{-\infty}\frac{dk_0}{(k_0 +|\mathbf{k}| -i\eps) (k_0 -|\mathbf{k}| +i\eps) (k_0 +k_z -i\eps) (k_0 -k_z +i\eps)},
\eea
by setting $n^{\mu} = (1, 0, 0, 1)$ and $\overline{n}^{\mu} = (1, 0, 0, -1)$,  
where $k_+=\n\cdot k = k_0+k_z$ and $k_- = n\cdot k = k_0 -k_z$.  Encircling the contour in the lower-half plane, the two 
poles at $k_0 = |\mathbf{k}| -i\eps$ and $k_0 = k_z -i\eps$ contribute to the integral. The result can be written as 
$I_{\mr{DY}}(k) = I_0 (k) + I_{\mr{T}} (k)$, where $I_0$ is the residue from the pole $k_0 = |\mathbf{k}| -i\eps$ and $I_{\mr{T}}$ 
is from the pole $k_0 = k_z -i\eps$. $I_0$ is given as 
\be
I_0 (k) = -2\pi i \cdot \frac{1}{2k_0}\frac{1}{k_+ k_-}\Bigl|_{k_0 = |\mathbf{k}|} = -2\pi i \int dk_0 \frac{\delta(k^2)}{k_+k_-}\Theta(k_0).
\ee 
Note that the contribution $I_0 (k)$ in Eq.~\eqref{MSDY0} is the same as $\tilde{M}_S$ in Eq.~\eqref{MS00} (or Eq.~\eqref{MS0} 
with $M^2=0$). So $I_0 (k)$, or $\tilde{M}_S$ is common to both the spacelike and the timelike 
processes.   
 
The residue $I_{\mr{T}}$ from the pole $k_0 =k_z-i\eps$ is present for the timelike process only, and it is given as 
\be
I_{\mr{T}}(k) = -2\pi i \frac{1}{-\blp{k}^2+i\eps}\cdot\frac{1}{2k_z - i\eps}= \frac{i\pi}{\blp{k}^2}\frac{1}{k_z - i\eps}.
\ee 
And the contribution from $I_{\mr{T}}$ in Eq.~\eqref{MSDY0} is given by
\bea
M_{S,\mr{T}} &=& \frac{\as C_F}{2\pi} \frac{(\mu^2 e^{\gamma_{\mathrm{E}}})^{\eps}}{\Gamma(1-\eps)} 
\int_0^{\infty} d\mathbf{k}_{\perp}^2
(\mathbf{k}_{\perp}^2)^{-1-\eps} \int_{-\infty}^{\infty}  \frac{dk_z}{k_z -i\eps}  %\nnb \\
\label{MST}
=\frac{\alpha_s C_F}{2\pi}  (-i\pi) \Bigl(\frac{1}{\euv} -\frac{1}{\eir}\Bigr). 
\eea  
There is no rapidity divergence here, and the factor $-i\pi$ gives a negative sign in the argument of the logarithm in $M_{S}$.  

As a result, applying the rapidity regulator in Eq.~\eqref{rsorig} to $\tilde{M}_S^{\mr{DY}}$, we obtain the soft  
contribution for the timelike process at one loop as 
\bea 
M_S^{\mr{DY}} &=& M_S + M_{S,\mr{T}} \nnb \\
\label{MSTf} 
&=& \frac{\alpha_s C_F}{2\pi} \Bigl[\frac{1}{\euv^2} -\frac{1}{\eir^2} -\Bigl(\frac{1}{\euv} -\frac{1}{\eir}\Bigr) 
\Bigl( \frac{1}{\eta_+} +\frac{1}{\eta_-} +\ln \frac{-\nu_+ \nu_-}{\mu^2} \Bigr) \Bigr].
\eea 
The above result can be generalized depending on the $i\varepsilon$ prescription of the soft Wilson lines. When  a collinear parton in the $n_i$ direction is 
incoming and the other parton in the $n_j$ direction is outgoing, the sign of the term $\nu_i \nu_j$ is positive. When the two collinear partons are 
both incoming or outgoing, the sign is negative.

\section{The $N$-jet operator\label{njet}}
The $N$-jet operator consists of $N$ collinear fields after integrating out the hard off-shell modes. It is constructed out of 
the collinear fermion fields $\chi_{n_i} = W_{n_i}^{\dagger} \xi_{n_i}$, and the collinear gauge fields 
$B_{n_i\mu}^{\perp a}= i\overline{n}_i^{\rho}g_{\perp}^{\mu\nu} G_{n_i, \rho\nu}^b 
\mathcal{W}_{n_i}$, where $W_{n_i}$ ($\mathcal{W}_{n_i}$) is the collinear Wilson line in the $n_i$ direction 
in the fundamental (adjoint) representation~\cite{Bauer:2001yt}.
They are invariant under each $n_i$-collinear gauge transformation. For example, the 2-jet operator with quarks
is $\overline{\chi}_{n_i} \Gamma \chi_{n_j}$, and the 3-jet operator is $\overline{\chi}_{n_i} 
\mathcal{B}_{n_j \perp}^{\mu} \Gamma_{\mu} \chi_{n_k}$ with some Dirac structures $\Gamma$ and $\Gamma_{\mu}$, and so on.

To be explicit, for an event with $2N_q$ quark and antiquark jets, and $N_g$ gluon jets with $N=2N_q+N_g$, the $N$-jet operator from the current 
can be schematically written as~\cite{Ellis:2010rwa}
\begin{equation} \label{njetop}
J^{\mu} = C (p_1,\cdots, p_N)  \prod_{i=1}^{N_q} \overline{\chi}_{n_i}  \prod_{j=1}^{N_q} \chi_{n_j} \prod_{k=1}^{N_g} 
\mathcal{B}_{n_k \perp}^{\mu_k},
\end{equation}
where the color and Dirac indices are suppressed.
After decoupling the soft interactions, each collinear field is redefined as
\begin{equation}
\chi_{n_i} \rightarrow S_{n_i} \chi_{n_i}, ~~~\mathcal{B}_{n_j \perp}^{\mu} \rightarrow \mathcal{S}_{n_i} 
\mathcal{B}_{n_j \perp}^{\mu},
\end{equation}
where $S_{n_i}$ ($\mathcal{S}_{n_i}$) is the soft Wilson line in the fundamental
(adjoint) representation. Here we require that all the $N$ collinear particles form $N$ jets, implying that
$n_i\cdot n_j \sim \mathcal{O} (1)$.
And we consider the $N$-jet singlet operators. In $e^+e^-$ annihilation, the $N$-jet operator is constructed out of 
the outgoing quarks and gluons, and the net color charge is zero. In hadronic collisions, we include the incoming particles to 
form an overall color singlet operator. For the $N$-jet operator, it involves $N$ collinear fields for the $N-2$ jets in the final state. 
 
Let us consider the amplitude (or the amputated Green's function)  by taking the matrix element of the $N$-jet operator. 
Employing the color-space formalism~\cite{Catani:1996jh,Catani:1996vz,Becher:2014oda}, the amplitude can be schematically written as 
\be
| \mc{M}_N \{p\}  \rangle = \sum_{\{a\}}\Bigl(\langle \{a\} |  \mc{M}_N \{p\}  \rangle \Bigr)\cdot | \{a\} \rangle, 
\ee
where $\{p\}\equiv (p_1,p_2,\cdots,p_N)$ denotes the external momenta of the $N$ collinear massless partons. 
$| \{a\} \rangle \equiv |a_1,a_2,\cdots,a_N \rangle $ is the orthonormal basis in the color space, where $a_i$ 
are the color indices for the external partons. 
Also $| \mc{M}_N \{p\}\rangle$ can be written as 
\be
| \mc{M}_N \{p\}  \rangle = F_N (\{p\}) | \mc{M}_N^{(0)} \{p\} \rangle,
\ee
where $|\mc{M}_N^{(0)} \{p\}\rangle$ is the $N$-jet amplitude at tree level with the contributions from the external 
on-shell spinors and polarization vectors. $F_N (\{p\})$ is the form factor that can be expanded in powers of $\as$.

The form factor $ F_N (\{p\})$ can be factorized as 
\be
\label{FNfact} 
F_N (\{p\}) = H_N (\{2\sigma_{ij} p_i\cdot p_j \},\mu) \Bigl[ \prod_k^N C_k (p_k^+,\mu,\nu_k) \Bigr] 
S_N (\{\sigma_{ij} n_i\cdot n_j/2 \},\mu,\nu_1,\cdots \nu_N),
\ee
where $H_N$ is the hard matching coefficient. The string $\{2\sigma_{ij} p_i\cdot p_j \}$ represents all the possible 
combinations of the hard momentum transfers with different $i$ and $j$ ($i,j=1,\cdots,N$). 
When both $p_i$ and $p_j$ are all incoming or all outgoing, $\sigma_{ij}=-1$, and otherwise $\sigma_{ij}=+1$. The corresponding string 
$\{\sigma_{ij} n_i\cdot n_j/2 \}$ appears in $S_N$. Since all the external partons are on-shell, the momentum 
can be written as $p_i^{\mu} = p_i^+ n_i^{\mu}/2$ with $p_i^+ = \n_i \cdot p_i$. 

\begin{figure}[t]
\begin{center}
\includegraphics[height=4.8cm]{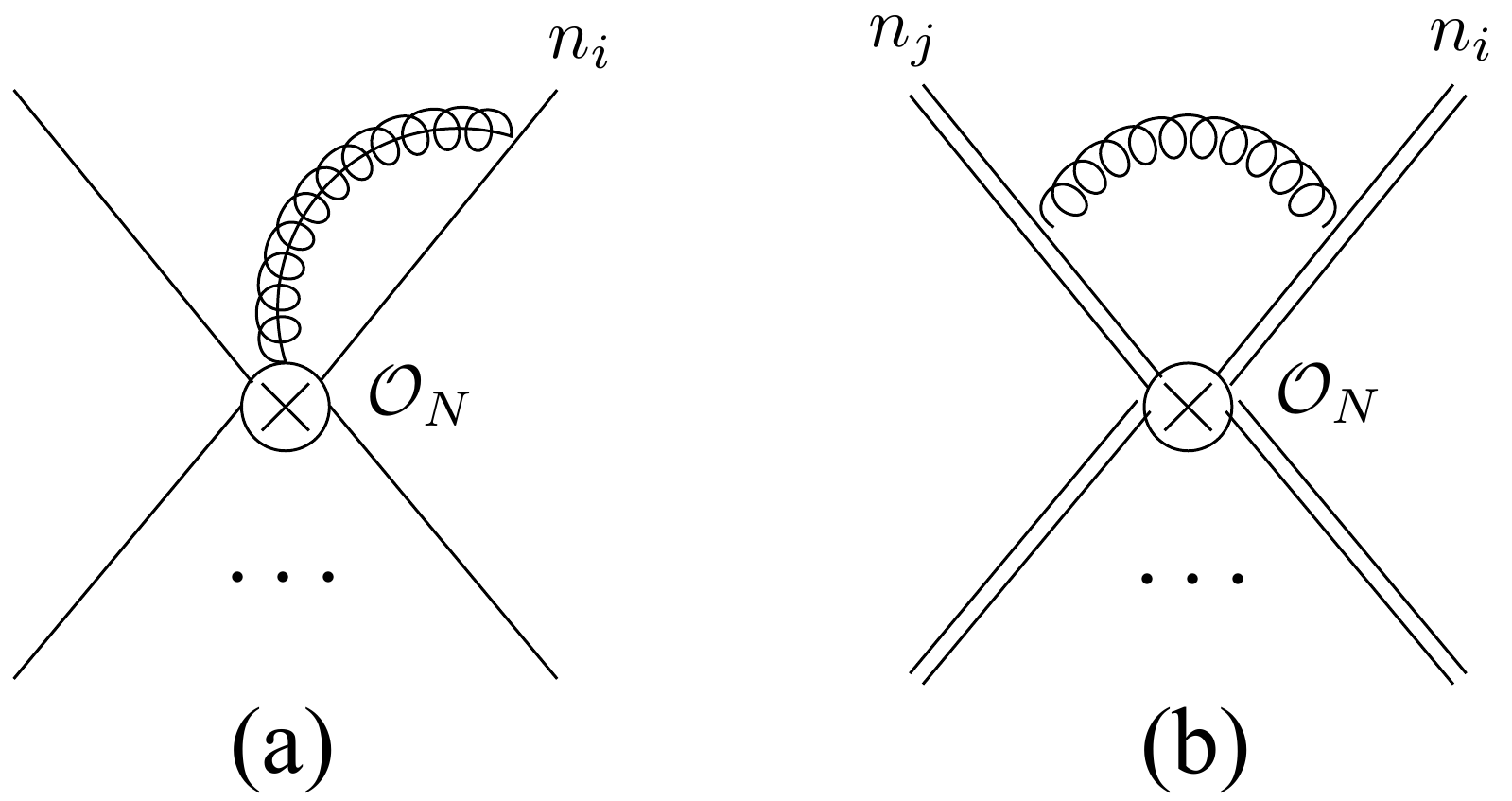}
\end{center}
\vspace{-0.8cm}
\caption{\label{fig5}\baselineskip 3.5ex
Feynman diagrams for the $N$-jet operator in SCET at one loop. (a) the $n_i$-collinear gluon exchange, (b)
 the soft gluon exchange between the soft Wilson lines $S_{n_i}$ and $S_{n_j}$.}
\end{figure}

Let us consider the one-loop contribution to $F_N (\{p\})$ in SCET. The $n_i$-collinear contribution at one loop  
for Fig.~\ref{fig5} (a) can be obtained from Eq.~\eqref{Mnbol0}, and is given by 
\begin{equation}
\label{loopCi}
C_i^{(1)} (p_i^+, \mu,\nu_i) = \frac{\alpha_s}{4\pi} \Bigl( \frac{1}{\euv} -\frac{1}{\eir}\Bigr) 
\Bigl[ \mathbf{T}_i^2 \Bigl( \frac{2}{\eta_i} + 2\ln \frac{\nu_i}{p_i^+} \Bigr)
+\frac{\hat{\gamma}_i}{2}\Bigr],
\end{equation}
where $\hat{\gamma}_i = 3C_F$ for a quark or an antiquark, and $\beta_0$ for a gluon.  $\mathbf{T}_i$ is the color charge of 
the $i$-th collinear particle, and $\bl{T}_i^2 = \bl{T}_i^a \cdot \bl{T}_i^a$ is $C_F$ for a quark or an antiquark  and $C_A$ for a gluon.    

Let us now consider the one-loop calculation for $S_N$. For a soft gluon exchange between $S_{n_i}$ 
and $S_{n_j}$  in Fig.~\ref{fig5} (b)\footnote{\baselineskip 3.0ex
In this section, we write the soft Wilson line in the $n$ direction collectively as $S_n$ for $S_n$ or $S_n^{\dagger}$.
It can be distinguished from the representation of the color charge operator $\bl{T}^a$ in the Wilson line.},
the amplitude contains the factor
\be
\frac{1}{n_i\cdot k}\frac{1}{n_j\cdot k}. 
\ee
The rapidity divergence can occur both in the $n_i$ and $n_j$ directions. To clarify this, we can apply the same reasoning employed in 
the back-to-back current. The rapidity divergence in the $n_i$ direction arises when $n_i\cdot k \to 0$ and 
$n_j\cdot k \approx (n_i\cdot n_j/2) ~\n_i \cdot k \to \infty$, and the rapidity divergence in the $n_j$ direction arises when 
$n_j\cdot k \to 0$ and  $n_i\cdot k \approx (n_j\cdot n_i/2) ~\n_j \cdot k \to \infty$ with $\mathbf{k}_{\perp}^2$ fixed.

As discussed in Section~\ref{sftreg}, we can employ the rapidity regulators in the soft Wilson lines in 
Eq.~\eqref{mSnp}. For the rapidity divergence in the $n_i$ direction, the rapidity regulator is inserted in $S_{n_j}$, and the regulator for 
the divergence in the $n_j$ direction, it is inserted in $S_{n_i}$. Then the contribution for the soft gluon exchange between $S_{n_i}$ and 
$S_{n_j}$ is given by 
\be
\label{Msij}
M_{S}^{ij} = ig^2 \mu_{\mr{\overline{MS}}}^{2\eps} ~\bl{T}_i \cdot \bl{T}_j \int \frac{d^Dk}{(2\pi)^D} 
\frac{n_i\cdot n_j}{k^2(n_i \cdot k)(n_j\cdot k)} \Bigl(\frac{\nu_i}{n_j\cdot k} \frac{n_i\cdot n_j}{2}\Bigr)^{\eta_i}
\Bigl(\frac{\nu_j}{n_i\cdot k} \frac{n_i\cdot n_j}{2}\Bigr)^{\eta_j}. 
\ee
It is written to make the expression look symmetric, and the simultaneous appearance of the two regulators may look confusing. 
But in extracting the rapidity divergence in the $n_i$ direction, the $\eta_j$ regulator can be dropped because there is no pole in $\eta_j$, 
and vice versa. 

We can directly compute Eq.~\eqref{Msij} by decomposing the momentum vector in the $n_i$-$n_j$ basis. Interestingly, there is another 
convenient way to recycle the result for the back-to-back current.
Let us boost the reference frame in order that two lightcone vectors $n_i$ and $n_j$ become 
back-to-back~\cite{Kasemets:2015uus}. With the lightcone vectors $n_i^{\mu}=(1, \hat{n}_i)$ and $n_j^{\mu}=(1,\hat{n}_j)$,
we find that the boost is obtained by the velocity $\bm{\beta} = (\hat{n}_i + \hat{n}_j)/2$. 
The lightcone vectors in the boosted frame are given as 
\be
\label{blv}
\tilde{n}_i^{\prime\mu} = \Bigl(\frac{1}{\gamma}, \frac{1}{2} (\hat{n}_i -\hat{n}_j)\Bigr), ~~~ \tilde{n}_j^{\prime\mu} = 
\Bigl( \frac{1}{\gamma}, \frac{1}{2} (\hat{n}_j -\hat{n}_i)\Bigr),
\ee
where the Lorentz factor $\gamma$ is given by 
\be
\gamma = \gamma_{ij}=\Bigl(\frac{n_i \cdot n_j}{2}\Bigr)^{-1/2}. 
\ee
In Eq.~\eqref{blv}, the boosted lightcone vectors are not normalized, but 
can be normalized by rescaling $n_i^{\prime\mu} = \gamma \tilde{n}_i^{\prime\mu}$, 
$n_j^{\prime\mu} = \gamma \tilde{n}_j^{\prime\mu}$ to satisfy $n_i^{\prime} \cdot n_j^{\prime}=2$. 
After the boost, $n_i \cdot k \rightarrow \tilde{n}_i^{\prime} \cdot k^{\prime} = n_i^{\prime} \cdot k^{\prime}/\gamma$, 
and $n_j\cdot k \to n_j^{\prime} \cdot k^{\prime}/\gamma$. 

In Eq.~\eqref{Msij}, the original amplitude without the rapidity regulators is boost invariant with 
$k\to k'$ and $n_{i,j} \to n'_{i,j}$. But the forms of the rapidity regulators are not. 
After the boost, the regulator transforms as
\bea
\frac{\nu_i}{n_j\cdot k} \frac{n_i\cdot n_j}{2} &\rightarrow& \frac{\nu_i}{\tilde{n}_j^{\prime} \cdot k^{\prime}} 
\frac{\tilde{n}_i^{\prime} \cdot\tilde{n}_j^{\prime}}{2}
=\frac{\nu_i/\gamma}{n_j^{\prime} \cdot k^{\prime}} = \frac{\nu_i/\gamma}{\overline{n}_i^{\prime} \cdot k^{\prime}},\\ 
\frac{\nu_j}{n_i\cdot k} \frac{n_i\cdot n_j}{2} &\rightarrow& \frac{\nu_j/\gamma}{n_i^{\prime} \cdot k^{\prime}}. 
\eea
Since $n'_i$ and $n'_j$ are back-to-back, we can use the regulator in Eq.~\eqref{rsorig}, and the soft rapidity 
regulator in the boosted frame is given by
\begin{equation}
R_S^{ij} \Bigl(\frac{\nu_i}{\gamma_{ij}},\frac{\nu_j}{\gamma_{ij}}\Bigr) = 
\Bigl( \frac{\nu_i/\gamma_{ij}}{n_j^{\prime}\cdot k^{\prime}} \Bigr)^{\eta_i} 
\theta (n^{\prime}_j \cdot k -n^{\prime}_i \cdot k) +\Bigl( \frac{\nu_j/\gamma_{ij}}{n_i^{\prime}\cdot k^{\prime}} \Bigr)^{\eta_j} 
\theta (n^{\prime}_i \cdot k -n^{\prime}_j \cdot k),
\end{equation}
with $n^{\prime}_j = \overline{n}^{\prime}_i$ and $n^{\prime}_i = \overline{n}^{\prime}_j$.

Therefore the soft function at one loop, Eq.~\eqref{Msij}, can be obtained from the back-to-back 
current  by replacing $\nu_i$ with $\nu_i/\gamma_{ij}$. For given $i$ and $j$, the result is given by
\begin{equation} \label{msij}
M_S^{ij} = \frac{\alpha_s}{4\pi} \mathbf{T}_i \cdot \mathbf{T}_j  \Bigl[ -2 \Bigl(\frac{1}{\euv^2} -\frac{1}{\eir^2}\Bigr) 
+\Bigl(\frac{1}{\euv} -\frac{1}{\eir}\Bigr)
\Bigl(\frac{2}{\eta_i} +\frac{2}{\eta_j}   -2\ln \frac{\mu^2 }{\sigma_{ij}\nu_i\nu_j/\gamma_{ij}^2}  \Bigr) \Bigr],
\end{equation}
where $\sigma_{ij}=-1$ if both collinear partons in the $n_i$, $n_j$ directions are either incoming or outgoing, and $\sigma_{ij}=+1$ otherwise. 
The source of the negative  sign is explained in section~\ref{timeS}. The total  contribution to $S_N$ at one loop is given as 
\begin{eqnarray}
S_N^{(1)} &=& \frac{1}{2} \sum_{(i,j)}^N M_S^{ij}  \nonumber \\
\label{SNol}
&=& \frac{\alpha_s}{4\pi}\sum_{(i,j)}^N    \mathbf{T}_i \cdot \mathbf{T}_j  \Bigl[ - \Bigl( \frac{1}{\euv^2} 
-\frac{1}{\eir^2}\Bigr) +\Bigl( \frac{1}{\euv}
-\frac{1}{\eir}\Bigr) \Bigl( \frac{1}{\eta_i}+\frac{1}{\eta_j} - \ln \frac{\mu^2 \gamma_{ij}^2}{\sigma_{ij}\nu_i \nu_j}  \Bigr) \Bigr],
\end{eqnarray}
where $(i,j)$ in the summation means that the case $i=j$ is excluded.

The total collinear contribution at one loop is given by
\begin{eqnarray}
C_N^{(1)} &\equiv& \sum_{i=1}^N C_i^{(1)} (p_i^+, \mu, \nu_i)  \\
&=& \frac{\alpha_s}{4\pi} \Bigl(\frac{1}{\euv} -\frac{1}{\eir}\Bigr)  
\label{CNol}
 \Bigl[ 
\sum_{(i,j)}^N  \mathbf{T}_i \cdot \mathbf{T}_j \Bigl( -\frac{1}{\eta_i} -\frac{1}{\eta_j} -\ln \frac{\nu_i \nu_j}{p_i^+ p_j^+} \Bigr) +
\sum_i^N \frac{\hat{\gamma}_i}{2}\Bigr].  \nonumber 
\end{eqnarray}
Here the color factor is reorganized using the color singlet condition, $\sum_i^N \mathbf{T}_i =0$. Therefore 
$\mathbf{T}_i^2 = -\sum_{j\neq i}^N \mathbf{T}_i \cdot \mathbf{T}_j$, and the sum of $\mathbf{T}_i^2$ is given as
\begin{equation}
\sum_i^N \mathbf{T}_i^2 = -\sum_i^N \sum_{j\neq i}^N \mathbf{T}_i\cdot \mathbf{T}_j \equiv -\sum_{(i,j)}^N 
\mathbf{T}_i\cdot \mathbf{T}_j.
\end{equation}

The total contribution to the $N$-jet operator at order $\as$ is given by
\begin{eqnarray}
  C_N^{(1)} + S_N^{(1)} 
&=& \frac{\alpha_s}{4\pi} \Bigl\{ \sum_{(i,j)}^N \mathbf{T}_i \cdot \mathbf{T}_j 
\Bigl[ -\Bigl( \frac{1}{\euv^2} -\frac{1}{\eir^2}\Bigr) -\Bigl(\frac{1}{\euv}
-\frac{1}{\eir}\Bigr) \ln \frac{\mu^2}{2\sigma_{ij}p_i\cdot p_j}\Bigr] \nonumber \\ \label{Neft}
&&+ \sum_{i}^N \Bigl(\frac{1}{\euv}
-\frac{1}{\eir}\Bigr) \frac{\hat{\gamma}_i}{2}\Bigr\},  
\end{eqnarray}
where $2p_i\cdot p_j = p_i^+ p_j^+ n_i\cdot n_j/2$. The dependence on the scales $p_i^+$, and $p_j^+$ in the collinear function and the directions 
$n_i\cdot n_j = 2/\gamma_{ij}^2$ in the soft function are combined to yield the relativistic invariant $p_i\cdot p_j$ in the final result. 
Note that all the dependence on the rapidity scales cancels when the collinear and the soft functions are added.
Also, from Eq.~\eqref{Neft}, it is straightforward to obtain the leading anomalous dimension of the hard part $H_N$ in the 
$N$-jet operator, using the fact that $F_N$ in the full theory is independent of the renormalization scale.
It is given by  
\be 
\gamma_H^N = \frac{\alpha_s}{4\pi} \Bigl\{ \sum_{(i,j)}^N \mathbf{T}_i \cdot \mathbf{T}_j 
\Bigl( 2 \ln \frac{\mu^2}{2\sigma_{ij}p_i\cdot p_j}\Bigr)-\sum_{i}^N  \hat{\gamma}_i\Bigr\},
\ee 
 and is the same as the results in Refs.~\cite{Becher:2009cu,Becher:2009qa}.

\section{Sudakov form factor from the soft-collinear current\label{sccurrent}} 

Let us consider the matrix element of the soft-collinear current in SCET
\be
\label{sccur}
\langle p|J^{\mu}|p'\rangle =  
 H(Q^2,\mu) \langle p| \overline{\xi}_n W_n \gamma^{\mu}  \tilde{S}_n^{\dagger}  q_s | p'\rangle,
\ee
where $q_s$ is the soft quark with momentum $p'$. The collinear momentum $p$ and the soft momentum $p'$ scale as 
$p=(p_+,p_{\perp},p_-)=p_+(1,\lambda,\lambda^2)$ and $p'=p_+(\lambda,\lambda,\lambda)$.   
The hard function $H(Q^2, \mu)$ to one loop is given as
\begin{equation} \label{hardh}
H(Q^2, \mu) = 1+\frac{\alpha_s C_F}{4\pi} \Bigl( -3 \ln \frac{\mu^2}{Q^2} -\ln^2  \frac{\mu^2}{Q^2} -8 +\frac{\pi^2}{6}\Bigr),  
\end{equation}
where $Q^2 = 2p\cdot p'$. Here the hard function can be directly computed in $\mr{SCET_I}$ with 
the hard-collinear mode, which scales as $p_{hc} = (p_{hc}^+, p_{hc}^{\perp},p_{hc}^-) = p_+(1,\lambda^{1/2},\lambda)$.
This mode can interact with the soft quark.  After integrating out 
the hard-collinear mode, we can obtain $H(Q^2,\mu)$ with $Q^2 \approx p_+ p'_-$. We refer to Ref.~\cite{Chay:2004va}
for the calculation of $H(Q^2,\mu)$ in $\mr{SCET_I}$.

For the current in Eq.~\eqref{sccur}, 
%and the the matrix element of the full-theory current $\langle p| J^{\mu}|p'\rangle = F(Q^2) \bar{u}_n (p) \gamma^{\mu} u_s (p')$,
the Sudakov form factor $F(Q^2)$ can be written in a factorized form as 
\be 
\label{Sudsc}
F(Q^2) = H(Q^2,\mu) C_n (p_+,\mu,\nu_+) S_q(p',\mu,\nu_+),
\ee
where $C_n$ and $S_q$ are the collinear and the soft functions for the soft-collinear current. The one-loop result 
for $C_n$ is given in Eq.~\eqref{Mnol} or \eqref{Mnol0}.

\begin{figure}[b]
\begin{center}
\includegraphics[height=4.2cm]{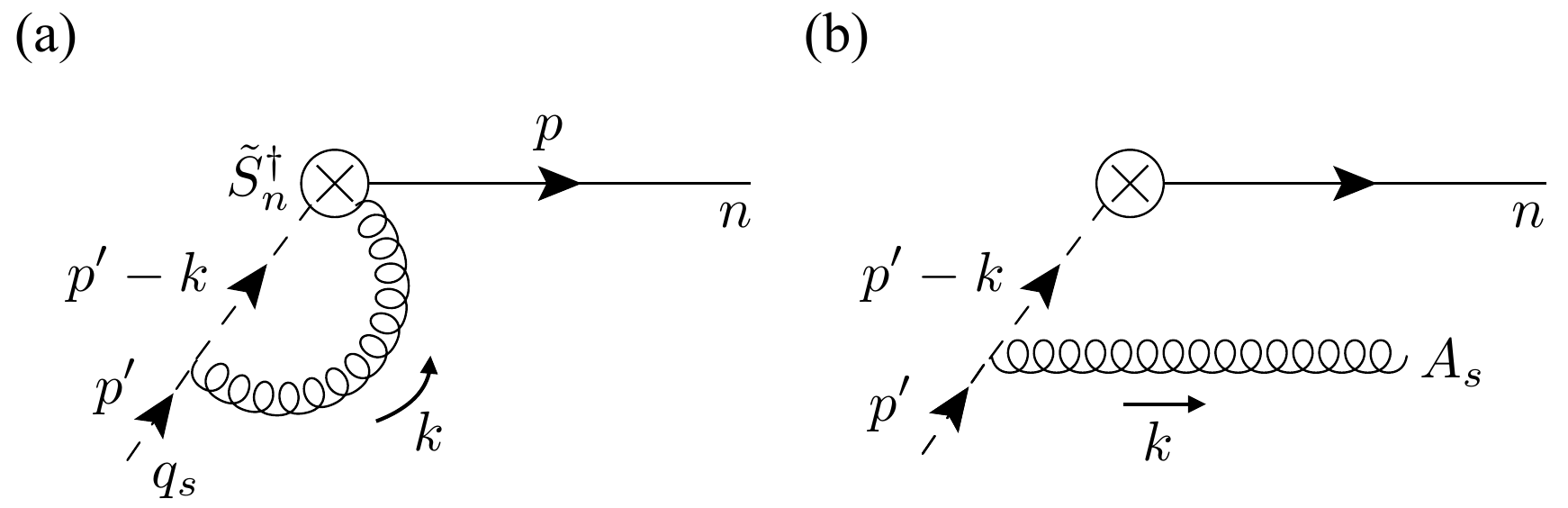}
\end{center}
\vspace{-0.8cm}
\caption{\label{fig6} \baselineskip 3.5 ex
(a) Feynman diagram of a soft gluon exchange  from a soft quark, (b) 
In the limit when the momentum of the soft gluon from the soft quark becomes collinear to the $n$ direction, 
the rapidity divergence arises.}
\end{figure}

Let us consider the rapidity divergence in $S_q$ including a soft quark [Fig.~\ref{fig6}-(a)].  
It arises when a soft gluon from the soft quark becomes $n$-collinear and its momentum 
$k$ reaches $\n\cdot k \to \infty$ (and $n\cdot k \to 0$), as shown in Fig.~\ref{fig6}-(b). (See Fig.~\ref{fig2} for comparison.)
In order to be consistent with the $n$-collinear sector, the rapidity regulator should be  $(\nu_+/\n\cdot k)^{\eta_+}$ as $\n\cdot k$ goes to infinity. 
On the other hand, since  the propagator in Fig.~\ref{fig6}-(b) is proportional to $1/p'\cdot k$, 
the rapidity regulator in the form $\nu_+/(p'\cdot k)$ is desired. 

As a result, we suggest the rapidity regulator for the soft quark sector at one loop as 
\begin{equation}
\label{regsq}
R_{S_q} (\nu_+;n^{\prime}\cdot k) = \Bigl(\frac{\nu_+}{p' \cdot k} \frac{n\cdot p'}{2}\Bigr)^{\eta_+} = 
\Bigl(\frac{\nu_+}{n' \cdot k}\frac{n\cdot n'}{2}\Bigr)^{\eta_+} 
\underset{\overline{n}\cdot k \rightarrow \infty}{\longrightarrow} \Bigl(\frac{\nu_+}{\overline{n}\cdot k}\Bigr)^{\eta_+},
\end{equation}
where $p^{\prime \mu} =E n^{\prime\mu}$, $E$ is the energy of the soft quark and $n^{\prime\mu}$ is the lightcone vector for
the soft massless quark. The last limit shows the consistency that we pick up the same rapidity regulator
in the soft quark sector as in the $n$-collinear sector when $\overline{n}\cdot k \rightarrow \infty$. 

At higher orders, it is complicated to set up a consistent rapidity regulator in the soft quark sector especially 
with the $n$-collinear regulator in Eq.~\eqref{wno}. That is because
multiple soft gluon radiations from the soft quark are not eikonalized. However, as discussed in Section \ref{sftreg}, 
the origin of the rapidity divergence from the collinear and soft gluon radiations is the same.
Therefore, once we set up the $n$-collinear rapidity regulator by modifying $W_n$ or the phase space,  
we can trace the corresponding factor in the multiple soft gluon radiations from the soft quark.

Before employing the regulator, the contribution to $S_q$ in Fig.~\ref{fig6}-(a) with the  pure on-shell dimensional regularization at one loop 
can be written as 
\bea
\tilde{M}_S^q &=& 2ig^2 C_F \mums \int \frac{d^D k}{(2\pi)^D} 
\frac{n\cdot (p' -k)}{(k^2 +i\veps) (k^2 -2k\cdot p' +i\veps)(n\cdot k -i\veps)} \nnb \\
\label{MSq0}
&=& 2ig^2 C_F \mums \Biggl[\frac{n\cdot p'}{n\cdot n'} \frac{1}{8\pi^2}\frac{i(4\pi)^{\eps}}{ \Gamma(1-\eps)} \int^{\infty}_0  
dn'\cdot k\int^{\infty}_0\frac{d\blp{k}^2(\blp{k}^2)^{-\eps}}{\blp{k}^2(\blp{k}^2+2En'\cdot k)} \\
&&~~~~~- \int \frac{d^D k}{(2\pi)^D} \frac{1}{k^2(k^2-2k\cdot p')}\Biggr].\nnb
\eea
Here the momentum $k^{\mu}$ is decomposed as
\begin{equation}
k^{\mu} =\frac{n^{\prime}\cdot k}{n\cdot n^{\prime}} n^{\mu} + \frac{n\cdot k}{n\cdot n^{\prime}} n^{\prime\mu} + k_{\perp}^{\mu},
\end{equation}
where $k^{\mu}_{\perp}$ is perpendicular to $n^{\mu}$ and $n^{\prime\mu}$. And $k^2$ is given by 
$k^2 =  2n\cdot k~ n^{\prime}\cdot k/n\cdot n^{\prime} -\mathbf{k}_{\perp}^2$.
In obtaining the first term in Eq.~\eqref{MSq0}, the integration measure is written as  
\begin{equation}
\frac{d^D k}{(2\pi)^D} =\frac{1}{16\pi^3} \frac{1}{n\cdot n^{\prime}} \frac{(4\pi)^{\eps}}{\Gamma (1-\eps)} 
dn\cdot k~ dn^{\prime} \cdot k~ d\mathbf{k}_{\perp}^2 (\mathbf{k}_{\perp}^2)^{-\eps},
\end{equation}
and the contour integral in the complex $n\cdot k$ plane is performed. 

In Eq.~\eqref{MSq0}, the rapidity divergence appears only in the first term. Applying the rapidity regulator, 
it can be written as 
\begin{equation}
\label{Ms1q}
M_{S1}^q = \frac{\alpha_s C_F}{2\pi} \frac{(\mu^2 e^{\gamma_{\mathrm{E}}})^{\eps}}{\Gamma (1-\eps)} 
\int_0^{\infty} \frac{dn^{\prime}\cdot k}{n^{\prime}\cdot k} R_{S_q} (\nu_+;n^{\prime}\cdot k)\int^{\infty}_0 
d\mathbf{k}_{\perp}^2 (\mathbf{k}_{\perp}^2)^{-\eps} 
\Bigl(\frac{1}{\mathbf{k}_{\perp}^2 + 2E n^{\prime} \cdot k} -\frac{1}{\mathbf{k}_{\perp}^2} \Bigr).
\end{equation}
In order to separate different types of divergences clearly, the integration region for $n^{\prime}\cdot k$ is divided into  
$0< n^{\prime}\cdot k <\Delta$  and $n^{\prime}\cdot k > \Delta$, where $\Delta$ is an arbitary soft energy scale. 
Then Eq.~\eqref{Ms1q} can be written into the three parts as 
\bea
\label{MSq1a}
M^q_{S1a} &=& \frac{\alpha_s C_F}{2\pi} \frac{(\mu^2 e^{\gamma_{\mathrm{E}}})^{\eps}}{\Gamma (1-\eps)} 
\int^\Delta_0 \frac{dn^{\prime}\cdot k}{n^{\prime}\cdot k} \int^{\infty}_0 d\mathbf{k}_{\perp}^2 (\mathbf{k}_{\perp}^2)^{-\eps} 
\Bigl(\frac{1}{\mathbf{k}_{\perp}^2 + 2E n^{\prime} \cdot k} -\frac{1}{\mathbf{k}_{\perp}^2} \Bigr),  \\
\label{MSq1b}
M^q_{S1b} &=& \frac{\alpha_s C_F}{2\pi} \frac{(\mu^2 e^{\gamma_{\mathrm{E}}})^{\eps}}{\Gamma (1-\eps)} 
\int_\Delta^{\infty} \frac{dn^{\prime}\cdot k}{n^{\prime}\cdot k} \int^{\infty}_0  
\frac{d\mathbf{k}_{\perp}^2 (\mathbf{k}_{\perp}^2)^{-\eps}}{\mathbf{k}_{\perp}^2 + 2E n^{\prime} \cdot k}, \\
\label{MSq1c}
M^q_{S1c} &=& -\frac{\alpha_s C_F}{2\pi} \frac{(\mu^2 e^{\gamma_{\mathrm{E}}})^{\eps}}{\Gamma (1-\eps)} 
\Bigl(\nu_+ \frac{n\cdot n'}{2}\Bigr)^{\eta_+} \int_\Delta^{\infty} \frac{dn^{\prime}\cdot k}{(n^{\prime}\cdot k)^{1+\eta_+}} \int^{\infty}_0  
d\mathbf{k}_{\perp}^2 (\mathbf{k}_{\perp}^2)^{-1-\eps}.
\eea
The rapidity divergence arises only in Eq.~\eqref{MSq1c}, and the rapidity regulator is inserted here. In Eqs.~\eqref{MSq1a} and 
Eq.~\eqref{MSq1b}, they have IR and UV divergences respectively, and there is no need for the rapidity regulator.

Combining all the results, we obtain 
\bea
M_{S1}^q &=&  M_{S1a}^q + M_{S1b}^q +M_{S1c}^q \nnb \\
&=& \frac{\alpha_s C_F}{2\pi} \Bigl[\frac{1}{\euv^2} -\frac{1}{\eir^2} +\Bigl( \frac{1}{\euv} -\frac{1}{\eir}\Bigr) 
\Bigl( -\frac{1}{\eta_+} +\ln \frac{\mu}{n\cdot p'} +\ln \frac{\mu}{\nu_+}\Bigr)\Bigr].
\eea 
The last term in Eq.~\eqref{MSq0} is given by
\begin{equation}
M_{S2}^q = \frac{\alpha_s C_F}{2\pi} \Bigl( \frac{1}{\euv} -\frac{1}{\eir}\Bigr) ,
\end{equation}
and the self-energy contribution for the soft quark are given by
\begin{equation}
Z_q^{(1)} + R_q^{(1)} = - \frac{\alpha_s C_F}{4\pi} \Bigl( \frac{1}{\euv} -\frac{1}{\eir}\Bigr). 
\end{equation}
Therefore the total soft contribution to $S_q$ at one loop is given as 
\begin{eqnarray}
S_q^{(1)}(n\cdot p',\mu,\nu_+) &=& M_{S1}^q + M_{S2}^q + \frac{1}{2} (Z_q^{(1)} + R_q^{(1)} ) \nnb \\
\label{Sqol}
&=& \frac{\alpha_s C_F}{2\pi} \Bigl[ \frac{1}{\euv^2} -\frac{1}{\eir^2} + \Bigl( \frac{1}{\euv} -\frac{1}{\eir}\Bigr) 
\Bigl( -\frac{1}{\eta_+} +\ln \frac{\mu^2}{\nu_+ n\cdot p'} +\frac{3}{4}\Bigr) \Bigr].
\end{eqnarray}

Interestingly, note that this result can be directly obtained from the factorized result of the back-to-back 
current in Eq.~\eqref{Sfscet}. Since the Sudakov form factor is invariant under boost, it should be the same either for the back-to-back current or for
the soft-collinear current. If we keep the $n$-collinear part in common in both cases, the soft function in the soft-collinear current should be
obtained as a product of the $\n$-collinear and the soft functions in the back-to-back current. It is expressed, with soft  $n\cdot p'$, as
\be
\label{factsq}
S_q (n\cdot p',\mu,\nu_+) = C_{\bar{n}} (n\cdot p',\mu,\nu_-) S_{n\bar{n}}(\mu,\nu_+,\nu_-) .
\ee 
 
In the back-to-back current, and in the $N$-jet operator, each factorized function 
has its own rapidity scale. Going further, if two factorized functions have the same rapidity scale, they can be combined, and can be regarded as a
single function. From this point of view, the soft function in the soft-collinear current can be acquired from the back-to-back current.  
The rapidity scale $\nu_-$ can be fixed near $n\cdot p'$ and no running is necessary in the combination. Finally the 
dependence on $\nu_-$ cancels when $C_{\bar{n}}$ and $S_{n\bar{n}}$ are combined. At one loop we easily reproduce Eq.~\eqref{Sqol} 
combining the results of $C_{\bar{n}}$ and $S_{n\bar{n}}$ in Eqs.~\eqref{Mnbol0} and \eqref{msoft} respectively.  
It is one of the advantages in using multiple rapidity scales.

The Sudakov form factor is derived from the back-to-back current and from the soft-collinear current, 
and it turns out to be the same. It is obvious
because the Sudakov form factor is a Lorentz invariant, and should be the same in all reference frames. 
Furthermore, since the Sudakov form factor is a
physical quantity, it should be independent of the factorization scales both in $\mu$ and $\nu$.  The main reason why we endeavor to
consider the Sudakov form factor in the two cases is that the evolution with respect to the rapidity scale looks seemingly 
different. In the back-to-back
current, there are two collinear directions involved and we introduce two rapidity scales with which 
each collinear part evolves using the RG equation.
On the other hand, in the soft-collinear current, there is only one rapidity scale associated with a single collinear direction. However, 
the evolution of the Sudakov form factor is the same, and independent of the factorization scales. It is also independent of
the order of evolution with respect to the renormalization scale and the rapidity scale. Because it is technical, the detail is deferred to Appendix~\ref{sudakov}.

\section{Sudakov form factor involving heavy quarks\label{sudheavy}}

We extend our discussion on the rapidity divergence from the massless case to the massive case. 
When a heavy quark is boosted and its energy is much larger than the mass $m$, the rapidity divergence also occurs 
in the collinear corrections. Like the massless case, the origin of the rapidity divergence is the radiation of the soft gluon, 
which cannot access the large rapidity region, hence it gives the divergence. Therefore the rapidity divergence
does not come from the naive collinear corrections to the boosted heavy quark sector, but from the zero-bin contribution. 

Let us consider the collinear loop correction of Fig.~\ref{fig1}-(a) for a boosted heavy quark. The naive collinear contribution is given by  
\begin{equation} \label{colmass}
\tilde{M}_n^m = -\frac{\alpha_s C_F}{2\pi} \frac{(\mu^2 e^{\gamma_{\mathrm{E}}} )^{\eps}}{\Gamma(1-\eps)} 
\int^1_0 dx \frac{1-x}{x} \int^{\infty}_0 d\mathbf{k}_{\perp}^2 \frac{(\mathbf{k}_{\perp}^2)^{-\eps}}{\mathbf{k}_{\perp}^2 +x^2 m^2}.
\end{equation}
Here the collinear divergence is absent due to the heavy quark mass, but the soft divergence remains as $x\to 0$. 
In the massless case, the $\mathbf{k}_{\perp}^2$-integral is independent of $x$, and we need a rapidity regulator to regulate the region 
near $x=0$ (though it is not the true rapidity divergence, which is obtained after the zero-bin subtraction). 
Even when we include the gluon mass $M$, the denominator becomes $\mathbf{k}_{\perp}^2 +(1-x) M^2$ and the same argument holds. [See Eq.~\eqref{naivec2}.]
However, in Eq.~\eqref{colmass}, the $\mathbf{k}_{\perp}^2$-integration affects the $x$ integral and there is no need to introduce the rapidity regulator.
Because of this, one may be tempted to say that there is no rapidity divergence in the collinear part, but the true collinear
contribution includes the zero-bin subtraction, from which the rapidity divergence arises.

As an illustration, let us consider the Sudakov form factor with a boosted heavy quark, in which the momentum transfer $Q$ is 
much larger than the quark mass $m$. The heavy quark momentum, collinear to the $n$-direction,  scales 
as $p^{\mu} = (p_+,p_{\perp},p_-) \sim Q(1,\lambda,\lambda^2)$, where $\lambda \sim m/Q$.
Now we perform the soft-collinear factorization with the soft mode scaling as
$p_s^{\mu} = (p_s^+, p_s^{\perp},p_s^-) \sim Q(\lambda,\lambda,\lambda)$. The decoupling of the soft gluons 
from the heavy quark sector yields the same Wilson line as in the case of a light quark. Therefore the zero-bin contribution 
to Eq.~\eqref{colmass} is also the same. As a result, the true collinear one-loop contribution with the zero-bin subtraction is given as 
\bea 
M_n^m = \tilde{M}_n^m-M_n^{\varnothing} &=& \frac{\as C_F}{2\pi}  \Bigl[\frac{1}{\UV} 
+\ln\frac{\mu^2}{m^2} +\frac{1}{2\IR}\Bigl(\frac{1}{\IR}+\ln\frac{\mu^2}{m^2}\Bigr) \nnb \\
\label{colol1}
&&+\Bigl(\frac{1}{\eta_+}+\ln\frac{\nu_+}{p_+}\Bigr) \Bigl(\frac{1}{\UV} -\frac{1}{\IR}\Bigr) 
+ \frac{1}{4} \ln^2\frac{\mu^2}{m^2}+2+\frac{\pi^2}{24} \Bigr].
\eea
Here the zero-bin contribution $M_n^{\varnothing}$ is the same as Eq.~\eqref{Mam0}, and the details of the calculation for Eq.~\eqref{colol1} 
are presented in Appendix~\ref{appBa}. Note that the collinear part for the boosted heavy quark has the same rapidity divergence 
as that of the light quark. This is not surprising because the rapidity divergence comes from the soft dynamics, i.e., the zero-bin 
contribution as far as the collinear sector is concerned.

\subsection{Heavy-to-heavy form factor}
 
Let us consider the situation in which the incoming and outgoing heavy quarks have large energy $\sim Q$ and they are moving in the opposite direction.  
In this case, the description of the form factor becomes more intriguing because the heavy quark mass $m$ enters into the system. 
Because the transferred momentum $Q$ is much larger than the quark mass $m$, the form factor involves a large logarithm of $Q/m$. 
So far, the resummation of this large logarithm has been considered in full QCD~\cite{Mitov:2006xs,Ahmed:2017gyt,Blumlein:2018tmz}, 
that is based on solving the so-called  ``KG integro-differential equation''~\cite{Collins:1980ih,Collins:1989bt}.  
In SCET the large logarithm of $Q/m$ can be fully resummed by the evolution with respect to the rapidity scale, in conjunction with the evolution
with respect to the renormalization scale.  

After integrating out hard interactions, the heavy-to-heavy form factor is matched onto SCET as 
\be 
\label{hhffmat}
F(Q^2,m^2) = H(Q^2,\mu) F_{\mr{SCET}} (Q^2,m^2,\mu).   
\ee
When $Q\gg m$, the mass dependence in hard interactions can be ignored. Therefore the hard function $H$ in Eq.~\eqref{hhffmat} is the same 
as that  of the light quark and the one-loop result is given by Eq.~\eqref{hardh}. 
Since the form factor is scale invariant, it can be computed at any scale. However, the structure of the complete factorization 
depends on which scale $\mu$ for $F_{\mr{SCET}}$ we consider.

\subsubsection{Factorization with $Q\gg m \sim \mu$}

When $Q\gg m \sim \mu$, the form factor in SCET, $F_{\mr{SCET}}$, can be factorized into the $n$- and 
$\n$-collinear parts and the soft part.
% like the case of the light quark.  --- I think it is redundant.
The boosted heavy quark sectors are described by the $n$- and $\n$-collinear interactions respectively, where the 
$n$- and $\n$-collinear momenta scale as $p_n^{\mu} = (\overline{n}\cdot p, p_{\perp}, n\cdot p) \sim  Q (1, \lambda, \lambda^2)$ and 
$p_{\bar{n}}^{\mu} = Q(\lambda^2,\lambda,1)$ with $\lambda \sim m/Q$. The collinear interactions are legitimately described 
by the massive version of SCET ($\mr{SCET_M}$)~\cite{Leibovich:2003jd,Rothstein:2003wh,Chay:2005ck}. 
The soft interactions with the momentum scaling $p_s^{\mu} \sim Q(\lambda,\lambda,\lambda)$
mediates the crosstalk between the two heavy, collinear sectors. As explained above, the soft part remains unchanged, compared to 
the case of the light quark. Therefore, $F_{\mr{SCET}}$ is written as 
\be 
\label{facthh1}
F_{\mr{SCET}} (Q^2,m^2,\mu) = C_{m,n} (p_+,m^2, \mu, \nu_+) C_{m,\bar{n}} (p_-^{\prime},m^2, \mu, \nu_-) 
S_{n\bar{n}}(\mu, \nu_+, \nu_-),
\ee
where the largest external momentum component in the $n$ $(\n)$-direction is given by $p_+$ $(p'_-)$. 
The soft function $S_{n\bar{n}}$ appears in Eq.~\eqref{Sfscet} with the non-zero gluon mass $M$.
Here we use the pure dimensional regularization by expressing the IR divergence as poles in $\eir$. 
The bare one-loop result for $S_{n\bar{n}}$ is given in Eq.~\eqref{msoft}.

The complete one-loop result for the collinear part $C_n$ can be obtained by the sum of  $M_n^{m}$ in Eq.~\eqref{colol1} and
the contribution of the self energy diagram as
\begin{equation}
C_{m,n}^{(1)} =M_n^m +\frac{1}{2}(Z_Q^{(1)} + R_Q^{(1)}). 
\end{equation}
Here the wavefunction renormalization and the residue of the heavy quark are given by 
\begin{equation}
Z_Q^{(1)} = -\frac{\alpha_s C_F}{4\pi}\frac{1}{\UV},~~ R_Q^{(1)} = -\frac{\alpha_s C_F}{4\pi} \Bigl(\frac{2}{\IR} +3 \ln \frac{\mu^2}{m^2} +4\Bigr).
\end{equation}
The bare one-loop result, $C_{m,n}^{(1)}$, is given as
\begin{align} \label{cnone}
C_{m,n}^{(1)} (p_+, \mu,\nu_+, m^2)& =\frac{\alpha_s C_F}{4\pi} \Bigl[ \frac{3}{2} \frac{1}{\euv} +\frac{1}{\eir^2} 
+\frac{1}{\eir} \Bigl( -1 +\ln \frac{\mu^2}{m^2}\Bigr)  \nonumber \\
&+ 2 \Bigl(\frac{1}{\euv} -\frac{1}{\euv} \Bigr) \Bigl( \frac{1}{\eta_+} +\ln \frac{\nu_+}{p_+}\Bigr)  
+ \frac{1}{2} \ln^2 \frac{\mu^2}{m^2} +\frac{1}{2} \ln \frac{\mu^2}{m^2} +2+\frac{\pi^2}{12}\Bigr], 
\end{align}
and $C_{m,\bar{n}}^{(1)}$ can be obtained from Eq.~\eqref{cnone} by replacing $\{p_+,\eta_+,\nu_+\}$ with $\{p'_-,\eta_-,\nu_-\}$. 

The heavy quark mass in the collinear parts regularizes the collinear divergence, but it does not affect the renormalization behavior 
with respect to the 
scales  $\mu$ and $\nu$. Hence the anomalous dimensions for $\mu$ and $\nu$ in each factorized function 
of Eq.~\eqref{facthh1} are the same as those in the case of the light quark. 
Also the exponentiated evolution kernel to resum large logarithms is given by the same form as the form factor with light quarks. 
The resummed result for the light form factor is given in Eqs.~\eqref{exp1} and \eqref{exp2}. 
In the exponentiation, 
The newly added term by applying the $\nu$-evolution is given by     
\be 
\label{added}
- \ln \frac{\nu_n^+ \nu_{\bar{n}}^-}{\nu_s^+ \nu_s^-}\cdot  a[\Gamma_C] (\mu_s,M). 
\ee
For the heavy-to-heavy form factor, the soft scales are given by $\nu_s^{\pm} \sim \mu_s \sim m$.  
The gluon mass $M$ to regularize the IR divergence also scales as $m$. Therefore, to NLL accuracy, 
Eq.~\eqref{added} is given as %can be here interpreted and approximated  as  
\be 
- \ln \frac{Q^2}{m^2}\cdot  a[\Gamma_C] (m,M) \approx - \frac{\as(m) C_F}{2\pi} \ln \frac{Q^2}{m^2} \ln \frac{m^2}{M^2}. 
\ee 

\subsubsection{Factorization with $Q\gg m \gg \mu$}
\label{hhff2} 

%We can consider the form factor in SCET, $F_{\mr{SCET}}$, at the scale 
When $\mu$ is much smaller than the heavy quark mass $m$,
the heavy quark mass $m$ and the collinear interactions with offshellnesses $p_n^2 \sim p_{\n}^2 \sim m^2$ are 
integrated out. Then the heavy quarks undergo the collinear-soft (csoft) interactions~\cite{Bauer:2011uc}. 
The csoft momenta $p_{cs}~(p_{\cs})$ in the $n~(\n)$ direction scale as  
\be 
p_{cs}^{\mu} = (p_{cs}^+,p_{cs}^{\perp},p_{cs}^-) = Q \zeta (1,\lambda,\lambda^2),~~p_{\cs}^{\mu} = Q\zeta (\lambda^2,\lambda,1),
\ee
where $\zeta$ is an another small parameter of order $\mu/m$.  

%After integrating out the collinear interactions, 
For the heavy quark with the csoft interaction, its momentum can be expressed with a fixed velocity  $v$  as 
\be 
p^{\mu} = mv^{\mu} + k^{\mu},  
\ee
with $v^2=1$. For the outgoing heavy quark moving in the $n$ direction, the velocity $v_n^{\mu}$ scales as 
$(v_n^+,v_n^{\perp},v_n^-)=(1/\lambda,1,\lambda)$. For the incoming quark in the $\n$ direction, the velocity scales 
as $v_{\bar{n}}^{\mu} = (\lambda,1,1/\lambda)$. These heavy quarks and their csoft interactions are  
described by the boosted heavy quark effective theory (bHQET)~\cite{Fleming:2007qr,Fleming:2007xt}, 
which can be directly obtained from $\mr{SCET_M}$~\cite{Kim:2020dgu,preparation}.  

%When the factorization scale $\mu$ is much lower than $m$, the heavy quarks, after integrating out the modes with the 
%offshellness $\sim m^2$,  are described by the csoft 
%interaction. As a result, t
% Redundant. See the first paragraph in this subsection.

The original collinear parts for the heavy quark sectors in Eq.~\eqref{facthh1} can be refactorized as 
\bea
\label{Cnref}
C_n (p_+,m^2, \mu, \nu_+)  &=& C_m (m^2,\mu) B_n (p_+,m^2,\mu,\nu_+), \\
\label{Cnbref}
C_{\bar{n}} (p_-^{\prime},m^2, \mu, \nu_-)  &=& C_m (m^2,\mu) B_{\bar{n}} (p'_-,m^2,\mu,\nu_-).
\eea
Here $C_m$ is the result of integrating out the collinear interactions. At NLO in $\as$, it is given as~\cite{Fleming:2007xt,Neubert:2007je,Fickinger:2016rfd}
\be 
\label{cm}
C_m (m^2, \mu) = 1+\frac{\alpha_s C_F}{4\pi} \Bigl(\frac{1}{2} \ln \frac{\mu^2}{m^2}+\frac{1}{2} \ln^2 \frac{\mu^2}{m^2}+2+\frac{\pi^2}{12}\Bigr).
\ee 
%Check.
In Eqs.~\eqref{Cnref} and \eqref{Cnbref}, $B_n$ and $B_{\bar{n}}$ are the $n$-csoft and $\n$-csoft functions in bHQET. 
As we will show, the csoft functions depend on the rapidity scales, which can be chosen to be of order $Q\zeta$ 
to minimize large rapidity logarithms. This choice of the scale corresponds to the size of the largest component of the csoft momentum.  

For $F_{\mr{SCET}}$ at $\mu \ll m$, we also have to introduce ultrasoft (usoft) interactions, with the usoft momentum scaling as 
$p_{us} \sim Q\zeta(\lambda,\lambda,\lambda)$. The usoft interactions are decoupled from both 
the csoft parts. And the resulting usoft function has the same form as the soft function $S_{n\bar{n}}$ in Eq.~\eqref{facthh1}. 
The only difference is that it consists of usoft gluons now.    
Let us consider the transition from the soft part to the usoft part systematically.
%However, since we originally have the soft mode with $p_s \sim Q(\lambda,\lambda,\lambda)$ that are decoupled from the 
%collinear parts, we need to separate the usoft part from the soft part systematically. 
% I changed these sentences since they look complicated.
In calculating the soft function 
$S_{n\bar{n}}$ at higher orders in $\as$, we have to subtract the usoft contributions to avoid double counting. 
Since the subtracted usoft contribution is always the same as the soft contribution, the overall contribution to 
the soft function should vanish to all orders in $\as$.  
Therefore, when $\mu \ll m$,  we have the usoft part, instead of the soft part.
As a result, the soft function $S_{n\bar{n}}$ in Eq.~\eqref{facthh1} is replaced with the usoft function. 

\begin{figure}[t]
\begin{center}
 \includegraphics[height=6cm]{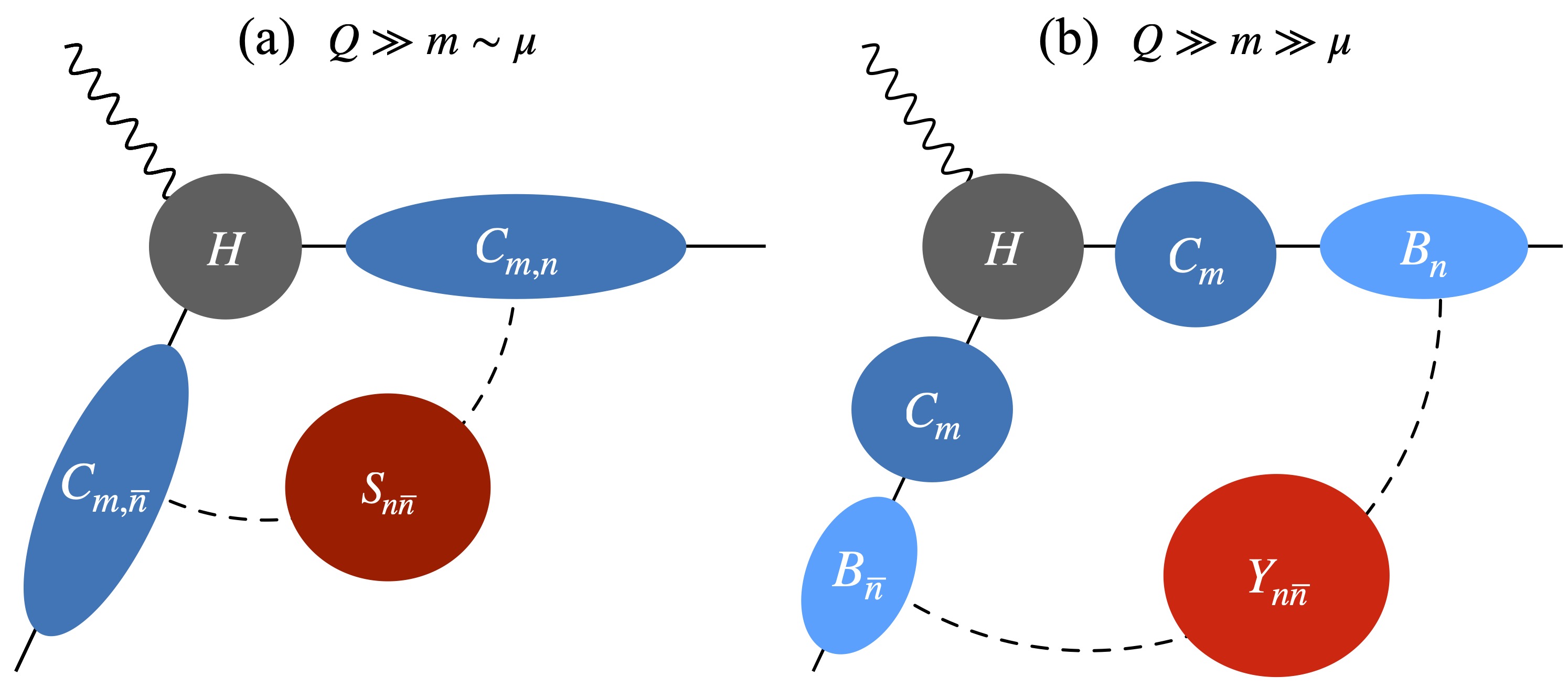}
\end{center}
\vspace{-0.5cm}
\caption{\label{hhfact}
Schematic picture of the factorization structure in the heavy-to-heavy form factor: 
(a) For $\mu\sim m$, the form factor consists of the hard, collinear and soft functions. 
%The soft interaction participates in the crosstalk between two collinear sectors. 
(b) For $\mu\ll m$, the collinear interactions with $p^2\sim m^2$ are integrated out to yield $C_m$. 
As a result, we have the csoft functions, $B_n$ and $B_{\bar{n}}$, and the usoft function $Y_{n\bar{n}}$.}
\end{figure}

Finally  $F_{\mr{SCET}}$ for $\mu \ll m$ can be factorized as 
\be 
\label{facthh2}
F_{\mr{SCET}} (Q^2,m^2,\mu) = [C_m (m^2,\mu)]^2 \cdot B_n (p_+,m^2, \mu, \nu_+) B_{\bar{n}} (p_-^{\prime},m^2, \mu, \nu_-) 
Y_{n\bar{n}}(\mu, \nu_+, \nu_-).
\ee
Here $Y_{n\bar{n}}$ is the usoft function, which is the same as $S_{n\bar{n}}$ in Eq.~\eqref{msoft}.
This factorization with $\mu \ll m$ can be also useful for the  elecroweak Sudakov form factor involving top quarks or 
the semi-inclusive deep inelastic scattering initiated by a heavy quark when an outgoing heavy quark (hadron) has  small $p_T$ 
compared with the heavy quark mass. 
The structure of the factorization for the form factor is illustrated in Fig.~\ref{hhfact} and is compared to the case for $\mu \sim m$. 

% Is the word ``irreducible'' necessary?
The irreducible one-loop contribution to csoft function $B_n$ can be inferred from the collinear one-loop calculation for 
the heavy quark in Eq.~\eqref{colmass}. Taking the gluon loop momentum to the csoft limit, $k^{\mu} \sim Q\zeta(1,\lambda,\lambda^2)$, 
the naive one-loop contribution to $B_n$ is written as\footnote{
This calculation should be performed in bHQET. If we set $v_n^{\perp} =0$,  
$v_n = (\frac{p_+}{m} \frac{n^{\mu}}{2} + \frac{m}{p_+} \frac{\n^{\mu}}{2})$, then the heavy quark 
propagator in bHQET becomes 
\be 
\frac{1}{v_n\cdot k} \propto \frac{1}{\blp{k}^2 + x^2 m^2},  \nnb 
\ee
where the loop momentum satisfies the on-shell condition, $k_- = \blp{k}^2/k_+$ after integraing over $k_-$ in the complex plane. 
}   
\be
\label{ncsoftol}
\tilde{M}^m_{cs} = -\frac{\alpha_s C_F}{2\pi} \frac{(\mu^2 e^{\gamma_{\mathrm{E}}} )^{\eps}}{\Gamma(1-\eps)} 
\int^{\infty}_0  \frac{dx}{x} \int^{\infty}_0 d\mathbf{k}_{\perp}^2 \frac{(\mathbf{k}_{\perp}^2)^{-\eps}}{\mathbf{k}_{\perp}^2 +x^2 m^2}.
\end{equation}
Here $x = k_+/p_+ \sim \zeta$ is a small quantity, hence the upper limit of $x$ is set to infinity. 
And $\blp{k}^2$ and $x^2 m^2$ is power-counted as $\mc{O}(Q^2\zeta^2 \lambda^2)$.
Eq.~\eqref{ncsoftol} seems to possess the rapidity divergence as $x\to \infty$. However, %the divergence is regularized 
it turns out to be the UV divergence, $1/\UV$, due to the presence of the heavy quark mass (the term $x^2m^2$ in the integral of $\blp{k}^2$). 
Therefore Eq.~\eqref{ncsoftol} does not contain the rapidity divergence as in the collinear case. 

Instead, the rapidity divergence comes from the zero-bin contribution when the loop momentum $k$ becomes usoft. 
Since the usoft momentum scales as $k_{us} \sim Q\zeta(\lambda,\lambda,\lambda)$, the momentum fraction $x$ in Eq.~\eqref{ncsoftol} 
gets further suppressed as $\mc{O}(\zeta \lambda)$ and the term $x^2m^2$ in the integral of $\blp{k}^2$ can be ignored. 
So the zero-bin contribution ends up with Eq.~\eqref{Mam0}, and it reads\footnote{
When matched onto bHQET to describe the csoft interactions, the collinear Wilson line $W_n$ becomes $W_n^{cs}$, which has the same form as $W_n$ 
with the collinear gluon $A_n$ replaced by the csoft gluon $A_{n}^{cs}$. So the same rapidity regulator for the csoft calculation can be used in $W_n^{cs}$.
}
\be 
\label{zbinus}
M_{cs}^{\varnothing} = - \frac{\alpha_s C_F}{2\pi} \frac{(\mu^2 e^{\gamma_{\mathrm{E}}})^{\eps}}{\Gamma (1-\eps)} \Bigl(\frac{\nu_+}{p_+}\Bigr)^{\eta_+} \int^{\infty}_0 
\frac{dx}{x^{1+\eta_+}}\int^{\infty}_0
d\blp{k}^2 (\blp{k}^2)^{-1-\eps}.   
\ee

Therefore, the regular csoft (irreducible) one-loop contribution is given as 
\bea
\label{mcsol}
M_{cs}^m &=& \tilde{M}_{cs}^m - M_{cs}^{\varnothing}  \\
&=& -\frac{\as C_F}{4\pi}\Bigl[\frac{1}{\UV^2} - \frac{1}{\IR^2} + \Bigl(\frac{1}{\UV} - \frac{1}{\IR}\Bigr) \ln\frac{\mu^2}{m^2} - \Bigl(\frac{2}{\eta_+} +2\ln \frac{\nu_+}{p_+} \Bigr) \Bigl(\frac{1}{\UV} - \frac{1}{\IR}\Bigr) \Bigl]. \nnb 
\eea
For the detailed calculation, we refer to Appendix.~\ref{appBb}. 
The self-energy contribution to $B_n$ is given as $(Z_h^{(1)} + R_{h}^{(1)})/2$, 
where the heavy quark wave function renormalization and the residue at one loop in bHQET are  given by 
\be 
Z_h^{(1)} = \frac{\alpha_s C_F}{2\pi} \frac{1}{\euv},~~ R_h^{(1)} = -\frac{\alpha_s C_F}{2\pi}\frac{1}{\eir}. 
\ee
Finally, the bare one-loop contribution to $B_n$ is given as 
\begin{align} 
\label{ncsoftL}
B_n^{(1)} (p_+,m^2,\mu, \nu_+) &= M_{cs}^m +\frac{1}{2} (Z_h^{(1)} +R_h^{(1)})   \\ 
&= \frac{\alpha_s C_F}{4\pi} \Bigl[ - \frac{1}{\euv^2} +\frac{1}{\eir^2} +\Bigl( \frac{1}{\euv} -\frac{1}{\eir}\Bigr) \Bigl( \frac{2}{\eta_+} 
+ 2 \ln\frac{\nu_+}{p_+} -\ln \frac{\mu^2}{m^2}+1\Bigr)\Bigr]. \nnb
\end{align}
The csoft function $B_{\bar{n}}^{(1)}$ can be obtained by replacing $\{p_+,\eta_+,\nu_+\}$ in $B_n^{(1)}$ with $\{p'_-,\eta_-,\nu_-\}$.  

If the IR divergence is regularized using the gluon mass $M$ that scales as $Q\zeta \lambda$, the renormalized NLO result for $B_n$ is expressed as   
\be 
\label{ncsoft1}
B_n (p_+,m^2,\mu, \nu_+) = 1+\frac{\alpha_s C_F}{2\pi} \Bigl[ \ln \frac{\mu^2}{M^2} \Bigl( 
\ln \frac{\nu_+ m}{p_+ M} - \ln \frac{\mu}{M} +\frac{1}{2} \Bigr) +\frac{1}{4} \ln^2 \frac{\mu^2}{M^2}  +\frac{\pi^2}{24}\Bigr],
\ee
from which, we can determine the characteristic rapidity scales for the csoft functions to minimize the large logarithms, 
and they are given by 
\be
\nu_{cs}^+ \sim \frac{p_+ M}{m} \sim Q\zeta,~~\nu_{\cs}^- \sim \frac{p'_- M}{m} \sim Q\zeta.
\ee
These scales correspond to the sizes of the largest components of the $n$- and $\n$-csoft momenta. 
For the NLO result of the usoft function with the gluon mass, we refer to Eq.~\eqref{Snnbm}. 

The anomalous dimensions for the factorized functions with respect to $\mu$ and $\nu$ scales are defined in Eq.~\eqref{RGEs}.  
%\begin{equation}
%\label{RGEs}
%\gamma^{\mu} = \frac{1}{f} \frac{d}{d\ln\mu} f,~~~ \gamma^{\nu} = \frac{1}{f} \frac{d}{d\ln\nu} f,
%\end{equation} 
We list the anomalous dimensions for the factorized functions in Eq.~\eqref{facthh2} as well as the hard function $H$ at order $\as$. 
The $\mu$-anomalous dimensions are given as
\begin{align} \label{hmuanom}
\gamma_H  &=   \frac{\alpha_s C_F}{2\pi} \Bigl( -2 \ln \frac{\mu^2}{Q^2} -3\Bigr), \ 
\gamma_m = \frac{1}{[C_m]^2} \frac{d[C_m]^2}{d\ln\mu}  =  \frac{\alpha_s C_F}{2\pi} \Bigl(2\ln \frac{\mu^2}{m^2} +1\Bigr), \nonumber \\
\gamma_{cs}^{\mu}&=\frac{1}{B_n} \frac{dB_n}{d\ln\mu} = \frac{\alpha_s C_F}{2\pi} ( 2 \ln \frac{\nu_+ m}{p_+ \mu} +1\Bigr), \
\gamma_{\cs}^{\mu}=\frac{1}{B_{\n}} \frac{dB_{\n}}{d\ln\mu} = \frac{\alpha_s C_F}{2\pi} ( 2 \ln \frac{\nu_- m}{p'_- \mu} +1\Bigr), \nonumber \\
\gamma_{us}^{\mu} &=     \frac{\alpha_s C_F}{\pi}\ln \frac{\mu^2}{\nu_+ \nu_-}.
\end{align}
The nonzero $\nu$-anomalous dimensions appear in the csoft and usoft functions, and they are given by
\be 
\label{hnuanom}
\gamma_{cs}^{\nu_+} =\gamma_{\cs}^{\nu_-}=\frac{\alpha_s C_F}{2\pi} \ln \frac{\mu^2}{M^2},~~~
\gamma_{us}^{\nu_+} = \gamma_{us}^{\nu_-} = -\frac{\alpha_s C_F}{2\pi} \ln \frac{\mu^2}{M^2}.
\ee
With $Q^2 = p_+p'_-$, the sum of the anomalous dimensions cancel:
\bea
&& \gamma_H^{\mu} + \gamma_m^{\mu} +\gamma_{cs}^{\mu} + \gamma_{\cs}^{\mu} +
\gamma_{us}^{\mu} =0, \\
&&\gamma_{cs}^{\nu_+}+\gamma_{us}^{\nu_+} =0,~~~\gamma_{\cs}^{\nu_-}+\gamma_{us}^{\nu_-} =0. 
\eea
This guarantees that the form factor is independent of $\mu$ and $\nu$.  
%we can determine the typical renomalization and rapidity scales:
%\begin{align}
%&\mu_H \sim Q, \ \mu_C \sim \mu_{\bar{C}} \sim m, \ \mu_{cs} \sim \mu_{us} \sim M,  \nonumber \\
%& \nu_{cs}^+ \sim p_+ M/m \sim Q\eta, \ \nu_{\bar{cs}}^- \sim p'_- M/m \sim Q\eta, \ \nu_{us}^{\pm} \sim M.
%\end{align}
The resummation of the large logarithms using the anomalous dimensions to NLL accuracy is presented 
in Appendix~\ref{resumhh}.

Combining all the ingredients of the factorized functions in Eqs.~\eqref{facthh1} and \eqref{facthh2}, 
the (bare) one-loop result of the form factor in SCET is written as 
\bea 
F_{\mr{SCET}}^{(1)}(Q^2,m^2,\mu) &=& C_{m,n}^{(1)} + C_{m,\bar{n}}^{(1)} + S_{n\bar{n}}^{(1)}  \nnb \\
&=& 2C_m^{(1)} +B_n^{(1)}+B_{\bar{n}}^{(1)} + Y_{n\bar{n}}^{(1)} \nnb \\
\label{hqffscetnlo} 
&=& \frac{\alpha_s C_F}{2\pi} \Bigl[ \frac{1}{\euv^2} 
+\frac{1}{\euv} \Bigl( \ln \frac{\mu^2}{p_+p'_-} +\frac{3}{2}\Bigr)
-\frac{1}{\eir} \Bigl( \ln \frac{m^2}{p_+ p'_-} +1\Bigr)  \\
&&~~~ + \frac{1}{2} \ln^2 \frac{\mu^2}{m^2} + \frac{1}{2} \ln \frac{\mu^2}{m^2}  + 2+\frac{\pi^2}{12} \Bigr], \nnb 
\eea 
where $p_+ p'_- = Q^2$ and we used a massless gluon. The rapidity divergences exactly cancel between the collinear and soft parts 
as well as between the csoft and the usoft parts. 
%The anomalous dimensions from thee UV divergences cancels when combined with $\gamma_H^{\mu}$ in Eq.~\eqref{hmuanom}. 
 % I think the above sentence is redundant.
  
Combining the hard function in Eq.~\eqref{hardh} and the renormalized result of Eq.~\eqref{hqffscetnlo}, in the limit $Q\gg m$, 
we also show the one-loop result of the form factor in full QCD  as\footnote{Here the $\mu$-dependence is due to the 
dimensional regularization of the IR divergence, not a renormalization effect.}  
\be
\label{hqffnlo}
F^{(1)} (Q^2,m^2) = \frac{\alpha_s C_F}{2\pi} \Bigl[-2\Bigl(1+\ln\frac{m^2}{Q^2} \Bigr) \Bigl( \frac{1}{\IR} + \ln\frac{\mu^2}{Q^2}\Bigr) - \ln\frac{m^2}{Q^2}+\ln^2\frac{m^2}{Q^2} -4 +\frac{\pi^2}{3}\Bigr]. 
\ee
This result coincides with  Ref.~\cite{Mitov:2006xs}. 
 
\subsection{Heavy-to-light form factor}

We can also describe the form factor in the weak transition of a heavy quark to a light quark. 
When the transferred momentum is much larger than the heavy quark mass, i.e, $Q \gg m$, 
the heavy-to-light form factor for the weak current such as $\overline{q}\gamma^{\mu}(1-\gamma_5) Q$ can be matched onto SCET as
\be 
\label{Fhl}
F_{hl} (Q^2,m^2) = H(Q^2,\mu) F_{\mr{SCET}}^{hl} (Q^2,m^2,\mu). 
\ee 
Here the hard function is the same as in the light form factor as well as in the heavy-to-heavy form factor.

The factorization of $F_{\mr{SCET}}^{hl}$ can be easily inferred from our analyses on the light-to-light and the heavy-to-heavy cases. 
For $\mu \sim m$, $F_{\mr{SCET}}^{hl}$ can be factorized as 
\be 
\label{hlfac1} 
F_{\mr{SCET}}^{hl} (Q^2,m^2,\mu\sim m) = C_n (p_+,\mu,\nu_+) C_{m,\bar{n}} (p'_-,m^2,\mu,\nu_-) S_{n\bar{n}} (\mu,\nu_+,\nu_-). 
\ee
Here $C_n$ is the collinear function for the light quark in Eqs.~\eqref{Sfscet} and \eqref{Sudsc},
and $C_{m,\n}$ is the ($\n$-)collinear function for the heavy quark in Eq.~\eqref{facthh1}. $S_{n\bar{n}}$ is the universal 
soft function for the back-to-back current. 

For $\mu \ll m$, we need to integrate out the collinear interaction with offshellness $\sim m^2$ in the heavy quark sector, which 
gives $C_m(m^2,\mu)$ in Eq.~\eqref{cm}. The SCET form factor can be written as 
\be 
\label{hlfac2} 
F_{\mr{SCET}}^{hl} (Q^2,m^2,\mu\ll m) = C_m(m^2,\mu)\cdot C_n (p_+,\mu,\nu_+) B_{\bar{n}} (p'_-,m^2,\mu,\nu_-) 
Y_{n\bar{n}} (\mu,\nu_+,\nu_-), 
\ee
where $B_{\bar{n}}$ and $Y_{n\bar{n}}$ are the csoft and the usoft functions introduced in Section~\ref{hhff2}. And the collinear function 
for the light quark here describes the interactions with the offshellness $p^2 \ll m^2$. The structure of the factorization is illustrated 
in Fig.~\ref{hlfact}. The resummed result for the form factor is presented in Appendix~\ref{reshl}. 

In Eq.~\eqref{hlfac2}, the behavior at the scale $\mu\ll m$ is described by $C_n$, $B_{\bar{n}}$ and $Y_{n\bar{n}}$. 
When we combine the three pieces as $f_{hl} = C_n\cdot B_{\bar{n}} \cdot Y_{n\bar{n}}$, the NLO result in $\as$ is given by 
\bea 
f_{hl} (Q^2, m^2,\mu) &=& 1+ C_{n}^{(1)}+ B_{\bar{n}}^{(1)} + Y_{n\bar{n}}^{(1)} \nnb \\
\label{fhlNLO}
&=& 1 + \frac{\alpha_s C_F}{4\pi} \Bigl[\frac{1}{\euv^2} -\frac{1}{\eir^2} +\Bigl( \frac{1}{\euv} 
-\frac{1}{\eir}\Bigr) \Bigl( 2\ln \frac{\mu m}{p_+ p'_-}  +\frac{5}{2}\Bigr)\Bigr],
\eea 
with $Q=p_+p'_-$.  There is no rapidity divergence in this combination, as expected.
%So this combination does not yield the rapidity divergence as we expect. 
%From the UV divergences in Eq.~\eqref{fhlNLO}, t
The anomalous dimension is given by  
\be
\label{fhlanom} 
\gamma_{f_{hl}} = \frac{\alpha_s C_F}{2\pi} \Bigl( 2 \ln \frac{\mu m}{p_+ p'_-} +\frac{5}{2}\Bigr)
= \frac{\alpha_s C_F}{2\pi} \Bigl( 2 \ln \frac{\mu m}{Q^2} +\frac{5}{2}\Bigr).  
\ee
Since the form factor in QCD is scale invariant, $\gamma_{f_{hl}}$ satisfies 
\be 
\gamma_H + \frac{1}{2} \gamma_m + \gamma_{f_{hl}} = 0,
\ee
where $\gamma_H$ and $\gamma_m$ at order $\as$ are given in Eq.~\eqref{hmuanom}. 

\begin{figure}[h]
\begin{center}
 \includegraphics[height=6cm]{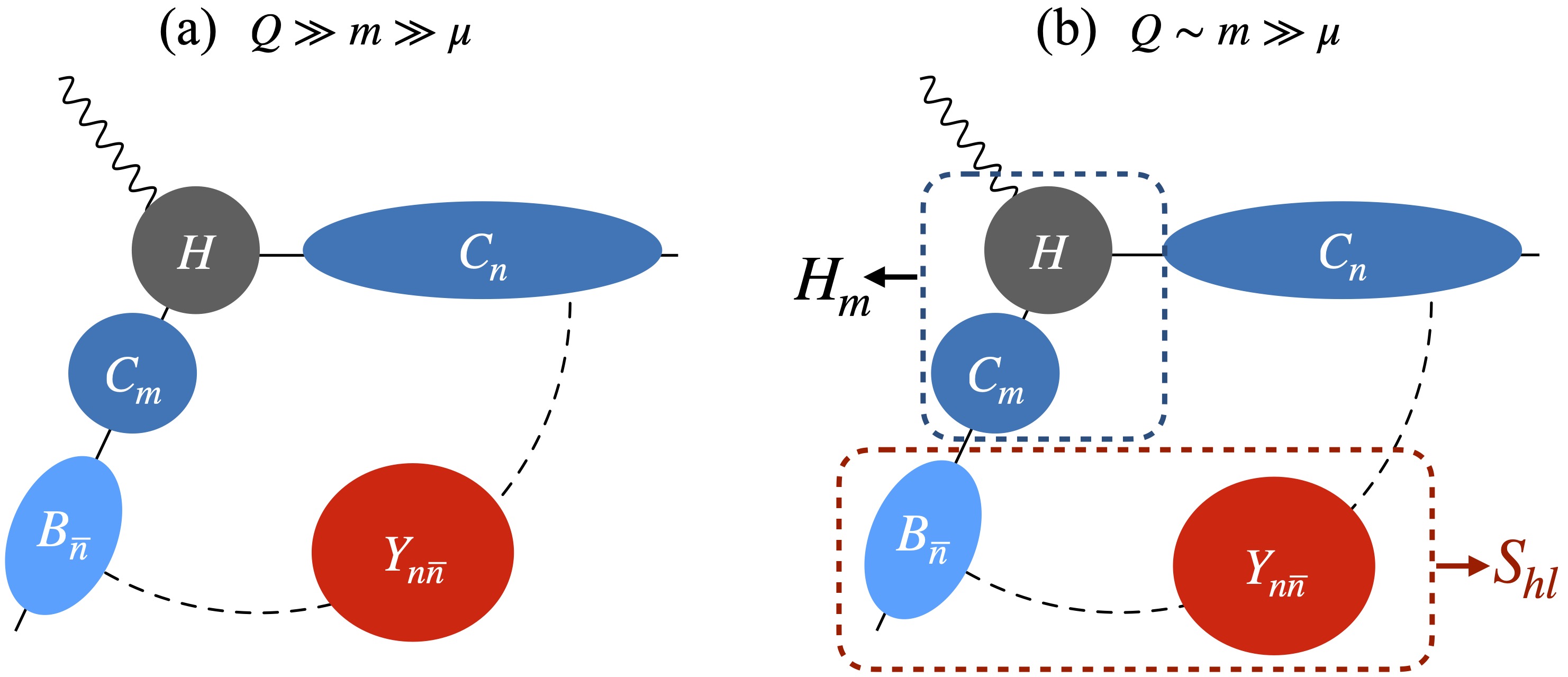}
\end{center}
\vspace{-0.5cm}
\caption{\label{hlfact}
Schematic structure of the factorization for the heavy-to-light form factor: (a) For   
$Q\gg m\gg \mu$, the incoming heavy quark is boosted and described by $C_m$ and $B_{\bar{n}}$. 
Here $B_{\bar{n}}$, $Y_{n\bar{n}}$ and $C_n$ have almost the same offshellness $(p^2 \sim \mu^2 \ll m^2)$, but they are separated 
by the rapidities. (b) The form factor with a static heavy quark. Here $Q\sim m$ and the heavy quark experiences
the soft interaction at lower energy scale $\mu$. So $H$ and $C_m$ can be merged into a new hard function $H_m$. 
And $B_{\bar{n}}$ and $Y_{n\bar{n}}$ can be combined into a new soft function $S_{hl}$. 
  }
\end{figure}

Interestingly enough, the factorized result of $F_{hl}$ in Eq.~\eqref{Fhl} with Eq.~\eqref{hlfac2} can be applied to the $B$ decay 
in the rest frame of the $B$ meson, where the factorization scale $\mu$ is much smaller than the heavy quark mass $m$.
Suppose that the static heavy quark with the momentum $p'$ decays to an energetic light quark with the momentum $p$ that has a maximal energy. 
In this case, the transferred momentum squared is given as
\be 
q^2 = (p-p')^2 = m^2 -2p\cdot p' = m^2 -Q^2 =0. 
\ee
Here we set $2p\cdot p' \equiv Q^2$, which we have so far denoted the transferred momentum squared for the back-to-back 
current with very energetic quarks. The produced light quark can be described by the $n$-collinear field and 
its largest momentum component is given by $p_+ \equiv \n\cdot p = m$.

Now we can find the factorized form factor for the static heavy quark decay from the energetic heavy-to-light form factor 
(Eq.~\eqref{Fhl} employing Eq.~\eqref{hlfac2}). As depicted in Fig.~\ref{hlfact}-(b), since $Q^2 =m^2$, $H$ in Eq.~\eqref{Fhl} and 
$C_m$ in Eq.~\eqref{hlfac2} can be merged into a new hard function $H_m$, which is given by 
\be 
\label{ham}
H_m (m^2,\mu) = H(Q^2=m^2,\mu) C_m(m^2,\mu) + \mr{finite~terms}. 
\ee
Here ``the finite terms" denote the higher-order corrections in powers of $m^2/Q^2$ in the energetic heavy-to-light form factor. 
They are neglected when $Q^2 \gg m^2$, which must be included here. 
However, the $\mu$-dependence resides genuinely in the combination of $H\cdot C_m$, 
hence the anomalous dimension is given as
%However, the dependence on $\mu$ genuinely reside in the combination of $H\cdot C_m$, 
%hence the renormalization behavior can be immediately found as 
\be 
\label{ghm}
\gamma_{H_m} = \frac{1}{H_m} \frac{d}{d\ln\mu} H_m = - \frac{\as C_F}{2\pi} \Bigl(\ln \frac{\mu^2}{m^2} + \frac{5}{2} \Bigr). 
\ee
It is consistent with the result of the direct computation of the static heavy-to-light current~\cite{Bauer:2000ew,Bauer:2000yr,Chay:2002vy}. 

Note that the overall description of the low energy behavior remains unchanged. 
%since the lower energy mode cannot recognize the high energy behavior.  -- Unnecessary 
Therefore, the combination of the factorized functions at the low scale, $f_{hl}$, is obtained from
%remains the same as 
Eq.~\eqref{fhlNLO} by replacing $Q^2( = p_+p'_-)$ with $m^2$. And the sum of the anomalous dimension for $f_{hl}$ and $H_m$ cancels. 
%in this case cancels Eq.~\eqref{ghm}.

In the static heavy quark sector, the residual interactions after integrating out the offshellness $\sim m^2$ 
become soft and the mode scales as $p_s \sim m(\zeta,\zeta,\zeta)$ with $\zeta \sim \mu/m \ll 1$. 
Therefore, $B_{\bar{n}}$ and $Y_{n\bar{n}}$ in Eq.~\eqref{hlfac2} with the same rapidity scale $\nu_- \sim m\zeta$ are combined 
into the new soft function $S_{hl}$ as\footnote{ 
Throughout this paper, we use the terminology ``ultrasoft'' to distinguish the mode from the soft mode with $p \sim (m,m,m)$ in 
describing the boosted heavy quark. In the static heavy quark decay, since the mode with 
the offshellness $\sim m^2$ is regarded as a hard mode, we  call the mode with $p\sim m(\zeta,\zeta,\zeta)$ as the soft mode. } 
 
\be
\label{shl}
S_{hl} (\mu,\nu_+) = B_{\bar{n}}(p'_-\to m, m^2, \mu, \nu_-) Y_{n\bar{n}} (\mu, \nu_+,\nu_-). 
\ee
This combination is illustrated in Fig.~\ref{hlfact}-(b). 
The bare one-loop result is given by 
\be 
S_{hl}^{(1)} (\mu,\nu_+) = \frac{\alpha_s C_F}{4\pi} \Bigl[ \frac{1}{\euv^2} -\frac{1}{\eir^2} +\Bigl( \frac{1}{\euv}
-\frac{1}{\eir} \Bigr) \Bigl( -\frac{2}{\eta_+} -2 \ln \frac{\nu_+}{\mu} +1\Bigr) \Bigr].
\ee

Finally, in terms of the quantities in Eqs.~\eqref{ham} and \eqref{shl}, we factorize the form factor for the static heavy quark decay as 
\be 
\label{Fhlst}
F_{hl}^{\mr{stat}} (m^2,\mu) = H_m (m^2,\mu) C_n (p_+,\mu,\nu_+) S_{hl} (\mu,\nu_+). 
\ee
When we sum over the soft and collinear contributions in Eq.~\eqref{Fhlst}, the rapidity divergence cancels, and it ends up with
the previous results~\cite{Bauer:2000ew,Bauer:2000yr,Chay:2002vy}.
However, the full resummation of the large logarithm of $\mu/m$ is completed by the rapidity evolution 
from $m$ to $\mu$, where $\mu <<m$. 
We show the resummed result in Appendix.~\ref{reshl} with the nonzero gluon mass $M \sim \mu$.

%Therefore the $\nu$ evolutionis not present in previous analyses though the resummation is numerically negligible for the evolution from $m(\sim  5$ GeV) to, say, $\mu\sim 1$ GeV. However small it is, it is theoretically consistent to include the $\nu$ evolution.

%As a result, the  final result of Eq.~\eqref{finfac} for the heavy-to-light current, which was directly obtained from SCET~\cite{Chay:2002vy}. And 
%each factorized part can be systematically related to the factorized parts in successive effective theories.
% 

\section{Conclusions\label{con}}
We have presented a new perspective regarding the origin of the rapidity divergence in SCET and its consistent treatment. The
approach is based on applying the very idea of the effective theory to a system with a hierarchy of rapidities, that is, a system 
of collinear particles with large rapidity
and soft particles with small rapidity, but with the same offshellness.  The effective theory with soft modes is obtained by integrating out the 
collinear modes with large rapidity. The matching procedure exactly amounts to the zero-bin subtraction for the collinear sector
in SCET. 
%In the naive collinear calculation, the rapidity divergence occurs only in the low-rapidity region, 
%but it is cancelled by the zero-bin subtraction converting the divergence from small rapidity to large rapidity. 
Since the rapidity divergence in the collinear sector comes from the subtraction of the overlapped soft contribution, 
the rapidity divergence in the soft sector has the opposite sign, which automatically guarantees the cancellation of the
rapidity divergence in the total contribution. 
This is comparable to the usual matching procedure in the conventional effective theory with the degrees of freedom separated 
by the renormalization scale $\mu$.
%that separates degrees of freedom by the ordinary renormalization scale $\mu$. 

In order to treat the rapidity divergences without any ambiguity, 
the main point in extracting the divergence in each sector is that we have to trace the same configurations both in the collinear and in the soft sectors. For example, the $n$-collinear Wilson line $W_n$ is obtained by the emission of the $n$-collinear gluons from the other part of the current, 
say, the $\overline{n}$-collinear sector from the back-to-back current. Therefore the corresponding soft sector in the matching, or the zero-bin 
subtraction, should come from the soft Wilson line $S_{\bar{n}}$ due to the emission of soft gluons from the $\overline{n}$-collinear sector,  not $S_n$ which 
is built by the emission of the soft gluons from the $n$-collinear sector.

We employ the same rapidity regulator of the form $(\nu/\overline{n}\cdot k)^{\eta}$ both in the soft sector and in the collinear sector
to regulate the rapidity divergence at large rapidity. When the current is not back-to-back, or when there is a soft quark involved, the 
same rapidity regulator is employed, but we use the appropriate expressions conforming to the regulator for the collinear part.
In this process, the directional dependence enters in the soft rapidity regulator and it is essential in extracting the 
Lorentz invariants in the full theory
when combined with the energy dependence 
from the collinear sector. Furthermore, unless the correct directional dependence
is incorporated, an additional UV divergence is induced because the directional dependence can appear in the coefficients 
of the $1/\euv$ pole~\cite{Bertolini:2017efs}. 
With our prescription, the directional dependence is correctly implemented without any problem. 

By extending the treatment of the rapidity divergence, we can associate independent rapidity scales for each collinear direction. Since physics
should be independent of the rapidity divergence in any collinear direction, its cancellation in the combination of the soft and the collinear 
contributions gives a severe constraint on the structure of the effective theory.  
On the practical side, when the factorized collinear and soft parts share the
same size of the rapidity scale, they can be combined to yield a new single function. It is illustrated in obtaining the soft function of the Sudakov 
form factor from the soft-collinear current, which is also confirmed by explicit calculations. This can be applied to various physical processes,
in which the rapidity scales can be varied depending on physics so that part of the factorized parts can be combined.  

We have extended the analysis on the rapidity divergence to the boosted heavy quark sector and found that the same mechanism of the divergence works. 
The resultant $\nu$ evolution enables us to fully resum large logarithms of $Q/m$ in the heavy quark form factors.   
Due to the presence of the heavy quark mass, the structure of the heavy quark form factors becomes more complicated and intriguing. 
We have consistently factorized the form factors matching onto $\mathrm{SCET_M}$ and bHQET successively. 

%We have extended the analysis on the rapidity divergence and the resultant resummation due to the $\nu$ evolution to the boosted heavy quark sector
%by considering the heavy-to-heavy and the heavy-to-light currents. Due to the presence of the new scale of the heavy quark mass, the structure of the effective
%theories becomes more interesting. We have shown the relationship between factorized quantities in $\mathrm{SCET_M}$ and bHQET, and HQET. 
%One important thing to note is that in studying the form factors, there exists rapidity divergence in the collinear and the soft sectors at each stage of the effective theory. 
%And the resummation
%of the large logarithms by the $\nu$ evolution should be performed to have a consistent effective theory, however small the effect %of the resummation may be. 

In SCET, it is important to include the evolution with respect to the rapidity scale because it yields the correct and the full 
resummation of the large logarithms when the rapidity divergence is involved.  Therefore the deep understanding of the rapidity 
divergence is essential. And we hope that our analysis will be a cornerstone in developing deeper insight on the rapidity divergence. 
%and we have explained the origin of the rapidity divergence, proposed how to extract it consistently in each sector.  
Our future plan is to apply this analysis systematically to various processes, with a variety of physical observables and beyond NLO calculations.

\appendix
 
\section{Resummation of large logarithms in Sudakov Form Factor\label{sudakov}}
We have presented the detailed computation of the Sudakov form factor for the back-to-back and the soft-collinear currents at one loop.
When pure dimensional regularization is employed with massless on-shell particles, the anomalous dimension with respect
to the rapidity scale is IR divergent, which means that the Sudakov form factor is not factorizable, and unphysical. 
But we can consider the Sudakov form factor  by putting a nonzero gauge boson mass.
We resum large logarithms by employing the RG evolution with respect to the renormalization and the rapidity scales. It will be explicitly 
shown that the order of the $\mu$-evolution and the $\nu$-evolution is irrelevant.

We analyze the Sudakov form factor in terms of the back-to-back collinear current, 
and the soft-collinear current respectively.  The second case can be obtained
from the first by boosting to one of the collinear directions. Due to the Lorentz invariance,
the Sudakov form factor and its evolution should be the same in both reference frames. 
However, the main theoretical interest here is that there are 
two different rapidity scales $\nu_{\pm}$ for 
the two collinear directions in the back-to-back current,\footnote{
Multiple rapidity scale evolution for the back-to-back case has been also discussed in Refs.~\cite{Jaiswal:2015nka,Scimemi:2018xaf}
}
 while there is only one rapidity scale $\nu_+$ for one collinear direction 
in the soft-collinear current. 
Therefore the RG evolution in the two cases look seemingly different. We illustrate how the RG evolution 
is performed, and show that the resummed results for both cases are the same.
 
We also perform the resummation of large logarithms in the form factor for the heavy-to-heavy current,
where the incoming and the outgoing heavy quarks are boosted and they move in opposite direction. We show that 
the full resummation of the large logarithm of $Q/m$ is completed by the rapidity evolution. 
Finally, we present the resummed result of the form factor for the heavy-to-light current.

\subsection{Sudakov form factor for the light back-to-back current}
\label{A1}

The amplitude of the back-to-back current at leading order is matched onto SCET as 
\be
\langle p|J^{\mu}|p'\rangle = \langle p | \bar{q}_n \gamma^{\mu} q_{\bar{n}} | p' \rangle = H(Q^2, \mu) V_{\mr{SCET}}^{\mu} (\mu),
\ee
where $V_{\mr{SCET}}^{\mu}$ is the SCET amplitude for the current in Eq.~\eqref{EFTcur}. 
The hard function $H(Q^2,\mu)$ at NLO in $\as$ is given by Eq.~\eqref{hardh}, which is the same in the soft-collinear current. 
Since $V_{\mr{SCET}}^{\mu}$ is factorized, the Sudakov form factor in full QCD is written as 
\be 
\label{Fbtb}
F(Q^2;M^2) = H(Q^2,\mu) C_{n} (p_+,\mu,\nu_+;M^2) C_{\bar{n}} (p'_-,\mu,\nu_-;M^2) S_{n\bar{n}}(\mu,\nu_+,\nu_-;M^2),
\ee
where $Q^2 = p_+ p'_-$ with the nonzero gauge boson mass $M$. From Eqs.~\eqref{MSc}, 
\eqref{Mnol} and \eqref{Mnbol}, the renormalized NLO results for $C_n$, $C_{\bar{n}}$ and $S_{n\bar{n}}$ to one loop read 
\bea 
\label{Cn}
C_{n} (p_+,\mu,\nu_+;M^2) &=& 1+ \frac{\as C_F}{4\pi} \Bigl[\ln\frac{\mu^2}{M^2} \Bigl(2\ln\frac{\nu_+}{p_+}+\frac{3}{2}\Bigr)
+\frac{9}{4} - \frac{\pi^2}{3} \Bigr], \\
\label{Cn}
C_{\bar{n}} (p^{\prime}_-,\mu,\nu_-;M^2) &=& 1+ \frac{\as C_F}{4\pi} \Bigl[\ln\frac{\mu^2}{M^2} \Bigl(2\ln\frac{\nu_-}{p'_-}
+\frac{3}{2}\Bigr)+\frac{9}{4} - \frac{\pi^2}{3} \Bigr], \\
\label{Snnbm}
S_{n\bar{n}}(\mu,\nu_+,\nu_-;M^2) &=&1+
\frac{\as C_F}{4\pi} \Bigl( \ln^2\frac{\mu^2}{M^2}-2\ln\frac{\mu^2}{M^2} \ln\frac{\nu_+\nu_-}{M^2}
-\frac{\pi^2}{6}\Bigr). 
\eea

The $\mu$ and $\nu$ anomalous dimensions are defined as
\begin{equation}
\label{RGEs}
\gamma^{\mu} = \frac{1}{f} \frac{d}{d\ln\mu} f,~~~ \gamma^{\nu} = \frac{1}{f} \frac{d}{d\ln\nu} f,
\end{equation}
for a given renormalized function $f$. 
Using Eq.~\eqref{RGEs}, we perform the RG evolutions to NLL accuracy, which resums large logarithms of order $\mc{O}(1)$ to 
all orders in $\as$.
%\footnote{Here we do power-counting on the large logarithm as $\mc{O}(1/\as)$.}
Here our power counting amounts to $\as L \sim \mc{O}(1)$,  where $L$ denotes a large logarithm.
The anomalous dimensions for the collinear and the soft parts are given as\footnote{\baselineskip 3.0ex 
Here the expressions for the $\nu$-anomalous dimensions $\gamma^{\nu}$ in Eq.~\eqref{gamnu} hold only for $\mu \sim M$. 
}
\begin{eqnarray}
&&\gamma_n^{\mu} = \frac{1}{2} \Gamma_C (\alpha_s) \ln \frac{\nu_+^2}{p_+^{2}} +\hat{\gamma}_c, \ 
\gamma_{\bar{n}}^{\mu} = \frac{1}{2} \Gamma_C (\alpha_s) \ln \frac{\nu_-^2}{p_-^{\prime 2}} +\hat{\gamma}_c, \
\gamma_s^{\mu} = \Gamma_C (\alpha_s) \ln \frac{\mu^2}{\nu_+ \nu_-}, 
  \label{gammu}\\
&&\gamma_n^{\nu_+} = 
\gamma_{\bar{n}}^{\nu_-} = \frac{1}{2} \Gamma_C (\alpha_s) \ln \frac{\mu^2}{M^2}, 
\  \ \gamma_s^{\nu_+} =\gamma_s^{\nu_-} = -\frac{1}{2} \Gamma_C (\alpha_s) 
\ln \frac{\mu^2}{M^2},  \label{gamnu}
\end{eqnarray}
where $\hat{\gamma}_c  = 3\alpha_s C_F/(4\pi)$. 
$\Gamma_C (\as)$ is the cusp anomalous dimension~\cite{Korchemsky:1987wg,Korchemskaya:1992je}, which can be expanded 
in powers of $\alpha_s$ as 
$\Gamma_{C} = \sum_{k=0} \Gamma_{k}(\as/4\pi)^{k+1}$. To NLL accuracy, we need its first two coefficients, 
\be
\Gamma_{0} = 4C_F,~~~\Gamma_{1} = 4C_F \Bigl[\bigl(\frac{67}{9}-\frac{\pi^2}{3}\bigr) C_A - \frac{10}{9} n_f\Bigr].
\ee
The anomalous dimension for the hard function $H(Q^2,\mu)$ is given by
\begin{equation}
\gamma_H = -\Gamma_c (\alpha_s) \ln \frac{\mu^2}{Q^2} +\hat{\gamma}_h,
\end{equation}
where $\hat{\gamma}_h = -6\alpha_s C_F/(4\pi) = -2\hat{\gamma}_c$. As a consistency check, note that the following sums 
of the anomalous dimensions are zero.
\bea
&&\gamma_H + \gamma_n^{\mu} + \gamma_{\bar{n}}^{\mu} +\gamma_s^{\mu} =0, \\ 
&&\gamma_n^{\nu_+} + \gamma_s^{\nu_+} =0,\  \ \gamma_{\bar{n}}^{\nu_-} + \gamma_s^{\nu_-} =0.
\eea

In Eq.~\eqref{Fbtb} the scales $\mu$ and $\nu_{\pm}$ in each factorized function are the factorization scales. 
In order to resum large logarithms, we need to consider the RG evolutions of the factorized functions from the factorization scales to their ``characteristic scales''. 
Here the characteristic scales mean the scales minimizing the large logarithms in the factorized function, 
and they are given by    
\bea
\mu_h \sim Q,\ \ \mu_n \sim \mu_{\bar{n}} \sim \mu_s \sim M, \\
\nu_n^{+} \sim p_+,\ \ \nu_{\bar{n}}^{-} \sim  p'_-,\ \ \nu_s^{\pm} \sim M. 
\eea

Evolving $\mu$ from the characteristic scale to the factorization scale $\mu_f$, the factorized functions are expressed as 
\begin{eqnarray}
H(Q^2, \mu_f) &=& U_H (\mu_f, \mu_h) H(Q^2, \mu_h),  \label{hevol}\\
C_n(p_+, \mu_f,\nu_+;M^2)  &=& U_n (\mu_f,\mu_n; \nu_+)C_n(p_+, \mu_n,\nu_+;M^2), \\ 
C_{\bar{n}}(p'_-, \mu_f,\nu_-;M^2)  &=& U_{\bar{n}} (\mu_f,\mu_{\bar{n}}; \nu_-) C_{\bar{n}} (p'_-, \mu_{\bar{n}},\nu_-;M^2)  , \\
S_{n\bar{n}}(\mu_f,\nu_{\pm};M^2) &=& U_s(\mu_f, \mu_s;\nu_{\pm}) S_{n\n}(\mu_s,\nu_{\pm};M^2).
\end{eqnarray}
The evolution kernels to NLL accuracy are given as
\begin{eqnarray} \label{evolfac}
U_H (\mu_f, \mu_h) &=& \exp \Bigl[ -2 S_{\Gamma} (\mu_f, \mu_h) - a[\Gamma_C] (\mu_f, \mu_f)  \ln \frac{\mu_f^2}{Q^2} 
+a[\hat{\gamma}_h] (\mu_f, \mu_f) \Bigr],  \label{uevol}\\
U_n (\mu_f,\mu_n; \nu_+)  &=& \exp \Bigl[a[\Gamma_C] (\mu_f, \mu_n) \ln \frac{\nu_+}{p_+} +a[\hat{\gamma}_c] (\mu_f, \mu_n) \Bigr],  \\
U_{\bar{n}} (\mu_f,\mu_{\n}; \nu_-)  &=& \exp \Bigl[a[\Gamma_C] (\mu_f, \mu_{\n}) \ln \frac{\nu_-}{p'_-} 
+a[\hat{\gamma}_c] (\mu_f, \mu_{\n}) \Bigr],\\
\label{softmevo}
U_s(\mu_f, \mu_s;\nu_{\pm}) &=& \exp \Bigl[2S_{\Gamma} (\mu_f, \mu_s) + a[\Gamma_C] (\mu_f, \mu_s) 
\ln \frac{\mu_f^2}{\nu_+ \nu_-}\Bigr],
\end{eqnarray}
where $S_{\Gamma}$ and $a[f]$ are defined as 
\be
S_{\Gamma} (\mu_1,\mu_2) = \int^{\alpha_s(\mu_1)}_{\as(\mu_2)} \frac{d\as}{\beta(\as)} \Gamma_{C}(\as) \int^{\as}_{\as(\mu_1)}
\frac{d\as'}{\beta(\as')}, ~~
a[f](\mu_1,\mu_2) = \int^{\as(\mu_1)}_{\as(\mu_2)} \frac{d\as}{\beta(\as)} f(\as).
\ee
Here $\beta(\as)$ is the QCD beta function $\beta(\as)=d\as/d\ln\mu=-2\as\sum_{k=0}\beta_k (\as/4\pi)^{k+1}$. 

Evolving $\nu$ to the factorization scale $\nu_f$, the factorized functions are given as  
\begin{eqnarray}
C_n(p_+, \mu,\nu_f^+;M^2)  &=& V_n (\nu_f^+,\nu_n^+; \mu) C_n(p_+, \mu,\nu_n^+;M^2)  , \\ 
C_{\bar{n}}(p'_-, \mu,\nu_f^-;M^2)  &=& V_n (\nu_f^-,\nu_{\bar{n}}^-; \mu) C_{\bar{n}} (p'_-, \mu,\nu_{\bar{n}}^-;M^2)  , \\
\label{softnevo}
S_{n\bar{n}}(\mu,\nu_f^{\pm};M^2) &=& V_S(\nu_f^{\pm}, \nu_s^{\pm};\mu) S_{n\bar{n}}(\mu,\nu_s^{\pm};M^2),
\end{eqnarray}
where the evolution kernels to NLL accuracy are given as
\begin{eqnarray}
V_n(\nu_f^+, \nu_n^+;\mu) &=&\exp \Bigl[\ln \frac{\nu_f^+}{\nu_n^+} \cdot a[\Gamma_C](\mu,M)\Bigr],  \\
V_{\n}(\nu_f^-, \nu_{\n}^-;\mu) &=&\exp \Bigl[ \ln \frac{\nu_f^-}{\nu_{\n}^-} \cdot a[\Gamma_C](\mu,M)\Bigr], \\
V_S (\nu_f^{\pm}, \nu_s^{\pm};\mu) &=& \exp \Bigl[ -\ln \frac{\nu_f^{\pm}}{\nu_s^{\pm}} \cdot a[\Gamma_C](\mu,M) \Bigr]. 
\end{eqnarray}
In order to fully resum the large logarithms of $\mu/M$, the following relation is used~\cite{Chiu:2012ir}: 
\be
\frac{d}{d\ln\mu} \gamma^{\nu} = \frac{d}{d\ln\nu} \gamma^{\mu} \propto \Gamma_C(\as).
\ee
The $\nu$-anomalous dimensions in Eq.~\eqref{gamnu} with a large logarithm is  modified as 
\be
\gamma_{n}^{\nu_+}=\gamma_{\bar{n}}^{\nu_-} = a[\Gamma_C](\mu,M), 
\  \ \gamma_s^{\nu_+} =\gamma_s^{\nu_-} = -a[\Gamma_C](\mu,M).
\ee

In resumming the large logarithms, we may choose various paths for the evolutions with respect to 
$\mu$ and $\nu$~\cite{Chiu:2012ir}. Let us first consider the  $\nu$ evolution with fixed $\mu_f$, and 
then the $\mu$ evolution at $\nu_i = \{\nu_n^+,\nu_{n}^-,\nu_s^{\pm}\}$. In this case the overall $\nu$-evolution kernel is given as 
\bea
\ln V (\mu_f) &\equiv& \ln [V_n(\nu_f^+,\nu_n^+;\mu_f)\cdot V_S(\nu_f^+,\nu_s^+;\mu_f)] 
+ \ln [V_{\bar{n}}(\nu_f^-,\nu_{\bar{n}}^-;\mu_f)\cdot V_S(\nu_f^-,\nu_s^-;\mu_f)] \nnb \\
\label{V1st}
&=& \ln \frac{\nu_s^+ \nu_s^-}{\nu_n^+\nu_{\bar{n}}^-} \cdot a[\Gamma_C](\mu_f,M).   
\eea 
Then we perform the $\mu$ evolution for the factorized functions at $\nu_i = \{\nu_n^+,\nu_{\bar{n}}^-,\nu_s^{\pm}\}$. The 
overall $\mu$-evolution kernel is given as 
\begin{eqnarray}
\ln U (\nu_i) &\equiv& \ln U_H (\mu_f, \mu_h) +\ln U_n (\mu_f, \mu_n;\nu_n^+) +\ln U_{\bar{n}} (\mu_f,\mu_{\bar{n}}; 
\nu_{\bar{n}}^-) +\ln U_s (\mu_f, \mu_s; \nu_s^{\pm}) \nonumber \\
\label{U1st}
&=& 2S_{\Gamma} (\mu_h, \mu_s) + a[\hat{\gamma}_c] (\mu_h, \mu_n) + a[\hat{\gamma}_c] (\mu_h, \mu_{\bar{n}}) 
+ \ln \frac{\mu_h^2}{Q^2}\cdot a[\Gamma_C] (\mu_h, \mu_s)  \\
&&+ \ln \frac{\nu_n^+ \nu_{\bar{n}}^-}{\nu_s^+ \nu_s^-}\cdot  a[\Gamma_C] (\mu_f, \mu_s)  - \ln \frac{\nu_n^+}{p_+}
\cdot a[\Gamma_C] (\mu_n, \mu_s)
- \ln \frac{\nu_{\bar{n}}^-}{p'_-}\cdot a[\Gamma_C] (\mu_{\bar{n}}, \mu_s). \nnb
\end{eqnarray}

Combining Eqs.~\eqref{V1st} and \eqref{U1st}, we  obtain the full exponentiation resumming the large logarithms to NLL accuracy. 
It is given as  
\bea 
&&\ln U(\nu_i) V(\mu_f) = 2S_{\Gamma} (\mu_h, \mu_s) + a[\hat{\gamma}_c] (\mu_h, \mu_n) + a[\hat{\gamma}_c] 
(\mu_h, \mu_{\bar{n}}) + \ln \frac{\mu_h^2}{Q^2}\cdot a[\Gamma_C] (\mu_h, \mu_s)  \nnb \\
\label{exp1}
&&~~~\ \ \ - \ln \frac{\nu_n^+ \nu_{\bar{n}}^-}{\nu_s^+ \nu_s^-}\cdot  a[\Gamma_C] (\mu_s,M)  - \ln \frac{\nu_n^+}{p_+}
\cdot a[\Gamma_C] (\mu_n, \mu_s) - \ln \frac{\nu_{\n}^-}{p'_-}\cdot a[\Gamma_C] (\mu_{\bar{n}}, \mu_s).
\eea
Note that the dependence on the factorization scales is shown to be cancelled explicitly. In Eq.~\eqref{exp1}, the first line is the result of 
the $\mu$ evolution only, and the second line reflects the rapidity scale evolution.

Since the renormalization scales $\mu_{n,\bar{n},s}$ are given by $\sim M$, we can ignore the last two terms in the second line of 
Eq.~\eqref{exp1} at NLL accuracy because they are of order $\mc{O}(\as)$. 
Then the second line in Eq.~\eqref{exp1} can be approximated as 
\be
\label{nulog}
- \ln \frac{\nu_n^+ \nu_{\n}^-}{\nu_s^+ \nu_s^-}\cdot  a[\Gamma_C] (\mu_s,M) \approx - \ln \frac{\nu_n^+ 
\nu_{\n}^-}{\nu_s^+ \nu_s^-} \cdot \Gamma_C (\as) \ln\frac{\mu_s}{M} \approx - \frac{\as (\mu_s) C_F}{2\pi} \ln \frac{Q^2}{M^2} 
\ln \frac{\mu_s^2}{M^2}.
\ee
This is the most important missing ingredient in the exponentiation when we do not perform the rapidity 
evolution. 

To check the path independence in the RG evolutions, let us consider a different path for the evolution. 
We perform the $\mu$ evolution at the scales $\nu_f^{\pm}$ first, and then the $\nu$ evolutions with 
the scales $\mu_i = \{\mu_h,\mu_{n},\mu_{\bar{n}},\mu_s\}$.  In this case, the overall $\mu$-evolution kernel is given by 
\begin{eqnarray}
\ln U (\nu_f^{\pm}) 
&=& 2S_{\Gamma} (\mu_h, \mu_s) + a[\hat{\gamma}_c] (\mu_h, \mu_n) + a[\hat{\gamma}_c] (\mu_h, \mu_{\bar{n}}) 
+ \ln \frac{\mu_h^2}{Q^2}\cdot a[\Gamma_C] (\mu_h, \mu_s)  \nnb \\
\label{U2nd}
&& - \ln \frac{\nu_f^+}{p_+}\cdot a[\Gamma_C] (\mu_n, \mu_s)
- \ln \frac{\nu_{f}^-}{p'_-}\cdot a[\Gamma_C] (\mu_{\bar{n}}, \mu_s). 
\end{eqnarray}
And the overall $\nu$-evolution kernel at $\mu_i$ is obtained as 
\bea
&&\ln V (\mu_i) \equiv \ln [V_n(\nu_f^+,\nu_n^+;\mu_n)\cdot V_S(\nu_f^+,\nu_s^+;\mu_s)] 
+ \ln [V_{\bar{n}}(\nu_f^-,\nu_{\bar{n}}^-;\mu_{\bar{n}})\cdot V_S(\nu_f^-,\nu_s^-;\mu_s)] \nnb \\
\label{V2nd}
&&~~~\ \ \ = \ln \frac{\nu_f^+}{\nu_n^+} \cdot a[\Gamma_C](\mu_n,M) + \ln \frac{\nu_f^-}{\nu_{\bar{n}}^-} 
\cdot a[\Gamma_C](\mu_{\bar{n}},M) -\ln \frac{\nu_f^+ \nu_f^-}{\nu_s^+\nu_s^-} \cdot a[\Gamma_C](\mu_s,M).   
\eea 

Combining Eqs.~\eqref{U2nd} and \eqref{V2nd}, the complete evolution kernel is given as 
\bea 
&&\ln V(\mu_i) U(\nu_f^{\pm}) = 2S_{\Gamma} (\mu_h, \mu_s) + a[\hat{\gamma}_c] (\mu_h, \mu_n) + a[\hat{\gamma}_c] 
(\mu_h, \mu_{\bar{n}}) + \ln \frac{\mu_h^2}{Q^2}\cdot a[\Gamma_C] (\mu_h, \mu_s)  \nnb \\
\label{exp2}
&&~~~\ \ \ - \ln \frac{\nu_n^+ \nu_{\bar{n}}^-}{\nu_s^+ \nu_s^-}\cdot  a[\Gamma_C] (\mu_s,M)  - \ln \frac{\nu_n^+}{p_+}
\cdot a[\Gamma_C] (\mu_n, \mu_s) - \ln \frac{\nu_{\bar{n}}^-}{p'_-}\cdot a[\Gamma_C] (\mu_{\bar{n}}, \mu_s).
\eea
It is the same as Eq.~\eqref{exp1}, confirming the path independence of the RG evolutions with respect to $\mu$ and $\nu$. 

\subsection{Sudakov form factor for the soft-collinear current}

The Sudakov form factor with the soft-collinear current is given in  Eq.~\eqref{Sudsc}, and is factorized as 
\be 
\label{Sudscm}
F(Q^2,M^2) = H(Q^2,\mu) C_n (p_+,\mu,\nu_+;M^2) S_q(p'_-,\mu,\nu_+;M^2),
\ee
where $Q^2 = p_+p'_-$,  $p_+ \sim Q^2/M$ and $p'_- \sim M$. The soft function $S_q$ at NLO in $\as$ is given by
\bea
\label{reSq} 
S_q(p'_-,\mu,\nu_+;M^2) = 1+\frac{\alpha_s C_F}{4\pi} \Bigl[  \ln^2 \frac{\mu^2}{M^2}
 -2\ln \frac{\mu^2}{M^2}  \ln \frac{\nu_+ p'_-}{M^2} +\frac{9}{4} -\frac{\pi^2}{2}\Bigr].
\eea

This soft function also can be obtained by combining $C_{\bar{n}}$ and $S_{n\bar{n}}$ from the back-to-back current as [See Eq.~\eqref{factsq}.] 
\be
S_q(p'_-,\mu,\nu_+;M^2) = C_{\bar{n}}(p'_-,\mu,\nu_-;M^2) S_{n\bar{n}}(\mu,\nu_+,\nu_-;M^2).
\ee
Therefore the resummed result for Eq.~\eqref{Sudscm} can be obtained from the back-to-back case,  
starting from Eq.~\eqref{exp1}. By identifying $\mu_{\bar{n}} = \mu_s$ and $\nu_{\bar{n}}^- = \nu_s^-$, we obtain the resummed 
result for the soft-collinear current. The exponentiation factor at NLL accuracy leads to 
\bea 
\ln F(Q^2,M^2) &=& 2S_{\Gamma} (\mu_h, \mu_s) + a[\hat{\gamma}_c] (\mu_h, \mu_n) + a[\hat{\gamma}_c] 
(\mu_h, \mu_{s}) + \ln \frac{\mu_h^2}{Q^2}\cdot a[\Gamma_C] (\mu_h, \mu_s)  \nnb \\
\label{exp3}
&& - \ln \frac{\nu_n^+ }{\nu_s^+ }\cdot  a[\Gamma_C] (\mu_s,M)  - \ln \frac{\nu_n^+}{p_+}\cdot a[\Gamma_C] (\mu_n, \mu_s),
\eea
where $\mu_n \sim \mu_s \sim M$. Here the last term is power counted as $\mc{O}(\as)$ and ignored, 
and the second line in Eq.~\eqref{exp3} can be approximated as 
\be
- \ln \frac{\nu_n^+ }{\nu_s^+ }\cdot  a[\Gamma_C] (\mu_s,M) \approx - \ln \frac{Q^2/M}{M} \cdot 
\Gamma_C (\as) \ln\frac{\mu_s}{M} \approx - \frac{\as C_F}{2\pi} \ln \frac{Q^2}{M^2} \ln \frac{\mu_s^2}{M^2}. 
\ee
This result is the same as the case for the back-to-back current in Eq.~\eqref{nulog}. Therefore we confirm that 
the resummed result for the soft-collinear current is identical to that for the back-to-back current. 
It should hold because the Sudakov form factor is Lorentz invariant, but here we have proved explicitly the relation between a single-scale evolution
and a two-scale evolution.
 
\subsection{Sudakov form factor for the heavy-to-heavy current}
\label{resumhh}

The factorized Sudakov form factor for the heavy-to-heavy current at the scale $\mu \ll m$ 
from Eqs.~\eqref{hhffmat} and \eqref{facthh2} is given as 
\bea
\label{FhhM}
F(Q^2,m^2;M^2) &=& H(Q^2,\mu) [C_m(m^2,\mu)]^2 \\
&\times& B_n (p_+,m^2, \mu, \nu_+;M^2) B_{\bar{n}} (p_-^{\prime},m^2, \mu, \nu_-;M^2) Y_{n\bar{n}}(\mu, \nu_+, \nu_-;M^2). \nnb
\eea
Here the nonzero gluon mass scaling as $M\ll m$ is employed to regularize the IR divergence. 
The evolution of the hard function $H$ from the hard scale $\mu_h \sim Q$ to the factorization scale $\mu$ is given by Eqs.~\eqref{hevol} and \eqref{uevol}. 
And the evolutions of the usoft function $Y_{n\bar{n}}$ are the same as the soft function $S_{n\bar{n}}$ 
for the form factor from the light quarks in Eq.~\eqref{Fbtb}. So the $\mu$- and $\nu$-evolution kernels 
are given by Eq.~\eqref{softmevo} and \eqref{softnevo} respectively. 

To NLL accuracy, the $\mu$-anomalous dimensions for $[C_m]^2$ and $B_{n,\bar{n}}$ are listed as 
\bea 
\gamma_m &=& \Gamma_C (\alpha_s) \ln\frac{\mu^2}{m^2} + \hat{\gamma}_m,  \\
\gamma_{cs}^{\mu} &=& \Gamma_C (\alpha_s) \Bigl(-\frac{1}{2} \ln\frac{\mu^2}{M^2}
+\ln \frac{\nu_+ m}{p_+ M}\Bigr)+\hat{\gamma}_{cs}, \\
\gamma_{\cs}^{\mu} &=& \Gamma_C (\alpha_s) \Bigl(-\frac{1}{2} \ln\frac{\mu^2}{M^2}
+\ln \frac{\nu_- m}{p'_- M}\Bigr)+\hat{\gamma}_{cs},
\eea
where $\hat{\gamma}_m = \hat{\gamma}_{cs} = \as C_F/(2\pi)$. 
The $\nu$-anomalous dimensions for $B_{n,\bar{n}}$ are given as 
\be
\gamma_{cs}^{\nu_+} = \gamma_{\cs}^{\nu_-} = a[\Gamma_C](\mu,M). 
\ee

Using the anomalous dimensions, we evolve the factorized functions from the factorization scales 
$(\mu_f,\nu_f^{\pm})$ to their characteristic scales to minimize large logarithms.  
The $\mu$-evolution kernels are given as
\bea
\label{Um}
\ln U_m (\mu_f, \mu_m) &=&  2S_{\Gamma} (\mu_f, \mu_m) +  \ln \frac{\mu_f^2}{m^2} a[\Gamma_C](\mu_f,\mu_m) + a[\hat{\gamma}_m](\mu_f,\mu_m),\\
\label{Ucs}
\ln U_{cs} (\mu_f, \mu_{cs}; \nu_+) &=& -S_{\Gamma} (\mu_f,\mu_{cs}) + \ln \frac{\nu_+ m}{p_+ \mu_f} a[\Gamma_C] (\mu_f, \mu_{cs}) +a[\hat{\gamma}_{cs}](\mu_f,\mu_{cs}),  \\
\label{Ucsb}
\ln U_{\cs} (\mu_f, \mu_{\cs};\nu_-) &=& -S_{\Gamma} (\mu_f,\mu_{\cs}) + \ln \frac{\nu_- m}{p'_- \mu_f} a[\Gamma_C] (\mu_f, \mu_{\cs}) +a[\hat{\gamma}_{cs}](\mu_f,\mu_{\cs}),  
\eea
where $\mu_m \sim m$ and $\mu_{cs}\sim \mu_{\cs} \sim M$.
And the $\nu$-evolution kernels are given by 
\be
\label{Vcs}
\ln V_{cs} (\nu_f^+, \nu_{cs}^+;\mu) =\ln \frac{\nu_f^+}{\nu_{cs}^+} a[\Gamma_C](\mu,M), ~~~
\ln V_{\cs}(\nu_f^-, \nu_{\cs}^-;\mu) = \ln \frac{\nu_f^-}{\nu_{\cs}^-} a[\Gamma_C](\mu,M), 
\ee
where $\nu_{cs}^+ \sim p_+ M/m$ and $\nu_{\cs}^- \sim p'_- M/m$. 

Applying Eqs.~(\ref{Um}-\ref{Vcs}) and using the evolved results of $H$ and $Y_{n\bar{n}}$, 
we obtain the resummed result of the heavy-to-heavy form factor. %, as in  Appendix.~\ref{A1}. 
The evolution of the form factor to NLL accuracy is given as 
\bea 
&&\ln F(Q^2,m^2;M^2) = 2S_{\Gamma} (\mu_h, \mu_m)+ S_{\Gamma} (\mu_{cs}, \mu_{us}) +S_{\Gamma} (\mu_{\cs}, \mu_{us}) +\ln\frac{\mu_h^2}{Q^2} ~a[\Gamma_C] (\mu_h,\mu_m) \nnb \\
&&~~~ + \ln\frac{m^2}{Q^2}~a[\Gamma_C] (\mu_m,\mu_{us})+\ln\frac{p_+ \mu_{cs}}{\nu_{cs}^+ m}~a[\Gamma_C]
 (\mu_{cs},\mu_{us})+ \ln\frac{p'_- \mu_{\cs}}{\nu_{\cs}^- m}~a[\Gamma_C] (\mu_{\cs},\mu_{us}) \nnb \\
\label{hhFres1}
&&~~~+\ln\frac{\nu_{us}^+\nu_{us}^-}{\nu_{cs}^+\nu_{\cs}^-}~a[\Gamma_C] (\mu_{us},M)
+\frac{C_F}{\beta_0} \Bigl(\ln \frac{\as(\mu_{m})}{\as(\mu_h)}+\frac{\as(\mu_{cs})}{\as(\mu_h)}+\frac{\as(\mu_{\cs})}{\as(\mu_h)}\Bigr). 
\eea
By suppressing the terms that are power-counted as $\mc{O}(\as)$ due to the fact that $\mu_{cs} \sim \mu_{\cs} \sim \mu_{us} \sim M$, 
Eq.~\eqref{hhFres1} reads 
\bea
&&\ln F(Q^2,m^2;M^2) = 2S_{\Gamma} (\mu_h, \mu_m) +\ln\frac{m^2}{Q^2}~a[\Gamma_C] (\mu_m,\mu_{us})+\ln\frac{\mu_h^2}{Q^2} ~a[\Gamma_C] (\mu_h,\mu_m) \nnb \\
\label{hhFres2}
&&~~~
+\frac{C_F}{\beta_0} \Bigl(\ln \frac{\as(\mu_{m})}{\as(\mu_h)}
+\ln\frac{\as(\mu_{cs})}{\as(\mu_h)}+\ln\frac{\as(\mu_{\cs})}{\as(\mu_h)}\Bigr)
+\ln\frac{\nu_{us}^+\nu_{us}^-}{\nu_{cs}^+\nu_{\cs}^-}~a[\Gamma_C] (\mu_{us},M),  
\eea
where the first two terms in the right side are $\mc{O}(1/\as)$ and the remaining terms are power-counted as $\mc{O}(1)$. 
The last term in Eq.~\eqref{hhFres2} is the result of the rapidity evolution. It can be written as 
\be 
\ln\frac{\nu_{us}^+\nu_{us}^-}{\nu_{cs}^+\nu_{\cs}^-}~a[\Gamma_C] (\mu_{us},M) \sim \ln\frac{m^2}{p_+p'_-} 
a[\Gamma_C] (\mu_{us},M) \approx \frac{\as C_F}{2\pi} \ln\frac{m^2}{Q^2} \ln \frac{\mu_{us}^2}{M^2}.  
\ee
It shows that the full resummation of $\ln Q/m$ in the form factor can be completed only after the rapidity  evolution. 

\subsection{Sudakov form factor for the heavy-to-light current}
\label{reshl}
With the nonzero gluon mass $M \sim \mu$, the resummation of the heavy-to-light form factor in
 Eqs.~\eqref{hlfac1} and \eqref{hlfac2} can be systematically performed by evolving with respec to $\mu$ and $\nu_{\pm}$. 
It is straightforward to follow the same procedure as in the light-to-light and the heavy-to-heavy form factors, 
 %It would be straightforward if we apply the similar resummation procedure with the case the light-to-light and 
 %the heavy-to-heavy form factor. So we do not discuss its detail. 

When the form factor in SCET is obtained at the scale $\mu \sim m$, the $\mu$- and $\nu$-renormalization behavior 
for each factorized function in Eqs.~\eqref{hlfac1} is the same as that of the light-to-light current. 
Therefore the resummed result is given by Eqs.~\eqref{exp1} and \eqref{exp2}. For  $\mu \ll m$, 
using the factorized result in Eq.~\eqref{hlfac2} with the gluon mass $M \ll m$, we obtain 
the resummed result at NLL accuracy as 
\begin{align}
\ln F_{hl} (Q^2,m^2;M^2) &= 2S_{\Gamma} (\mu_h, \mu_{us}) - S_{\Gamma} (\mu_m, \mu_{\cs})
+\ln\frac{\mu_h^2}{Q^2}~a[\Gamma_C] (\mu_h,\mu_{us}) \nnb \\
&-
\frac{1}{2}\ln\frac{\mu_m^2}{m^2} ~a[\Gamma_C] (\mu_m,\mu_{\cs})+\frac{C_F}{\beta_0} \Bigl(\frac{1}{2}\ln \frac{\as(\mu_m)}{\as(\mu_h)}+\ln\frac{\as(\mu_{\cs})}{\as(\mu_h)}+\frac{3}{2}\ln\frac{\as(\mu_{n})}{\as(\mu_h)}\Bigr) \nnb \\
\label{hlres1}
&+
\ln\frac{\nu_{us}^+\nu_{us}^-}{\nu_{n}^+ p'_-}~a[\Gamma_C] (\mu_{us},M)
-\ln\frac{\nu_{\cs}^-}{p'_-}~a[\Gamma_C] (\mu_{\cs},M),  
\end{align}
where the characteristic scales for the factorized functions are given by 
\bea 
&& \mu_n\sim \mu_{us} \sim \mu_{\cs} \sim M, \nnb \\
&&\nu_n^+ \sim p_+ \sim Q,~~\nu_{\cs}^- \sim \frac{p'_- M}{m} \sim \frac{Q M}{m},~~\nu_{us}^{\pm} \sim M.  
\eea
The last two terms in Eq.~\eqref{hlres1} result from the rapidity evolution. Because $\mu_{us} \sim \mu_{\cs}\sim M$, 
they can be simplified further as 
\bea 
&&\ln\frac{\nu_{us}^+\nu_{us}^-}{\nu_{n}^+ p'_-}~a[\Gamma_C] (\mu_{us},M)
-\ln\frac{\nu_{\cs}^-}{p'_-}~a[\Gamma_C] (\mu_{\cs},M) \nnb \\
&&~~~
\sim \Bigl(\ln\frac{\nu_{us}^+}{\nu_{n}^+}+\ln\frac{\nu_{us}^-}{\nu_{\cs}^-}\Bigr)
a[\Gamma_C] (\mu_{us},M) 
\approx \frac{\as(\mu_{us}) C_F}{2\pi}  
\Bigl(\ln\frac{M}{Q}+\ln\frac{m}{Q}\Bigr) \ln \frac{\mu_{us}^2}{M^2}.
\eea

From the resummed result in Eq.~\eqref{hlres1}, we can consider the resummation of the form factor for the static heavy quark decay. 
Suppose that the light quark has a maximal energy, i.e., $p_+ = m$. Then $Q^2 = 2p\cdot p'$ is 
given by $m^2$, and $\mu_{h,m}$ in Eq.~\eqref{hlres1} becomes $\mu_h =\mu_m \sim m$.
 The characteristic usoft and csoft scales in Eq.~\eqref{hlres1} are determined as 
 $\mu_{us} = \mu_{\cs} \sim M$ and $\nu_{us}^- = \nu_{\cs}^- \sim M$. 
And the scales are unified as the soft scales for the static heavy quark decay.   
Therefore the resummed result for the static heavy quark is given as 
\bea
\ln F_{hl} (m^2;M^2) &=& S_{\Gamma} (\mu_h, \mu_s) 
+\frac{1}{2} \ln\frac{\mu_h^2}{m^2}~a[\Gamma_C] (\mu_h,\mu_{s}) \nnb \\
\label{hlres2}
&&+\frac{C_F}{\beta_0} \Bigl(\ln \frac{\as(\mu_s)}{\as(\mu_h)}+\frac{3}{2}\ln\frac{\as(\mu_{n})}{\as(\mu_h)}\Bigr)
+\ln\frac{\nu_{s}^+}{\nu_{n}^+}~a[\Gamma_C] (\mu_{s},M).  
\eea
Here $\mu_s \sim M$, and the last term is the missing ingredient unless we consider the rapidity evolution. 
Since $\nu_n^+ \sim m$ and $\nu_s^+ \sim M$, it can be expressed as 
\be 
\ln\frac{\nu_{s}^+}{\nu_{n}^+}~a[\Gamma_C] (\mu_{s},M) \approx \frac{\as(\mu_s) C_F}{2\pi}  
\ln\frac{M}{m} \ln \frac{\mu_{s}^2}{M^2}. 
\ee

\section{One-loop calculations in the boosted heavy quark sector}% to extract rapidity divergence}
\label{appB}

\subsection{Collinear one-loop calculation with the soft zero-bin mode $p_s^{\mu} \sim (m,m,m)$} 
\label{appBa}

When the soft mode with $p_s^{\mu} \sim (m,m,m)$ is decoupled from the boosted heavy quark sector, 
we have to subtract the soft contribution from the naive collinear calculation for the heavy quark. 
The regular collinear one-loop calculation is given by $M_n^{m} = \tilde{M}_n^m - M_n^{\varnothing}$, 
as shown in Eq.~\eqref{colol1}. It reads 
\bea 
M_n^{m} &=& - \frac{\as C_F}{2\pi} \frac{(\mu^2 e^{\gamma_{\mathrm{E}}} )^{\eps}}{\Gamma(1-\eps)} \Biggl[\int^1_0 \frac{dx}{x} (1-x)  \int^{\infty}_0 \frac{d\blp{k}^2 (\blp{k}^2)^{-\eps}}{\blp{k}^2+x^2 m^2} 
- \int^1_0 \frac{dx}{x} \int^{\infty}_0 d\blp{k}^2 (\blp{k}^2)^{-1-\eps} \Biggr] \nnb \\
\label{Mam} 
&&+\frac{\as C_F}{2\pi} \frac{(\mu^2 e^{\gamma_{\mathrm{E}}} )^{\eps}}{\Gamma(1-\eps)} \Bigl(\frac{\nu_+}{p_+}\Bigr)^{\eta_+} \int^{\infty}_1 dx x^{-1-\eta_+} \int^{\infty}_0 d\blp{k}^2 (\blp{k}^2)^{-1-\eps}.  
\eea 
In the right side of Eq.~\eqref{Mam}, the first term in the square bracket is the naive contribution~$(\tilde{M}_n^m)$, and the 
remaining terms are the contributions from the zero-bin subtraction~($-M_n^{\varnothing}$). 
Since the rapidity divergence arises as $x\to \infty$, we put the regulator only in the term in the second line of Eq.~\eqref{Mam}. 

We denote the first line and the second line of the right side in Eq.~\eqref{Mam} as $M_n^{m,A}$ and $M_n^{m,B}$ respectively. Then  
$M_n^{m,A}$ can be reorganized, and can be written as 
\bea 
\label{MamA}
M_n^{m,A} &=& - \frac{\as C_F}{2\pi} \frac{(\mu^2 e^{\gamma_{\mathrm{E}}} )^{\eps}}{\Gamma(1-\eps)} \Biggl[
- \int^1_0 dx \int^{\infty}_0 \frac{d\blp{k}^2 (\blp{k}^2)^{-\eps}}{\blp{k}^2+x^2 m^2} 
\\
&&~~~~~~ +\int^1_0 \frac{dx}{x} \int^{\infty}_0 d\blp{k}^2 (\blp{k}^2)^{-\eps} \Bigl(\frac{1}{\blp{k}^2+x^2m^2} - \frac{1}{\blp{k}^2} \Bigr) \Biggl]. \nnb 
\eea 
Here the first term in the square bracket gives a UV pole as $\blp{k}^2$ goes to infinity. 
In the second term,  only the IR divergence survives. 
%Both the UV and the IR divergences are regularized as poles in $\eps$, but they can be clearly separated. 
Then  $M_n^{m,A}$ is given by 
\be
\label{MamA1}
M_n^{m,A} = \frac{\as C_F}{2\pi} \Bigl[\frac{1}{\UV} +\ln\frac{\mu^2}{m^2} +\frac{1}{2\IR}\Bigl(\frac{1}{\IR}+\ln\frac{\mu^2}{m^2}\Bigr) 
+ \frac{1}{4} \ln^2\frac{\mu^2}{m^2}+2+\frac{\pi^2}{24} \Bigr].
\ee
It is straightforward to compute the term in the second line of Eq.~\eqref{Mam}, and is given as 
\be 
\label{MamB}
M_n^{m,B} = \frac{\as C_F}{2\pi} \Bigl(\frac{1}{\eta_+} + \ln \frac{\nu_+}{p_+}\Bigr)\Bigl(\frac{1}{\UV}-\frac{1}{\IR}\Bigr). 
\ee
Eq.~\eqref{colol1} is obtained by combining Eqs.~\eqref{MamA1} and \eqref{MamB}. 

\subsection{Csoft one-loop calculation with the usoft zero-bin mode $p_{us}^{\mu} \sim \zeta(m,m,m)$} 
\label{appBb}

In this subsection, we compute the irreducible one-loop contribution to the csoft function $B_n$. 
The naive one-loop contribution ($\tilde{M}_{cs}^m$) and the zero-bin contribution ($M_{cs}^{\varnothing}$) are 
listed in Eqs.~\eqref{ncsoftol} and \eqref{zbinus}. And the regular contribution with the zero-bin subtraction, 
$M_{cs}^m = \tilde{M}_{cs}^m - M_{cs}^{\varnothing}$,  can be written as 
\bea
\label{cs1} 
M_{cs}^m &=& - \frac{\as C_F}{2\pi} \frac{(\mu^2 e^{\gamma_{\mathrm{E}}} )^{\eps}}{\Gamma(1-\eps)} \Biggl\{
\int^1_0 \frac{dx}{x} \Biggl[\int^{\infty}_0 \frac{d\blp{k}^2 (\blp{k}^2)^{-\eps}}{\blp{k}^2+x^2 m^2} 
-  \int^{\infty}_0 d\blp{k}^2 (\blp{k}^2)^{-1-\eps} \Biggr] \  \\
&& +\int^{\infty}_1 \frac{dx}{x} \int^{\infty}_0 \frac{d\blp{k}^2 (\blp{k}^2)^{-\eps}}{\blp{k}^2+x^2 m^2} 
-\Bigl(\frac{\nu_+}{p_+}\Bigr)^{\eta_+} \int^{\infty}_1 dx x^{-1-\eta_+} \int^{\infty}_0 d\blp{k}^2 (\blp{k}^2)^{-1-\eps}\Biggr\},
\nnb \\
&\equiv& M_{cs}^{m,A} +  M_{cs}^{m,B}+M_{cs}^{m,C}. \nnb 
\eea
Here we divide the integration region of $x$ into $x\in [0,1]$ and $x\in [1,\infty]$. 
And we denote $M_{cs}^{m,A}$,  $M_{cs}^{m,B}$ and $M_{cs}^{m,C}$ as the contributions from the first, 
the second and the third terms in the curly brackets respectively.

The contribution $M_{cs}^{m,A}$ for $x\in [0,1]$ involves only the IR divergence 
since the UV divergence as $\blp{k}^2 \to \infty$ is cancelled. It is given by
\be
\label{cs2}
M_{cs}^{m,A} = \frac{\as C_F}{4\pi} \frac{1}{\IR} \Bigl(\frac{1}{\IR} + \ln \frac{\mu^2}{m^2} \Bigr). 
\ee
The contribution $M_{cs}^{m,B}$ is UV-divergent, and it is given by 
\be
\label{cs3}
M_{cs}^{m,B} = -\frac{\as C_F}{4\pi} \frac{1}{\UV} \Bigl(\frac{1}{\UV} + \ln \frac{\mu^2}{m^2} \Bigr). 
\ee
The contribution $M_{cs}^{m,C}$ comes from the zero-bin subtraction and includes the rapidity divergence. 
Since we compute it with on-shell regularization and the massless gluon, the result is the same as Eq.~\eqref{MamB}. 
Finally, combining Eqs.~\eqref{cs2}, \eqref{cs3} and \eqref{MamB}, we obtain Eq.~\eqref{mcsol}.

 \begin{acknowledgments}
J. Chay is supported by Basic Science Research Program through the National Research Foundation of Korea (NRF) funded by 
the Ministry of Education(Grant No. NRF-2019R1F1A1060396). C.~Kim was supported by Basic Science Research Program through the 
National Research Foundation of Korea (NRF) funded by the Ministry of Science and ICT (Grant No. NRF-2017R1A2B4010511).
\end{acknowledgments}

%%%%%%%%%%%%%%%%%%%%%%%%%%%%%%%%%%%%%%%%%%%%%%%%%%%%%%%%%%%%%%%%%%%%%%

%\phantomsection
%\addcontentsline{toc}{section}{References}

\bibliographystyle{JHEP1}
\bibliography{rapidity}

%%%%%%%%%%%%%%%%%%%%%%%%%%%%%%%%%%%%%%%%%%%%%%%%%%%%%%%%%%%%%%%%%%%%%%

\end{document}